\documentclass[12pt,a4paper,final]{iopart}

\usepackage{iopams}  
\usepackage{graphicx}
\usepackage[breaklinks=true,colorlinks=true,linkcolor=blue,urlcolor=blue,citecolor=blue]{hyperref}
\usepackage{braket}
\usepackage{xcolor}

\begin{document}

\title[The Uniform Electron Gas in the High Density Regime]{\textit{Ab initio} path integral Monte Carlo simulation of the Uniform Electron Gas in the High Energy Density Regime}

\author{Tobias Dornheim}
\address{Center for Advanced Systems Understanding (CASUS), G\"orlitz, Germany}
\ead{t.dornheim@hzdr.de}

\author{Zhandos A. Moldabekov$^{1,2}$}
\address{$^1$Institute for Experimental and Theoretical Physics, Al-Farabi Kazakh National University, 050040 Almaty, Kazakhstan}
\address{$^2$Institute of Applied Sciences and IT, 050038 Almaty, Kazakhstan}
%\ead{author.two@mail.com}

\author{Jan Vorberger}
\address{Helmholtz-Zentrum Dresden-Rossendorf (HZDR), 01328 Dresden, Germany}
%\eads{\mailto{author.three@mail.com}, \mailto{author.three@gmail.com}}

\author{Simon Groth}
\address{Institut f\"ur Theoretische Physik und Astrophysik, Christian-Albrechts-Universit\"at, 24098 Kiel, Germany}

\begin{abstract}
The response of the uniform electron gas (UEG) to an external perturbation is of paramount importance for many applications. Recently, highly accurate results for the static density response function and the corresponding local field correction have been provided both for warm dense matter [\textit{J.~Chem.~Phys.}~\textbf{151}, 194104 (2019)] and strongly coupled electron liquid [\textit{Phys.~Rev.~B}~\textbf{101}, 045129 (2020)] conditions based on exact \textit{ab initio} path integral Monte Carlo (PIMC) simulations. In the present work, we further complete our current description of the UEG by exploring the high energy density regime, which is relevant for, e.g., astrophysical applications and inertial confinement fusion experiments. To this end, we present extensive new PIMC results for the static density response in the range of $0.05 \leq r_s \leq 0.5$ and $0.85\leq\theta\leq8$. These data are subsequently used to benchmark the accuracy of the widely used random phase approximation and the dielectric theory by Singwi, Tosi, Land, and Sj\"olander (STLS). Moreover, we compare our results to configuration PIMC data where they are available and find perfect agreement with a relative accuracy of $0.001-0.01\%$.
All PIMC data are available online.
\end{abstract}

\vspace{2pc}
\noindent{\it Keywords}: Uniform Electron Gas, Path Integral Monte Carlo, Density Response
%\submitto{\}
%\maketitle

\section{Introduction}

The uniform electron gas (UEG), also known as \textit{jellium} or quantum one-component plasma, is defined as a system of Coulomb interacting electrons in a neutralizing homogeneous background and constitutes one of the most import model systems in physics and chemistry~\cite{quantum_theory,loos,review}.
Having been introduced as a simple model description for conduction electrons in metals~\cite{mahan}, the UEG exhibits a variety of interesting effects, such as Wigner crystallization~\cite{wigner,drummond_wigner,ichimaru_wigner} and a possible incipient excitonic mode which has been predicted to appear at low density~\cite{takada1,takada2,dornheim_dynamic,dynamic_folgepaper}.

Moreover, the availability of highly accurate quantum Monte Carlo (QMC) data~\cite{gs2,moroni2,spink,ortiz1,ortiz2} at metallic densities and zero-temperature has sparked remarkable developments in many fields, most notably the hitherto unequaled success of density functional theory (DFT) regarding the description of real materials~\cite{dft_review,burke_perspective}.

Recently, the interest in matter under extreme conditions~\cite{fortov_review} has led to the need for analogous quantum Monte Carlo data at finite temperature. Of particular importance is the so-called warm dense matter (WDM) regime, which is defined by two characteristic parameters that are both of the order of one: i) the density parameter $r_s=\overline{r}/a_\textnormal{B}$ (with $\overline{r}$ and $a_\textnormal{B}$ being the average inter-particle distance and first Bohr radius) and ii) the reduced temperature $\theta=k_\textnormal{B}T/E_\textnormal{F}$ (with $k_\textnormal{B}$ and $E_\textnormal{F}$ being the Boltzmann constant and Fermi energy~\cite{quantum_theory}), cf.~Fig.~\ref{fig:overview} below. These conditions occur in astrophysical objects like giant planet interiors~\cite{knudson,militzer3,manuel} and are expected to manifest on the pathway towards inertial confinement fusion~\cite{hu_ICF}. Furthermore, WDM is nowadays routinely realized in large research facilities around the globe like the Linac Coherent Light Source (LCLS)~\cite{lcls1}, the National Ignition Facility (NIF)~\cite{moses}, and the European XFEL~\cite{xfel1} which came into operation only recently. A topical review article where different experimental techniques on WDM are introduced can be found in Ref.~\cite{falk_wdm}.

From a theoretical perspective~\cite{new_POP,wdm_book}, the description of WDM is rendered difficult by the simultaneous presence of Coulomb correlations (ruling out weak-coupling expansions like Green functions~\cite{vanLeeuwen,niclas,kwong,kas1,kas2,kas3}), thermal excitations (ruling out ground-state methods like diffusion Monte Carlo~\cite{wmc_review}), and quantum degeneracy effects (ruling out classical and semi-classical methods like molecular dynamics~\cite{MD}), leaving computationally expensive thermal QMC methods as the most promising choice~\cite{review,dornheim_pop,brown_chapter}. Yet, QMC simulations of warm dense electrons have been prevented for decades by the infamous fermion sign problem (FSP, see Ref.~\cite{dornheim_sign_problem} for a topical review article), which leads to an exponential increase in computation cost with increasing system size or decreasing temperature. In particular, the widely used path integral Monte Carlo (PIMC) method~\cite{imada,pollock,cep} cannot access significant parts of the WDM regime, which sparked the need for new developments.

This demand was met recently by a surge of new developments in the field of fermionic QMC simulations at finite temperature, e.g., Refs.~\cite{brown_ethan,schoof_prl,malone1,malone2,dornheim,dornheim2,groth,dornheim3,dubois,dornheim_prl}. On the one hand, Dornheim and co-workers~\cite{dornheim,dornheim2,dornheim_neu} have introduced the so-called permutation blocking PIMC (PB-PIMC) method, which significantly alleviates the FSP in the standard PIMC method, and, thus, is more efficient when quantum degeneracy effects are important. On the other hand, two independent groups have simultaneously developed two QMC methods that are complementary to PIMC and PB-PIMC, as they are efficient at high density and strong degeneracy, but break down with increasing coupling strength, where (PB-)PIMC simulations are comparably easy. Both methods are similar in spirit, since they are formulated in antisymmetric Fock space and use determinants of plane waves as basis sets. The density matrix QMC method~\cite{blunt,malone1} constitutes the extension of the full configuration interaction QMC method~\cite{booth} to finite temperature and can be viewed as a realization of diffusion Monte Carlo in second quantization. In contrast, the configuration PIMC (CPIMC) approach~\cite{schoof_cpp,schoof_prl,groth} can be viewed as a Metropolis Monte Carlo~\cite{metropolis} evaluation of the exact infinite perturbation expansion around the ideal system, and is more similar to real-space PIMC.

The bottom line is that the combination of CPIMC at high density with PB-PIMC at strong coupling has allowed for the first full description of the warm dense UEG in the range of $0\leq r_s \leq 20$ with an unprecedented accuracy of $\sim0.3\%$~\cite{groth_prl,review}. More specifically, Groth and co-workers~\cite{groth_prl} have parametrized the exchange-correlation free energy $f_\textnormal{xc}$ of the UEG, which, in principle, gives access to all thermodynamic quantities of the UEG and can be used for thermal DFT calculations~\cite{mermin_dft,rajagopal_dft} on the level of the local density approximation~\cite{kushal}. Yet, one crucial piece regarding the behaviour of warm dense electrons was still missing: the response of the UEG to an external perturbation, which is described by the density response function~\cite{kugler1}
\begin{eqnarray}\label{eq:define_LFC}
\chi({q},\omega) = \frac{ \chi_0({q},\omega) }{ 1 - 4\pi/q^2\big[1-G({q},\omega)\big]\chi_0({q},\omega)} \quad .
\end{eqnarray}
Here $\chi_0(q,\omega)$ denotes the density response function of the ideal (i.e., noninteracting) system that is known from theory~\cite{quantum_theory}, and $G(q,\omega)$ is the local field correction (LFC) that includes the complete, wave number-resolved description of exchange--correlation effects. Consequently, the LFC is of paramount importance for many applications, such as the calculation of electrical and thermal conductivities~\cite{Desjarlais:2017,Veysman:2016}, the construction of effective potentials~\cite{ceperley_potential,zhandos1,zhandos2}, the incorporation of correlation effects into quantum hydrodynamics~\cite{diaw1,diaw2,zhandos_hydro,new_POP}, and the development of advanced exchange--correlation functionals for DFT~\cite{burke_ac,lu_ac,patrick_ac,goerling_ac}.

\begin{figure}\centering
\includegraphics[width=0.8\textwidth]{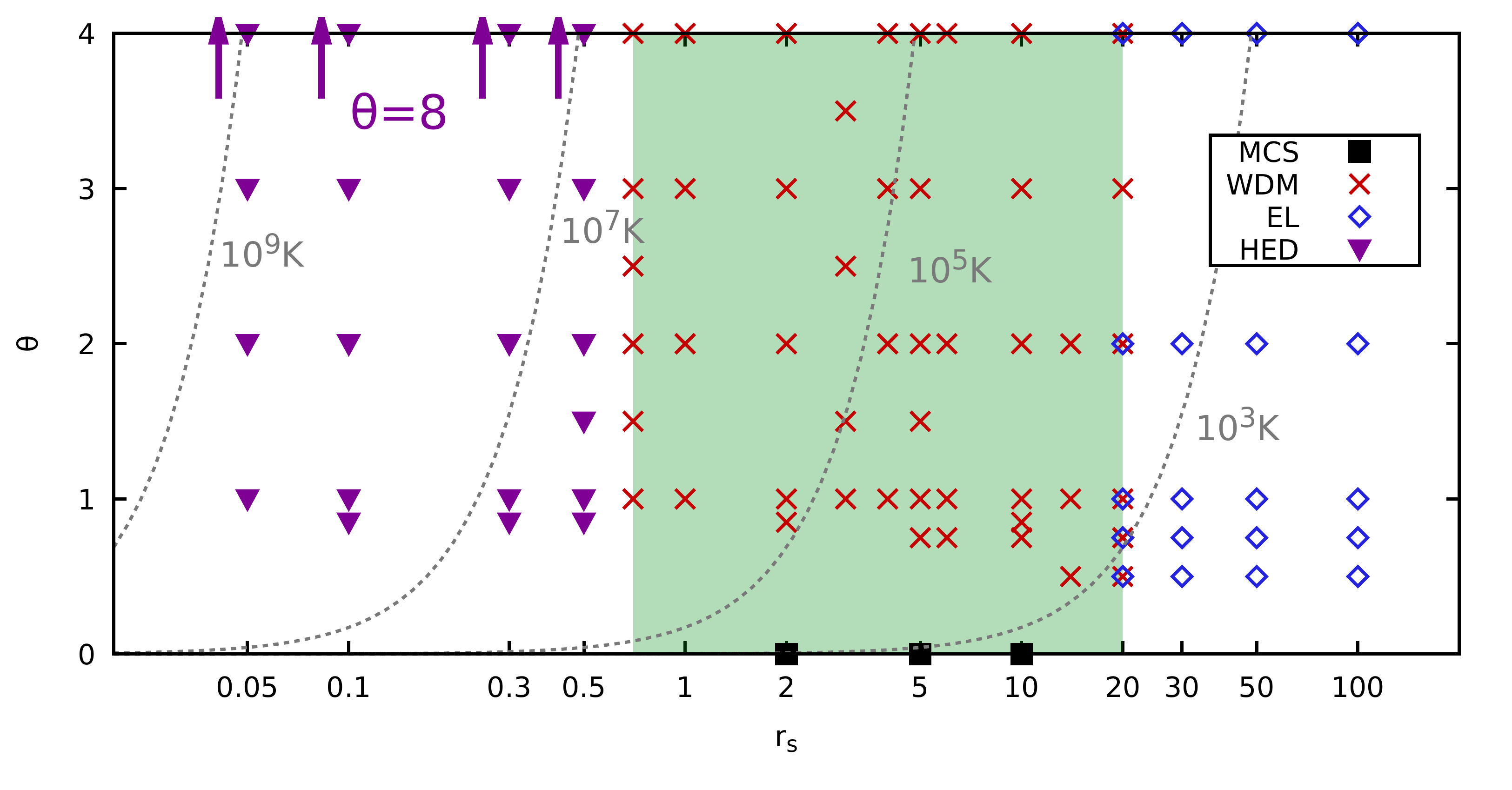}
\caption{\label{fig:overview}
Parameter overview of quantum Monte Carlo data for the static density response of the uniform electron gas in the $r_s$-$\theta$ plane. The shaded green area depicts the validity range of the recent machine-learning representation of the static local field correction of the warm dense electron gas (warm dense matter, WDM) by Dornheim \textit{et al.}~\cite{dornheim_ML} that combines PIMC data (red crosses) with the ground-state results by Moroni \textit{et al.}~\cite{moroni2}. The blue diamonds correspond to the strongly coupled electron liquid (EL) studied in Ref.~\cite{dornheim_electron_liquid}.
The purple triangles depict our new simulation data for the high-energy-density (HED) regime, with the upward arrows indicating PIMC results for $\theta=8$ that is available at $r_s=0.05, 0.1, 0.3$, and $0.5$. Moreover, the dashed grey lines correspond to lines of constant temperature.
}
\end{figure}

Very recently, we have achieved a major breakthrough by presenting a machine-learning based representation of the static LFC $G(q)=G(q,0)$ covering the entire relevant WDM regime, $0.7\leq r_s \leq 20$. This is illustrated in Fig.~\ref{fig:overview}, where we show the $r_s$-$\theta$-plane for the relevant parameters. The black squares at $r_s=2,5$ and $10$ correspond to the ground-state results by Moroni \textit{et al.}~\cite{moroni2} (MCS) and the red crosses to the new PIMC data from Ref.~\cite{dornheim_ML}. These two data sets (in addition to an analytic representation of the MCS data by Corradini \textit{et al.}~\cite{cdop}) were then used to train a deep neural network, which accurately predicts $G(q;r_s,\theta)$ over the entire shaded green area, i.e., the full WDM regime.

Shortly thereafter, two of us~\cite{dornheim_electron_liquid} have extended these efforts to even stronger coupling strength, $20 \leq r_s \leq 100$ (blue diamonds), where the UEG forms an electron liquid. In particular, these new data have been used to benchmark different dielectric theories~\cite{stls,stls2,tanaka_hnc}, with no approach being able to qualitatively reproduce all features of the QMC data.

Considering the density parameter $r_s$, this leaves two directions for further investigations: on the one hand, the UEG is expected to form a Wigner crystal~\cite{drummond_wigner} at even lower densities with $r_s\sim200$. While these conditions are currently beyond the capability of experiments, the onset of crystallization is interesting in its own right, and might help to shed light upon open questions such as the formation of a charge-density wave~\cite{dornheim_electron_liquid,iyetomi_cdw} and a possible excitonic collective mode~\cite{takada1,takada2,dornheim_dynamic,dynamic_folgepaper}. 
On the other hand, going into the other direction ($r_s\lesssim0.5$) brings us to the high-energy density (HED) regime, which is important for many applications.
For example, neutron star envelopes contain quantum plasmas of electrons and ions with $10^{-4}\leq r_s\leq 2$  and with the degeneracy parameter in the range of $0\leq\theta\lesssim 10$ \cite{Gudmundsson, Potekhin}. For instance, the outer envelope of a neutron star with a density $\rho=10^2~{\rm g~cm^{-3}}$,  surface temperature $10^6~\rm K$, and a surface gravity of $10^{14}~{\rm cm~s^{-2}}$ contains partially degenerate electrons with $r_s\simeq 0.4$ and $\theta\simeq 3$ \cite{Gudmundsson}. 
Another astrophysical example is the Solar core with $r_s\approx 0.2$ and $\theta\approx 1$ \cite{Fortov_book}. In addition, we mention ICF experiments, where dense plasmas with $0.1\lesssim r_s\lesssim 2$ and $\theta\sim 1$ are realized\cite{Atzeni, Fortov_book}. For example, in the high density shell around the hot spot, deuterium-tritium (DT)  plasmas with $r_s\simeq 0.4$ and  $\theta \simeq 0.9$, as well as $r_s\simeq 0.2$ and $\theta \simeq 0.84$ have been reported in experiments on a cryogenic DT implosion at the OMEGA facility~\cite{Hu, Boehly}.

To understand the evolution and physics of the aforementioned astrophysical objects and ICF plasmas, various approximate models have been developed to describe electronic thermodynamic \cite{Gudmundsson, Potekhin, tanaka_hnc} and transport properties \cite{Potekhin1999, Reinholz95, Reinholz2000, Fortmann2010} in the HED regime.
While perturbative expansions like the random phase approximation~\cite{pines} are expected to be accurate, benchmarks against more accurate \textit{ab initio} date are highly desirable. To this end, we have carried out extensive PIMC simulations of the UEG in the range of $0.05 \leq r_s \leq 0.5$ (cf.~the purple triangles in Fig.~\ref{fig:overview}) to compute the static density response function and, in this way, the static local field correction. These data are then compared against RPA and the finite-temperature extension~\cite{stls,stls2} of the dielectric approximation by Singiwi \textit{et al.}~\cite{stls_original} (STLS), which allows for an unambiguous quantification of the systematic error in both methods. Overall, we find systematic errors of up to a few per cent around intermediate wave numbers, which vanish both towards high temperature and small $r_s$, where our PIMC data and the dielectric results cannot be distinguished. This further illustrates the consistency of our PIMC approach to the static density response and corroborates the high quality of the results presented in Refs.~\cite{dornheim_ML,dornheim_electron_liquid,dynamic_folgepaper}.

The paper is organized as follows: in Sec.~\ref{sec:theory}, we introduce the required theoretical background, starting with a brief introduction to the standard PIMC method (Sec.~\ref{sec:pimc}), and how it can be used to compute $\chi(q)$ and $G(q)$ (Sec.~\ref{sec:response_theory}). The presentation of our simulation results in Sec.~\ref{sec:results} begins with a comprehensive discussion of finite-size effects (Sec.~\ref{sec:fsc}), where we also briefly touch upon the manifestation of the fermion sign problem over the different physical regimes of the UEG.
In Sec.~\ref{sec:data}, we present our extensive new PIMC data for $\chi(q)$ and $G(q)$, and investigate the systematic deviations to RPA and STLS. Lastly, we compare PIMC data for the static structure factor $S(q)$ against previous highly accurate CPIMC data~\cite{dornheim_cpp} and find excellent agreement with a statistical uncertainty of partly below $0.1\%$. A similar comparison for the density response function of noninteracting fermions is given in the same section.
The paper is concluded by a brief summary and outlook in Sec.~\ref{sec:summary}.

\section{Theory\label{sec:theory}}

\subsection{Path integral Monte Carlo\label{sec:pimc}}

Let us consider a system of $N=N_\uparrow+N_\downarrow$ unpolarized electrons (with $N_\uparrow=N/2$ and $N_\downarrow=N/2$ being the number of spin-up and -down electrons, respectively) at an inverse temperature $\beta=1/T$ in a cubic simulation cell of volume $V=L^3$ in thermodynamic equilibrium (we assume Hartree atomic units throughout this work).
In this case, all thermodynamic expectation values can be computed from the canonical partition function, which can be expressed in coordinate space as
\begin{eqnarray}\label{eq:Z}
 Z = \frac{1}{N_\uparrow!N_\downarrow!} \sum_{\sigma_\uparrow\in S_{N_{\uparrow}}}\sum_{\sigma_\downarrow\in S_{N_{\downarrow}}} \textnormal{sgn}^\textnormal{f}(\sigma_\uparrow,\sigma_\downarrow) 
 &
 \int \textnormal{d}\mathbf{R}\ \bra{ \mathbf{R} } e^{-\beta\hat{H}} \ket{ \hat{\pi}_{\sigma_\uparrow}\hat{\pi}_{\sigma_\downarrow}\mathbf{R}} \quad .
\end{eqnarray}
Note that the double sums and the corresponding permutation operators $\hat{\pi}_{\sigma_\downarrow}$ ensure the proper antisymmetry of Eq.~(\ref{eq:Z}) under the exchange of particle coordinates, with the sign function $\textnormal{sgn}^\textnormal{f}(\sigma_\uparrow,\sigma_\downarrow)$ being positive (negative) for an even (odd) number of pair exchanges.
For the standard path integral Monte Carlo method~\cite{imada,pollock,cep} as it is used throughout this work, one performs a Trotter decomposition of the density operator $\hat\rho=e^{-\beta\hat H}$ and eventually recasts Eq.~(\ref{eq:Z}) into the form
\begin{eqnarray}\label{eq:Z_W}
 Z = \int \textnormal{d}\mathbf{X}\ W(\mathbf{X}) \quad ,
\end{eqnarray}
which allows for the usual interpretation in terms of the classical isomorphism~\cite{chandler}: the partition function is given as the sum over all possible paths $\mathbf{X}$ in the imaginary time, and each path has to be taken into account with the appropriate configuration weight $W(\mathbf{X})$, which is a function that can be readily evaluated. We note that a more detailed derivation of Eq.~(\ref{eq:Z_W}) in particular and the PIMC method in general has already been presented elsewhere~\cite{cep,review,dornheim_permutation_cycles} and need not be repeated here.

The basic idea of the standard PIMC approach is to stochastically sample the paths $\mathbf{X}$ according to $P(\mathbf{X})=W(\mathbf{X})/Z$ using the Metropolis algorithm~\cite{metropolis}. For bosons (such as ultracold atoms, see, e.g., Refs.~\cite{supersolid,dornheim_superfluid,dynamic_alex2}) and boltzmannons (i.e., distinguishable particles~\cite{clark_casula,dornheim_analyzing}), $P(\mathbf{X})$ is strictly positive and quasi-exact simulations of up to $N\sim10^4$ particles are possible~\cite{boninsegni1,boninsegni2}.
For fermions (such as electrons, which we simulate in this work), on the other hand, $P(\mathbf{X})$ can be both positive and negative and, thus, cannot be interpreted as a probability distribution.

In order to still use the PIMC method, one must evaluate the ratio
\begin{eqnarray}\label{eq:fermionic_expectation_value}
 \braket{\hat A} = \frac{\braket{\hat A\hat S}'}{\braket{\hat S}'} \quad ,
\end{eqnarray}
where $\braket{\dots}'$ indicates the expectation value that is obtained by using the absolute value of the weight function $W(\mathbf{X})$. The denominator is typically simply denoted as the average sign $S$ and constitutes a measure for the degree of cancellations due to positive and negative contributions to $Z$. Moreover, it is straightforward to see that the sign exponentially decreases both with $\beta$ and the system size $N$, which is the origin of the notorious fermion sign problem ~\cite{troyer,loh}, see Ref.~\cite{dornheim_sign_problem} for a topical review article.
More specifically, the relative statistical uncertainty from a fermionic Monte Carlo calculation is inversely proportional to $S$~\cite{ceperley_fermions},
\begin{eqnarray}\label{eq:FSP}
 \frac{\Delta A}{A} \sim \frac{1}{S \sqrt{N_\textnormal{MC}}} \quad ,
\end{eqnarray}
which leads to an \textit{exponential wall} for large systems and low temperatures, which can only be compensated by increasing the number of Monte Carlo samples as $1/\sqrt{N_\textnormal{MC}}$.
In fact, Eq.~(\ref{eq:FSP}) constitutes the main obstacle in our work, and limits our simulations to $0.85 \leq \theta$ and $N\leq 34$, cf.~Fig.~\ref{fig:SIGN}.

For completeness, we mention that all simulation results shown in this work have been obtained using a canonical adaption~\cite{mezza} of the worm algorithm developed by Boninsegni \textit{et al.}~\cite{boninsegni1,boninsegni1}.

\subsection{Static density response\label{sec:response_theory}}

The central goal of the present work is to use the PIMC method to compute highly accurate data for the static density response function $\chi(q)$ and, if possible, the static local field correction $G(q)$, cf.~Eq.~(\ref{eq:define_LFC}).

In the ground state, this is done by actually simulating a harmonically perturbed system and subsequently quantifying the deviation towards the unperturbed uniform system~\cite{moroni2,bowen,moroni,bowen2,neutron}. Recently, this idea was extended to finite temperature in Refs.~\cite{dornheim_pre,groth_jcp} using the novel PB-PIMC and CPIMC methods. While the simulation of the inhomogeneous system does indeed give exact results for $\chi(q)$ and can be used even beyond the validity range of linear response theory, it suffers from a major drawback:
one has to perform separate simulations of the perturbed system for each individual value of the wave number $q$. In addition, one should vary the perturbation amplitude as well to ensure that linear response theory is still accurate.

For this reason, the first extensive results for the static density response function of the warm dense UEG~\cite{dornheim_ML,dynamic_folgepaper} were obtained by invoking the imaginary-time version of the fluctuation--dissipation theorem~\cite{bowen},
\begin{eqnarray}\label{eq:static_chi}
\chi({q}) = -n\int_0^\beta \textnormal{d}\tau\ F({q},\tau) \quad ,
\end{eqnarray}
with $F(q,\tau)$ being the usual intermediate scattering function~\cite{siegfried_review} evaluated at an imaginary time argument $\tau\in[0,\beta]$,
\begin{eqnarray}\label{eq:scattering_function}
F({q},\tau) &=& \frac{1}{N} \braket{\rho({q},\tau)\rho(-{q},0)} \quad .
\end{eqnarray}
In particular, Eq.~(\ref{eq:static_chi}) allows to compute the full wave-number dependence of the static density response function from a single PIMC simulation of the unperturbed UEG, which leads to a substantial speed-up.

As a side-note, we mention that the estimation of imaginary-time correlation functions such as Eq.~(\ref{eq:scattering_function}) is routinely done within the standard PIMC method, see Refs.~\cite{berne1,berne2,boninsegni_ceperley,dynamic_alex1} for a few examples. On the other hand, no estimator for $F(q,\tau)$ does yet exist for the CPIMC method, although first results for the Matsubara Green function were recently presented by Groth~\cite{groth_thesis}. Moreover, the PB-PIMC method, too, is not well suited to exploit Eq.~(\ref{eq:static_chi}), as it is only efficient when the number of imaginary-time slices is small. In that case, however, the integral along the $\tau$-axis is afflicted with a large discretization error, thereby substantially limiting the accuracy of the results. Therefore, the standard PIMC method as described in Sec.~\ref{sec:pimc}
constitutes the method of choice despite the more severe sign problem.

Once PIMC data for $\chi(q)$ is available, the static local field correction can be directly computed by solving Eq.~(\ref{eq:define_LFC}),
\begin{eqnarray}\label{eq:Get_G}
G(q) = 1 - \frac{q^2}{4\pi}\left( \frac{1}{\chi_0(q)} - \frac{1}{\chi(q)} \right).
\end{eqnarray}
We note that both the PIMC data for $\chi^N(q)$ and $\chi_0^N(q)$ are subject to a system-size dependence and, hence, cannot be used to describe the thermodynamic limit
\begin{eqnarray}
\chi(q) = \lim_{N\to\infty} \chi^N(q) \quad .
\end{eqnarray}
Luckily, Moroni \textit{et al.}~\cite{moroni,moroni2} have reported for the ground state that the corresponding finite-size local field correction $G^N(q)$ [which is obtained by inserting both $\chi_0^N(q)$ and $\chi^N(q)$ into Eq.~(\ref{eq:Get_G})] does not significantly depend on $N$, $G(q)\approx G^N(q)$. Recently, this assumption was empirically confirmed at finite temperature~\cite{groth_jcp}, and it does indeed hold in the HED regime studied here, see Figs.~\ref{fig:LFC_rs0.3_theta4} and \ref{fig:real_LFC_rs0.5_theta2}.

The corresponding static density response function in the thermodynamic limit can then be computed as
\begin{eqnarray}\label{eq:chi_FSC}
\chi(q) = \frac{ \chi_0(q) }{ 1 - 4\pi/q^2[1-G^N(q)]\chi_0(q)} \quad ,
\end{eqnarray}
with $\chi_0(q)$ being the ideal response function for $N\to\infty$, see, e.g., Ref.~\cite{quantum_theory}.

%\textcolor{red}{Mention momentum quantization in a finite simulation cell!}

\section{Results\label{sec:results}}

\subsection{Finite-size effects\label{sec:fsc}}

\begin{figure}\centering
\includegraphics[width=0.6\textwidth]{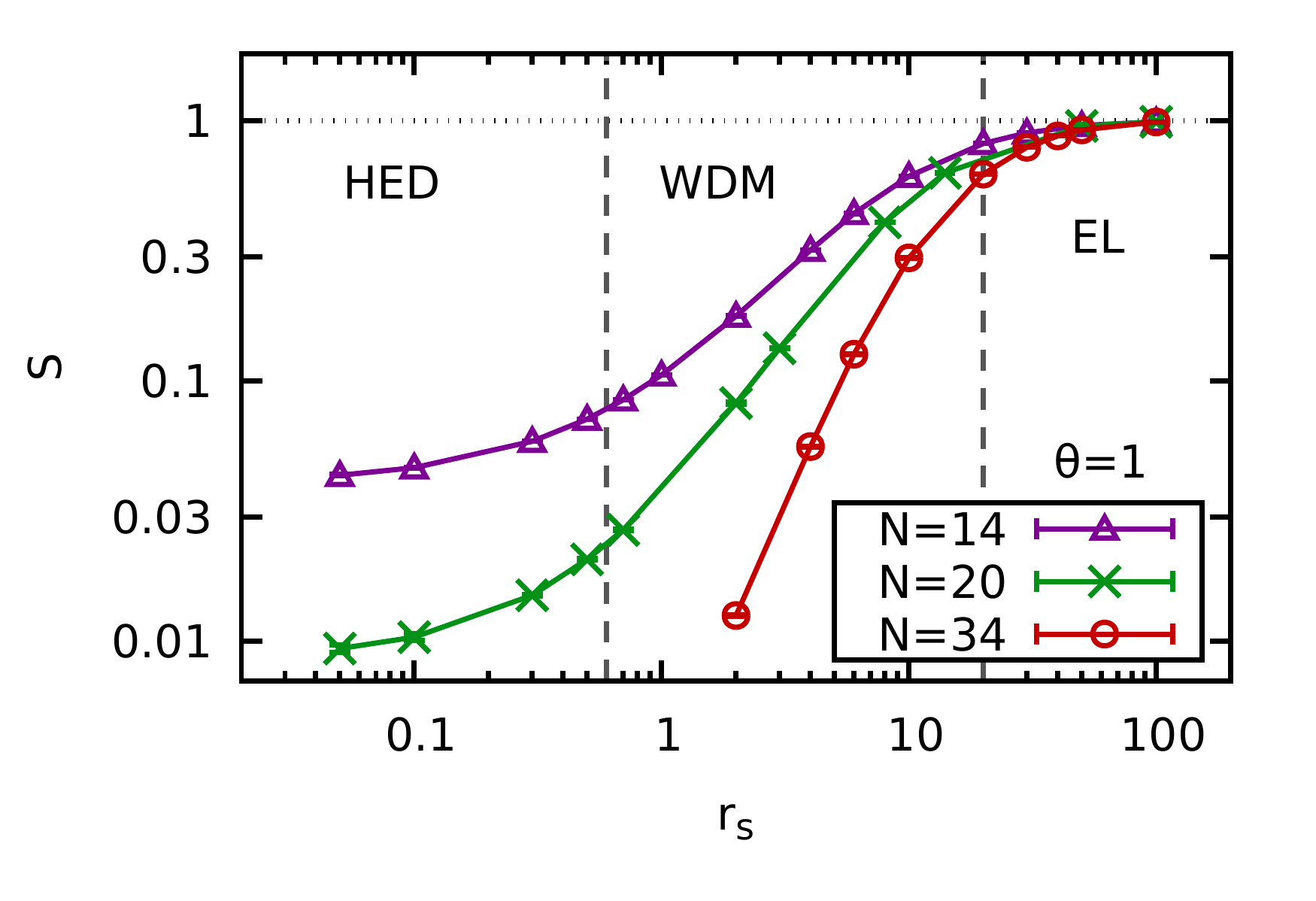}
\caption{\label{fig:SIGN}
PIMC results for the average sign $S$ plotted versus the density parameter $r_s$ at $\theta=1$ for $N=14$ (purple triangles), $N=20$ (green crosses), and $N=34$ (red circles). The dashed vertical lines indicate the different physical regimes denoted as HED (high-energy density, $0\leq r_s \lesssim 0.6$), WDM (warm dense matter, $0.6\lesssim r_s \lesssim 20$), and EL (electron liquid, $20\lesssim r_s \lesssim 100$).
}
\end{figure}

Let us start our discussion of the PIMC data with an examination of finite-size effects. Recall that in this context the fermion sign problem constitutes the main obstacle and limits feasible system sizes to $N\leq 34$ electrons. This is illustrated in Fig.~\ref{fig:SIGN}, where we show the $r_s$-dependence of the average sign over several orders of magnitude at the Fermi temperature ($\theta=1$) for $N=14$ (purple triangles), $N=20$ (green crosses), and $N=34$ (red circles).

At low density ($r_s\gtrsim 20$), Coulomb coupling effects dominate and the UEG forms an electron liquid (EL)~\cite{dornheim_electron_liquid}. In practice, paths of individual particles within a standard PIMC simulation are then separated by the strong repulsion, and quantum degeneracy effects are of limited relevance. Consequently, $S$ remains large at these conditions, and the sign problem is not that severe.

The central region of Fig.~\ref{fig:SIGN} ($0.6\lesssim r_s \lesssim 20$) is commonly known as warm dense matter (WDM)~\cite{review,new_POP} and encompasses typical metallic densities. This regime is characterized by the complicated interplay of i) Coulomb interaction, ii) thermal excitations, and iii) quantum degeneracy effects. For PIMC simulations of electrons, the WDM regime thus constitutes the transition region in which the sign drops from around one (classical limit) and approaches the noninteracting limit with decreasing $r_s$. Moreover, we note that PIMC simulations of, e.g., $N=34$ electrons already become unfeasible at $\theta=1$ and $r_s=2$, with an average sign of $S\sim10^{-2}$. In particular, Eq.~(\ref{eq:FSP}) implies a ten-thousand fold increase in computation time as compared to a Boltzmann system at the same conditions, which is not possible at present.

Finally, the left part of Fig.~\ref{fig:SIGN} corresponds to the high-energy-density (HED) regime in which we are interested in the present work. At these conditions, the UEG is only weakly coupled, with the Coulomb interaction acting as a relatively small perturbation. As a side note, we mention that this makes the HED regime ideally suited for the CPIMC method due to its formulation as a Monte Carlo evaluation of the exact infinite perturbation series around the ideal system~\cite{schoof_prl,review}. Moreover, the density matrix QMC method~\cite{malone1,malone2}, too, excels under the conditions for similar reasons, see Ref.~\cite{dornheim_pop} for a recent juxtaposition of fermionic QMC methods at finite temperature. For standard PIMC, on the other hand, the HED regime poses a serious challenge as the average sign is small and eventually attains the respective ideal limit. For example, at $r_s=0.05$ and $\theta=1$ our PIMC simulations are limited to $N=14$ unpolarized electrons.

Due to these limitations, it is of paramount importance to correct for potential finite-size effects in our PIMC results, which will be analyzed in detail below.

\begin{figure}\centering
\includegraphics[width=0.6\textwidth]{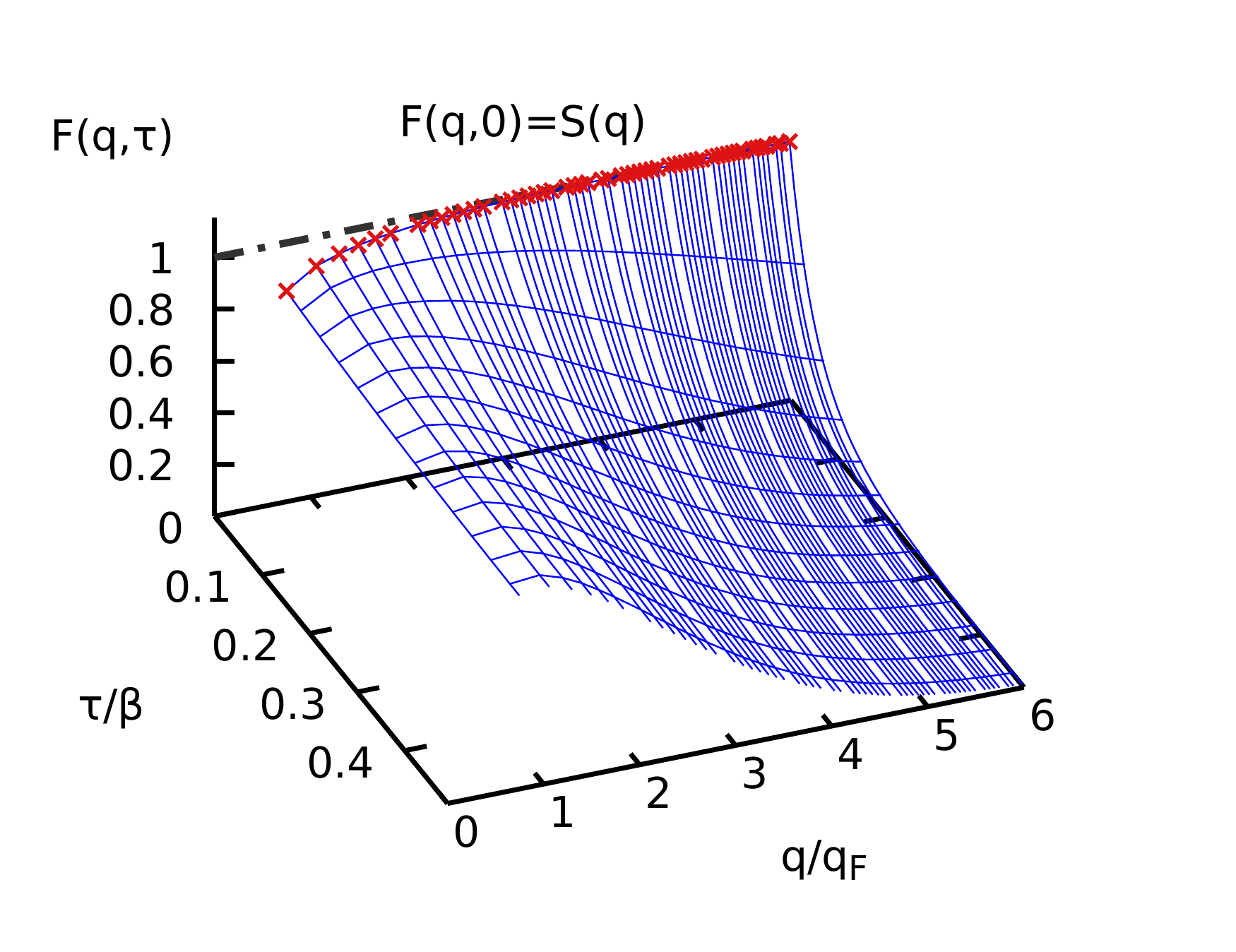}
\caption{\label{fig:FPLOT}
PIMC results for the imaginary-time density--density correlation function $F(q,\tau)$ for $N=20$ electrons at $r_s=0.5$ and $\theta=2$.
}
\end{figure}

Let us start this investigation by briefly considering the raw PIMC data for the imaginary-time density--density correlation function $F(q,\tau)$ [cf.~Eq.~(\ref{eq:scattering_function})], which are shown for $N=20$ electrons at $r_s=0.5$ and $\theta=2$ in Fig.~\ref{fig:FPLOT}. First and foremost, we note that $F$ is symmetric with respect to $\tau$ around $\tau=\beta/2$, so that a depiction of the half-plane is sufficient. In addition, the $\tau=0$ limit is given by the static structure factor $S(q)$, which is shown as the red crosses in Fig.~\ref{fig:FPLOT}. Although a direct physical interpretation of $F(q,\tau)$ is rather difficult, we find that it somewhat reflects the degree of structure within the system, and is smooth for the HED regime, but exhibits a pronounced structure in the electron liquid regime, see Ref.~\cite{dornheim_electron_liquid}.

For completeness, we also mention the relation
\begin{eqnarray}\label{eq:FS}
F(\mathbf{q},\tau) = \int_{-\infty}^\infty \textnormal{d}\omega\ S(\mathbf{q},\omega) e^{-\tau\omega} \quad ,
\end{eqnarray}
which implies that $F(q,\tau)$ is connected to the dynamic structure factor $S(q,\omega)$ by a Laplace transform. In particular, Eq.~(\ref{eq:FS}) can be used as the starting point for a \textit{reconstruction} of $S(q,\omega)$ (i.e., an inverse Laplace transform), which is a well-known, but notoriously hard problem~\cite{jarrell,schoett}. While being beyond the scope of the present work, the numerical inversion of Eq.~(\ref{eq:FS}) has recently allowed for the first exact results for $S(q,\omega)$ of warm dense electrons going from warm dense matter up to the electron liquid regime, see Refs.~\cite{dynamic_folgepaper,dornheim_dynamic} for details.

\begin{figure}\centering
\includegraphics[width=0.47\textwidth]{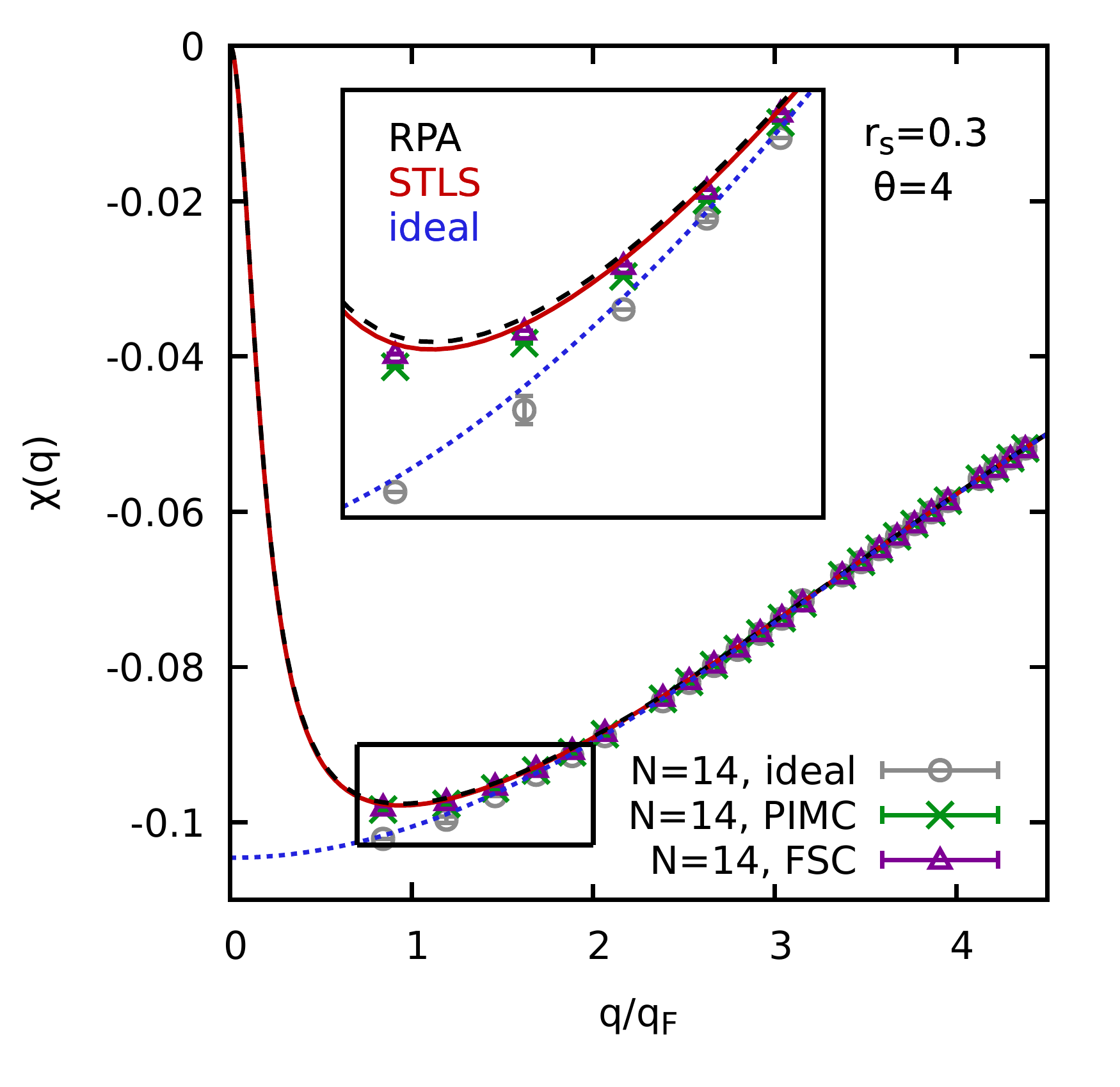}
\includegraphics[width=0.47\textwidth]{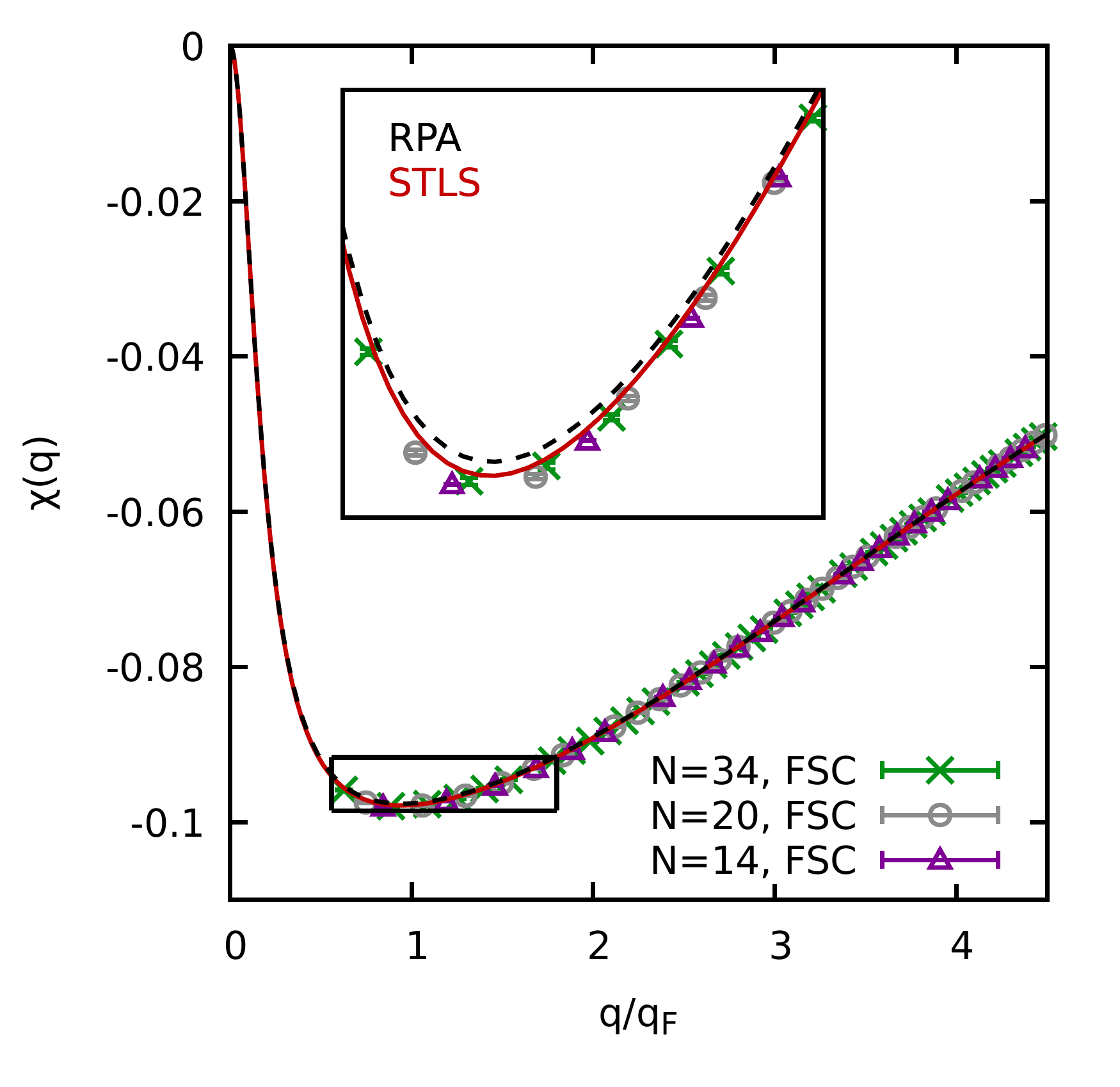}
\caption{\label{fig:FSC_rs0.3_theta4}
Wave-number dependence of the static density response function $\chi(q)$ for $r_s=0.3$ and $\theta=4$. The left panel shows QMC results for $N=14$ electrons, with the green crosses and purple triangles corresponding to raw and finite-size corrected PIMC data, and the grey circles to CPIMC results for the noninteracting case. The dashed black and solid red curves depict RPA and STLS results, and the dotted blue curve $\chi_0(q)$ in the thermodnynamic limit. The right panel shows finite-size corrected data for $N=14$ (triangles), $N=20$ (circles), and $N=34$ (crosses).
}
\end{figure}

Let us now proceed to the topic at hand, i.e., the calculation of the static density response function $\chi(q)$ via Eq.~(\ref{eq:static_chi}). The results are shown in the left panel of Fig.~\ref{fig:FSC_rs0.3_theta4} for $N=14$ electrons at $r_s=0.3$ and $\theta=4$ as the green crosses. In addition, the dashed black and solid red lines correspond to the RPA and STLS data and are shown as a reference. Note that at such high density and temperature, the two curves almost coincide and the effect of the approximate local field correction from the STLS formalism is small.

Overall, the uncorrected PIMC data and the two dielectric curves exhibit the same qualitative behavior. Still, the PIMC simulation predicts a stronger response around intermediate wave numbers $q\sim q_\textnormal{F}$ (see also the inset showing a magnified segment). To ensure that this effect is not solely an artifact of the finite-size in the PIMC simulation, we apply the finite-size correction from Eq.~(\ref{eq:chi_FSC}) as explained in Sec.~\ref{sec:response_theory}. The results are shown as the purple triangles and are located nearly halfway between the raw PIMC data and the theoretical curves. 

The origin of the finite-size effects is illustrated by the grey circles, which depict CPIMC data for the static density response function of the noninteracting system at the same system size ($\chi_0^N(q)$), see Ref.~\cite{groth_jcp} for a detailed discussion. Evidently, these data are systematically lower than the corresponding results in the thermodynamic limit (dotted blue curve), which explains the finite-size effects in $\chi^N(q)$.

\begin{figure}\centering
\includegraphics[width=0.47\textwidth]{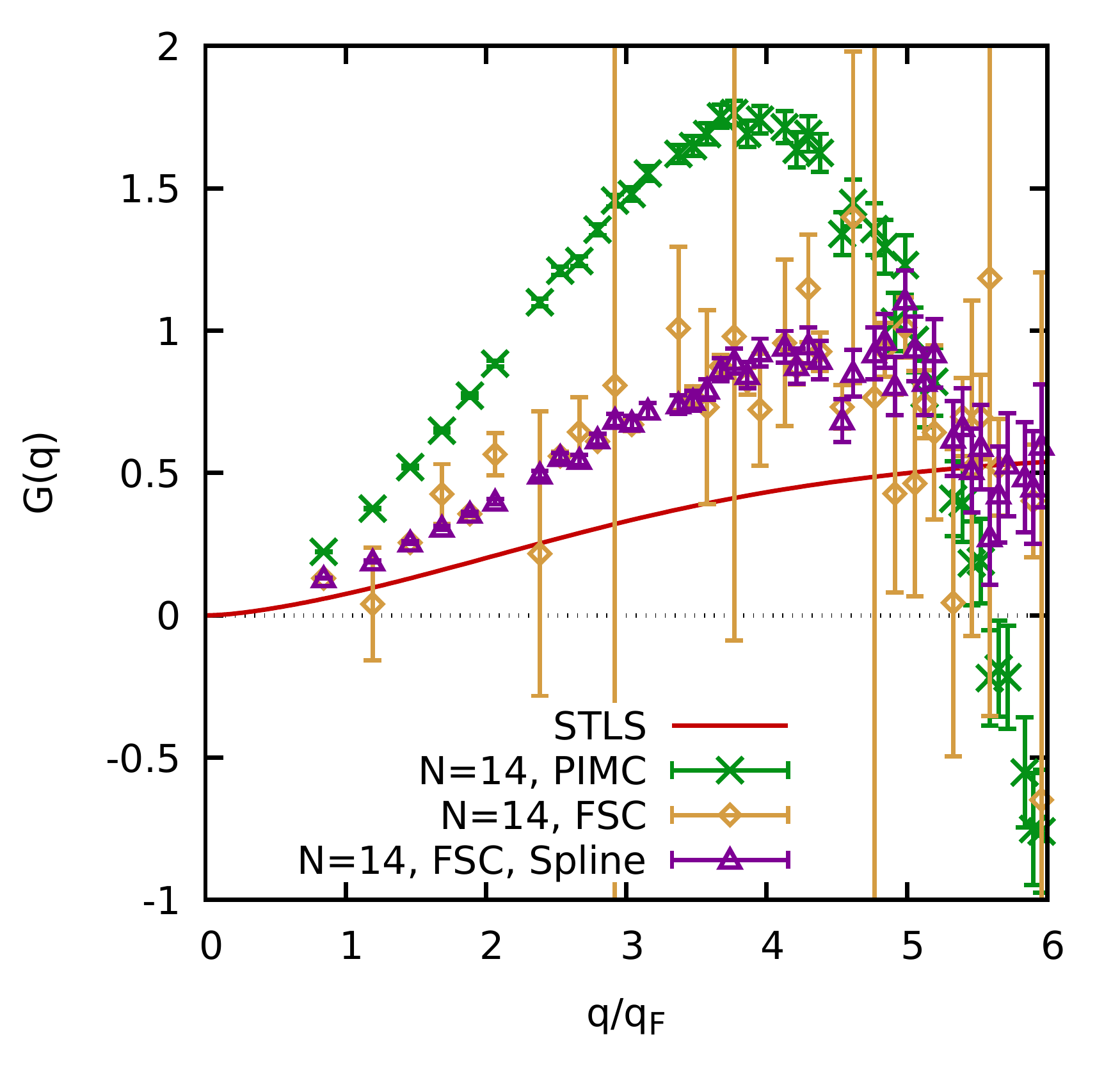}
\includegraphics[width=0.47\textwidth]{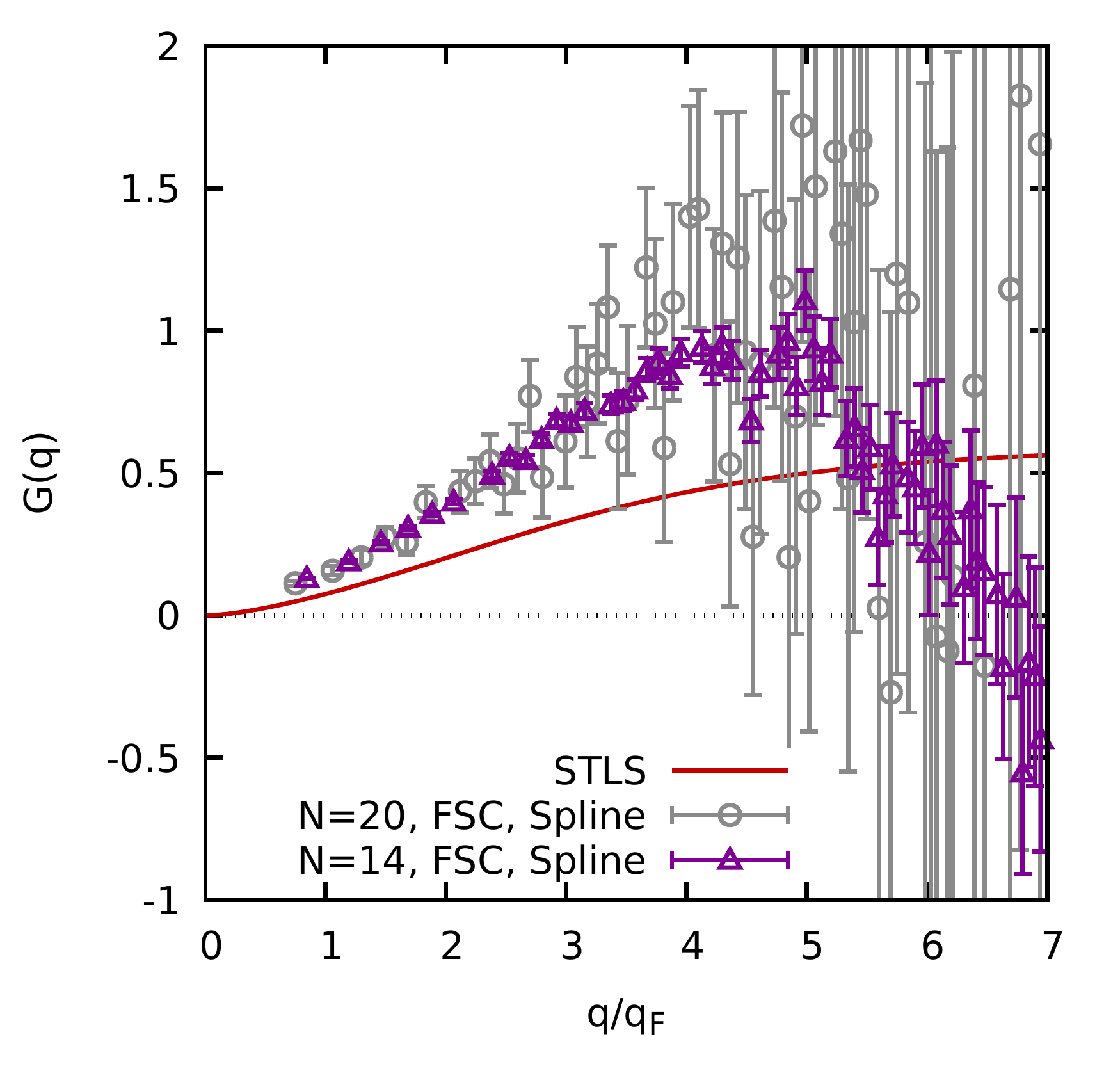}
\caption{\label{fig:LFC_rs0.3_theta4}
Wave-number dependence of the static local field correction $G(q)$ for $r_s=0.3$ and $\theta=4$. The left panel depicts PIMC data for $N=14$, with the green crosses having been obtained using $\chi(q)$ in Eq.~(\ref{eq:Get_G}), the yellow diamonds using the CPIMC data for $\chi_0^N(q)$, and the purple triangles using a cubic basis spline interpolation of the latter. The solid red line depicts STLS and has been included as a reference. The right panel also includes finite-size corrected data for $N=20$ electrons (grey circles).
}
\end{figure}

This can be seen particularly well in the left panel of Fig.~\ref{fig:LFC_rs0.3_theta4}, where we show the corresponding results for the static local field correction $G(q)$. Again, the solid red curve shows the STLS data and has been included as a reference, whereas it holds $G(q)=0$ in RPA. The green crosses depict the "raw" PIMC data which have been obtained by evaluating Eq.~(\ref{eq:Get_G}) using the PIMC data for $\chi^N(q)$, but $\chi_0(q)$ in the thermodynamic limit. In contrast, the yellow diamonds have been obtained by consistently using both $\chi^N(q)$ and $\chi^N_0(q)$, which leads to a significantly shifted curve. We note that the large error bars for some wave numbers are a consequence of the CPIMC calculation for $\chi_0^N(q)$, which is fairly nontrivial and requires the application of boundary conditions with a small twist angle, see Ref.~\cite{groth_jcp} for a detailed explanation. This can also be seen by the grey circles in the left panel of Fig.~\ref{fig:FSC_rs0.3_theta4}, which, for example, exhibit small errorbars for the first and third $q$-value, but a much larger uncertainty for the second.

To mitigate these fluctuations, we use a cubic basis spline to interpolate the CPIMC data for $\chi_0^N(q)$. Using the spline to evaluate Eq.~(\ref{eq:Get_G}) finally leads to the purple triangles in the left panel of Fig.~\ref{fig:LFC_rs0.3_theta4}, which exhibit the same progression as the yellow diamonds, but are significantly smoother. We note that all data for $G(q)$ and the corresponding finite-size corrected results for $\chi(q)$ [see Eq.~(\ref{eq:chi_FSC})] in this work have been obtained in this way.

The only assumption yet to be verified in order to confirm our data for $\chi(q)$ is the absence of finite-size effects in $G(q)$.
This is investigated in the right panel of Fig.~\ref{fig:LFC_rs0.3_theta4}, where we also include PIMC data for $N=20$ electrons (grey circles). First and foremost, we note that the statistical uncertainty is significantly increasing with $q$, which is a direct consequence of Eq.~(\ref{eq:Get_G}): for large wave numbers, $G(q)$ is given by the small difference between two nearly identical functions ($\chi(q)$ and $\chi_0(q)$ converge for large $q$), which is subsequently enhanced by the inverse of the Coulomb interaction $4\pi/q^2$. Therefore, the relative statistical error of $G(q)$ monotonically increases, until it cannot be resolved since its effect on $\chi(q)$ effectively vanishes.

Still, a meaningful comparison between the PIMC data for the two particle numbers is possible for $q\lesssim4q_\textnormal{F}$, and we find perfect agreement between the two data sets. This further illustrated the importance of the CPIMC data for $\chi^N_0(q)$, since a substitution of $\chi_0(q)$ would introduce large finite-size effects in $G(q)$ as well, cf.~the green crosses in the left panel of Fig.~\ref{fig:LFC_rs0.3_theta4}.

Finally, we demonstrate the effect of the finite-size correction for $\chi(q)$ [Eq.~(\ref{eq:chi_FSC})] in the right panel of Fig.~\ref{fig:FSC_rs0.3_theta4}. Evidently the three finite-size corrected PIMC data sets are in perfect agreement with each other within the given statistical uncertainty, which further validates our approach, and reveals the difference towards both RPA and STLS as a real effect.

\begin{figure}\centering
\includegraphics[width=0.47\textwidth]{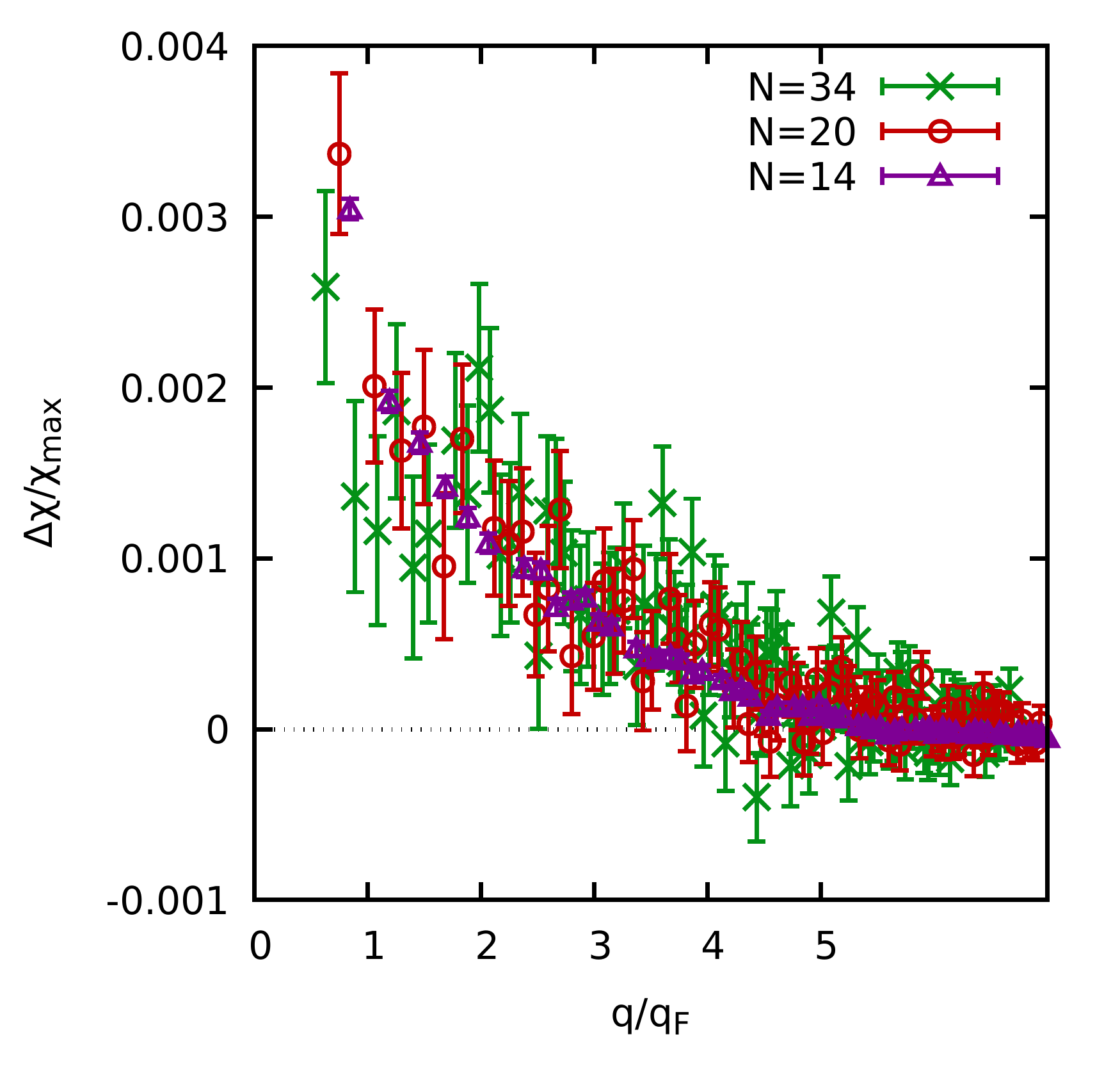}
\caption{\label{fig:delta_Chi_rs0.3_theta4}
Relative deviation in the static density response function $\Delta\chi/\chi_\textnormal{max}$ between our finite-size corrected PIMC data and STLS for $r_s=0.3$ and $\theta=4$. The purple triangles, red circles, and green crosses have been obtained using PIMC results for $N=14$, $N=20$, and $N=34$, respectively.
}
\end{figure}

This is further investigated in Fig.~\ref{fig:delta_Chi_rs0.3_theta4}, where we show the relative difference between STLS and our PIMC data for all three particle numbers. All curves exhibit the same expected trend, that is, a vanishing difference for large $q$ and maximum deviation of $0.3\%$ around the Fermi wave number. For completeness, we mention that the deviations also vanish for $q\to 0$, which we have empirically verified for larger $r_s$-values, where data for smaller $q$ are available. Yet, the most important point in the context of finite-size corrections is that the three curves are in excellent agreement for all $q$, which, again, illustrates the validity of our approach.

\begin{figure}\centering
\includegraphics[width=0.47\textwidth]{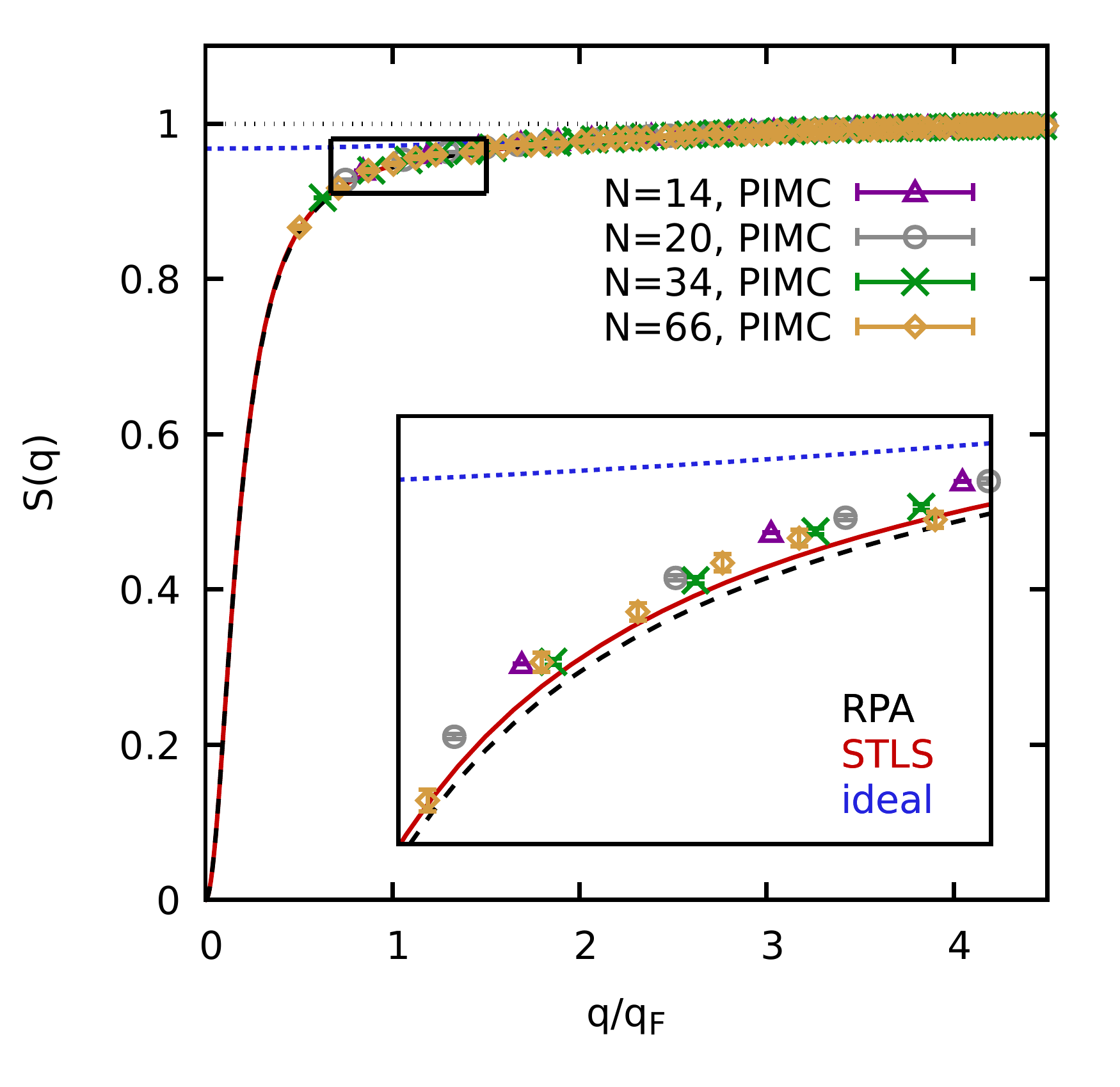}
\caption{\label{fig:Sq_rs0.3_theta4}
Wave-number dependence of the static structure factor $S(q)$ for $r_s=0.3$ and $\theta=4$. The symbols depict our raw PIMC results for $N=14,20,34,$ and $66$. The dashed black, solid red, and dotted blue lines correspond to RPA, STLS, and ideal results and have been included as a reference.
}
\end{figure}

Let us conclude this section with a brief look onto the static structure factor $S(q)$, which is shown in Fig.~\ref{fig:Sq_rs0.3_theta4}. Firstly, we note that both RPA and STLS exhibit the correct asymptotic behavior predicted by the perfect screening sum-rule~\cite{kugler_bounds},
\begin{eqnarray}
\lim_{q\to 0}S(q) = \frac{q^2}{2\omega_\textnormal{p}} \textnormal{coth}\left( \frac{\beta\omega_\textnormal{p}}{2} \right) \quad ,
\end{eqnarray}
with $\omega_\textnormal{p}=\sqrt{3/r_s^3}$ being the usual plasma frequency, whereas the ideal curve attains substantially larger values. Moreover, we show PIMC data for $S(q)$ for four different particle numbers $N$, which do exhibit small, yet significant finite-size effects, as can be seen particularly well for $N=14$ (purple triangles) and $N=34$ (green crosses) in the inset.

This, however, is of no important consequence, since in the HED regime highly accurate CPIMC data for $S(q)$ are available for significantly larger particle numbers, see Refs.~\cite{dornheim_prl,dornheim_cpp,review} for details.

A comparison between our new PIMC results and previous CPIMC data is shown in Sec.~\ref{sec:CPIMC}.

\subsection{PIMC data for the static density response\label{sec:data}}

%\begin{figure}\centering
%\includegraphics[width=0.47\textwidth]{{INSET_Chi_FSC_rs0.05_theta2}.pdf}
%\caption{\label{fig:CHI_rs0.05_theta2}
%Static density response function $\chi(q)$ at $r_s=0.05$ and $\theta=2$, for $N=20$.
%}
%\end{figure}  

\begin{figure}\centering
\includegraphics[width=0.47\textwidth]{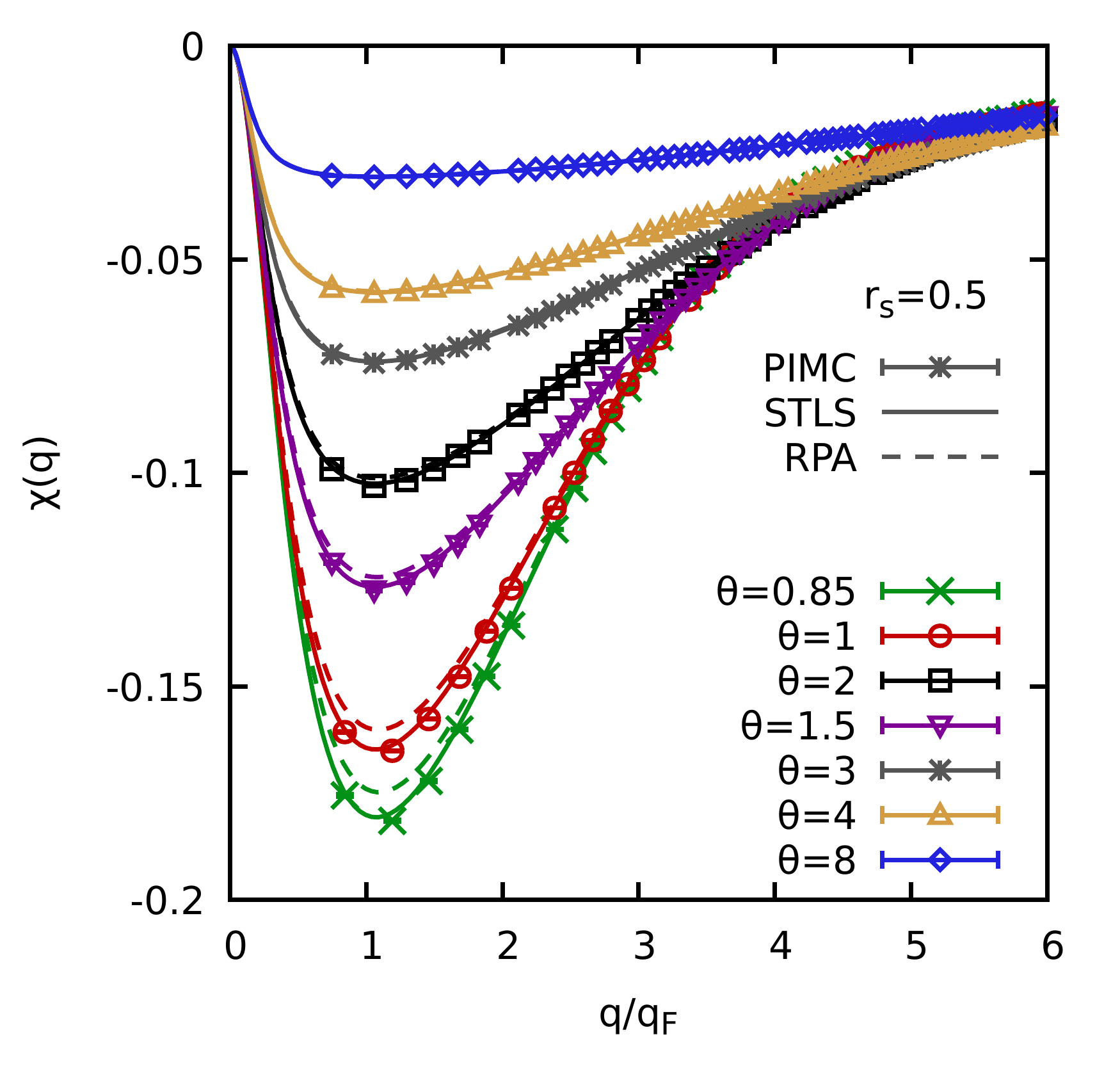}
\includegraphics[width=0.47\textwidth]{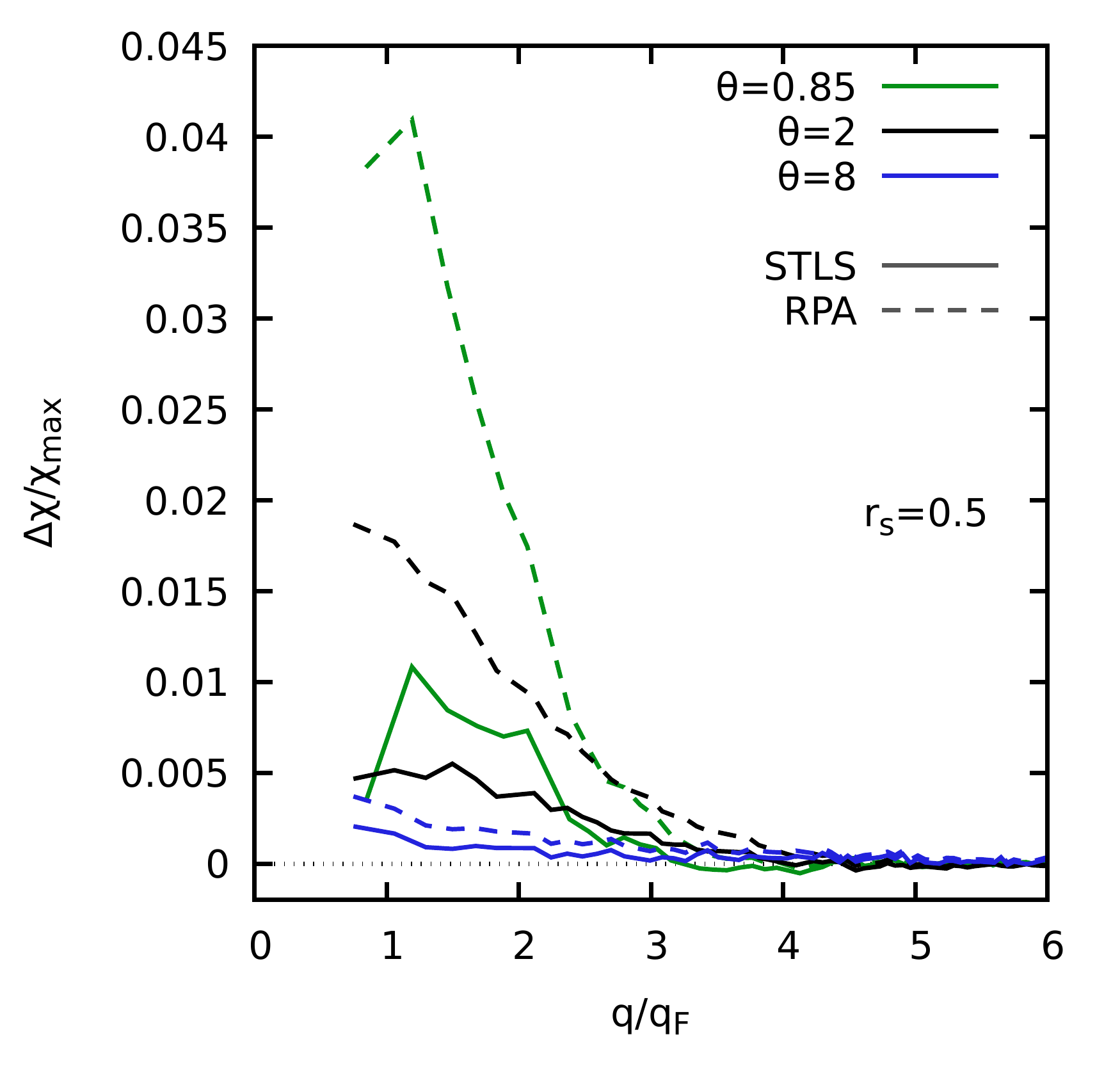}
\includegraphics[width=0.47\textwidth]{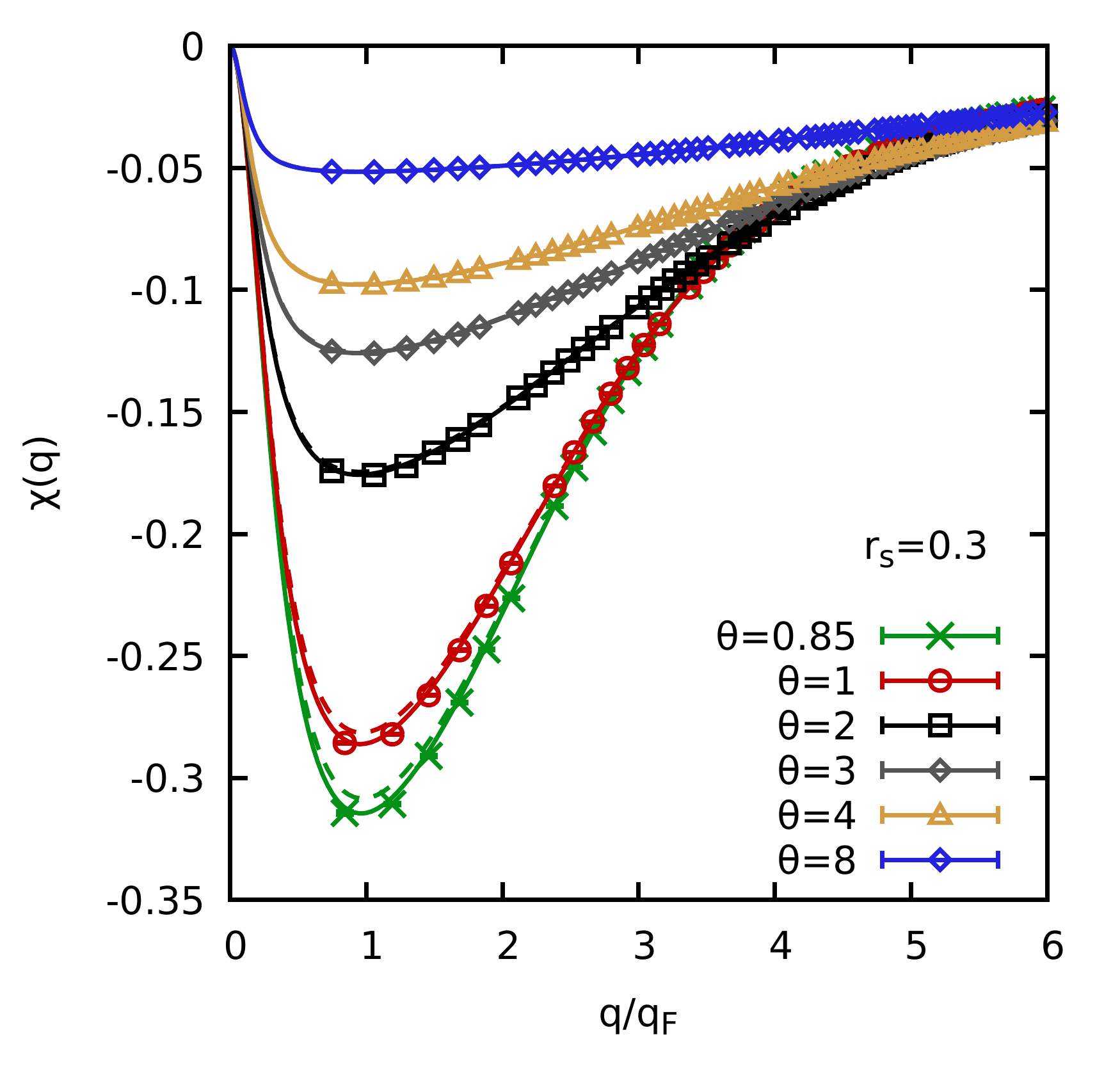}
\includegraphics[width=0.47\textwidth]{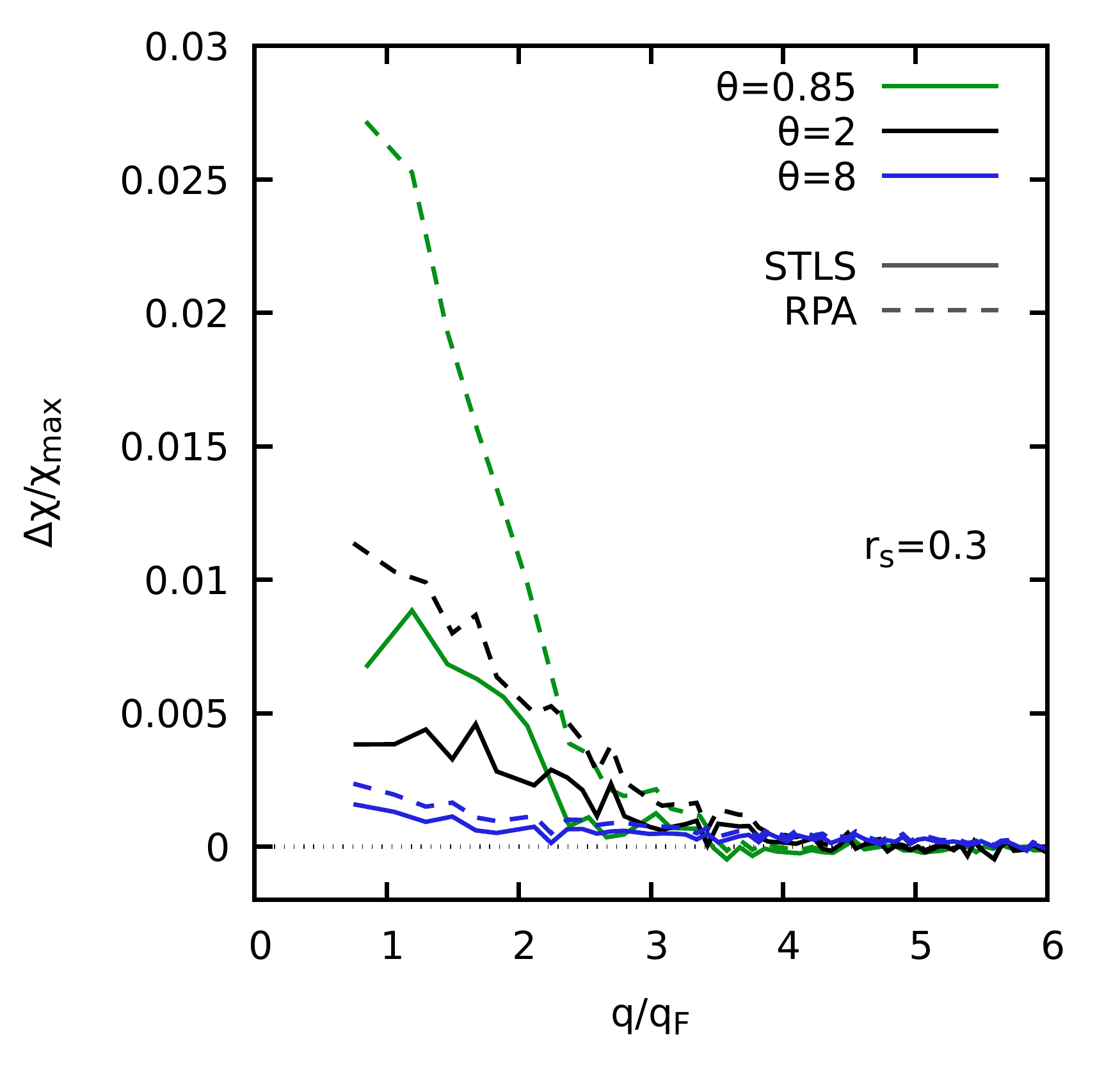}
\caption{\label{fig:CHI_panel_rs}
The static density response function for $r_s=0.3$ and $0.5$ for different temperatures. The colored symbols depict our finite-size corrected PIMC data for $\chi(q)$, and the solid (dashed) lines the corresponding results from STLS (RPA) at the same conditions. The right panel shows the relative deviation between our PIMC data and STLS (RPA), we a few temperatures have been omitted for better visibility.
}
\end{figure}

Let us now use our PIMC approach to the static density response of the UEG to systematically investigate the systematic errors in the dielectric formalism. 
In Fig.~\ref{fig:CHI_panel_rs}, we show $\chi(q)$ for $r_s=0.5$ (top row) and $r_s=0.3$ (bottom row) for several values of the reduced temperature $\theta$. The different symbols correspond to the finite-size corrected PIMC data and the solid (dashed) lines to the corresponding predictions from STLS (RPA). First and foremost, we note that the curves from all three methods are in qualitative agreement with each other for all parameter combinations: in the limit of small wave numbers, the density response function exhibits a parabolic behaviour~\cite{kugler_bounds}
\begin{eqnarray}~\label{eq:parabola}
\lim_{q\to 0} \chi(q) = - \frac{q^2}{4\pi} \quad ,
\end{eqnarray}
followed by a peak around $q_\textnormal{F}$, and a slow convergence towards zero in the large $q$ limit. Recall that the PIMC data are limited to a minimum wave number of $q_\textnormal{min}=2\pi/L$ by the finite simulation cell and, therefore, cannot access the regime of Eq.~(\ref{eq:parabola}) under the present conditions. This is different at large $r_s$, where, due to the small density, the simulation cells are larger and PIMC data are available in the small-$q$ limit, too.

Comparing the curves for different temperatures at a constant density, we find that the UEG reacts more pronounced at low temperature, as the electrons are more strongly correlated. %\textcolor{red}{Does anybody know, why that would be? At large $T$, coupling is weak, and so is the reaction. But with decreasing $r_s$ (see Fig.~\ref{fig:CHI_panel_theta}), the reaction becomes larger. Is this discussed somewhere?}
%\textcolor{violet}{Here the relation for temperature $T\simeq (\theta/r_s^2)\cdot 0.58\times 10^6~{\rm K}$. Therefore, at constant $\theta$, the decrease of $r_s$ means increase of density and also increase of temperature. With decrease of $r_s$  an inter-electronic distance decreases as $a \sim r_s$ and, in contrast,  temperature increases as $T\sim 1/r_s^2$.  For $\theta>1$, the inter-electronic coupling at constant $\theta$ is $\Gamma\sim 1/(aT)\sim r_s$ and decreases with decrease in $r_s$ $->$ weaker coupling leads to stronger reaction.}
In order to more systematically analyze the systematic errors of dielectric theory, we show the relative deviations between PIMC and STLS (RPA) as the solid (dashed) lines in the right column of Fig.~\ref{fig:CHI_panel_rs}. Note that we choose to divide the difference $\Delta \chi(q)$ by the maximum in $\chi(q)$,
\begin{eqnarray}
\chi_\textnormal{max} = \max_{q} \chi(q) \quad ,
\end{eqnarray}
and not by $\chi(q)$ itself. The latter would lead to large systematic deviations when $\chi(q)$ is small, which is potentially misleading, as one is typically interested in wave-number integrals over the density response, so that $\chi_\textnormal{max}$ is a more meaningful scale.

Overall, we always find the largest deviations around the Fermi wave numbers, as it is expected~\cite{dornheim_electron_liquid,dynamic_folgepaper,dornheim_ML}, whereas they vanish in the limit of large wave numbers, where even RPA becomes exact.  Furthermore, the relative size of the deviations is largest at low temperature where RPA (STLS) deviates from the PIMC data by a few percent ($\sim1\%$), but nearly vanishes for the largest depicted temperature, $\theta=8$, where we find $\Delta\chi/\chi_\textnormal{max}\sim0.1\%$. This is expected, as exchange--correlation effects become less important with increasing $\theta$.

\begin{figure}\centering
\includegraphics[width=0.47\textwidth]{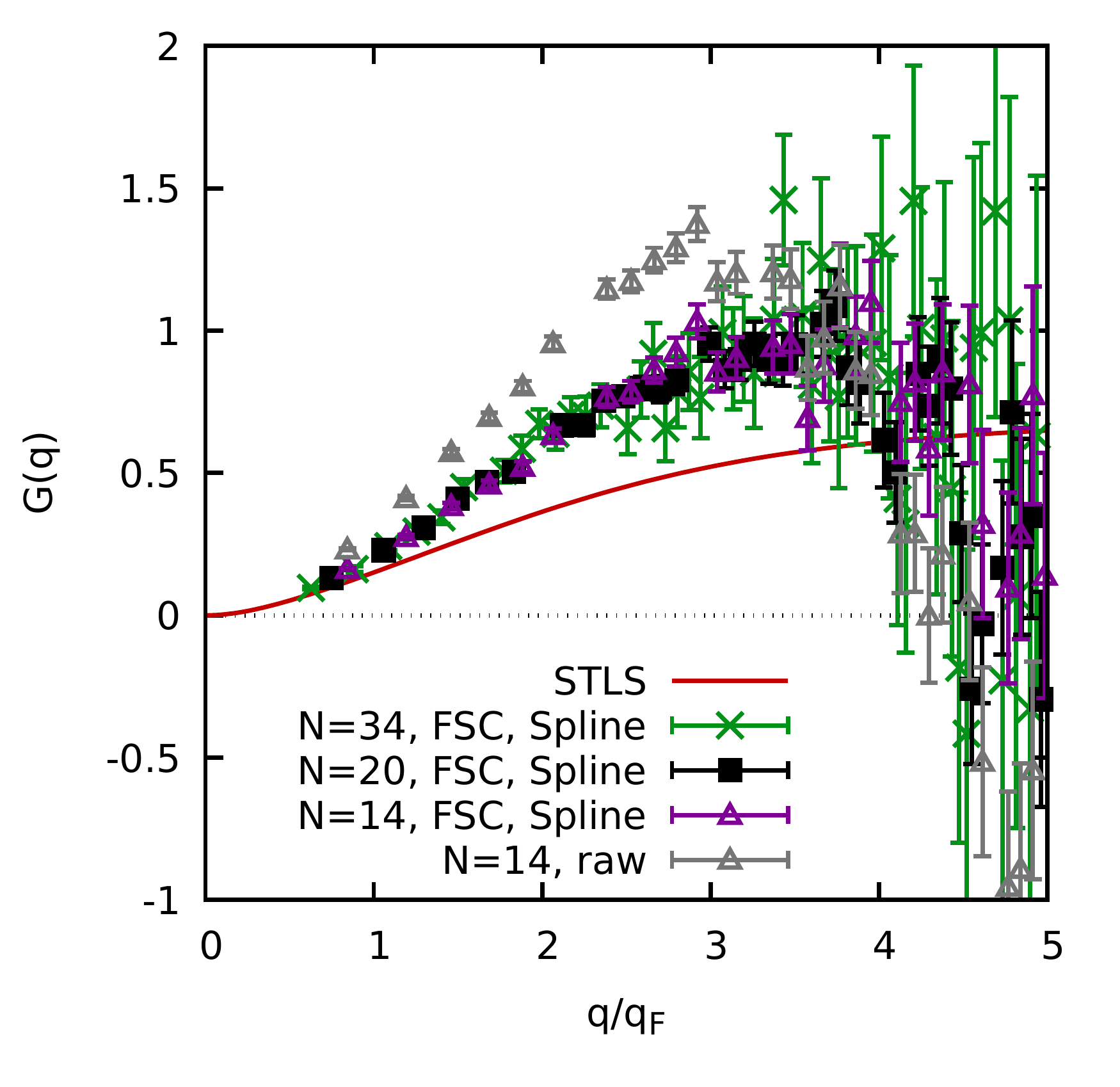}
\caption{\label{fig:real_LFC_rs0.5_theta2}
Static local field correction at $r_s=0.5$ and $\theta=2$. The green crosses, black squares, and purple triangles correspond to finite-size corrected PIMC data for $N=34,20,$ and $14$, respectively. The grey triangles show the uncorrected PIMC data for $N=14$, and the solid red line the results from STLS.
}
\end{figure}

\begin{figure}\centering
\includegraphics[width=0.47\textwidth]{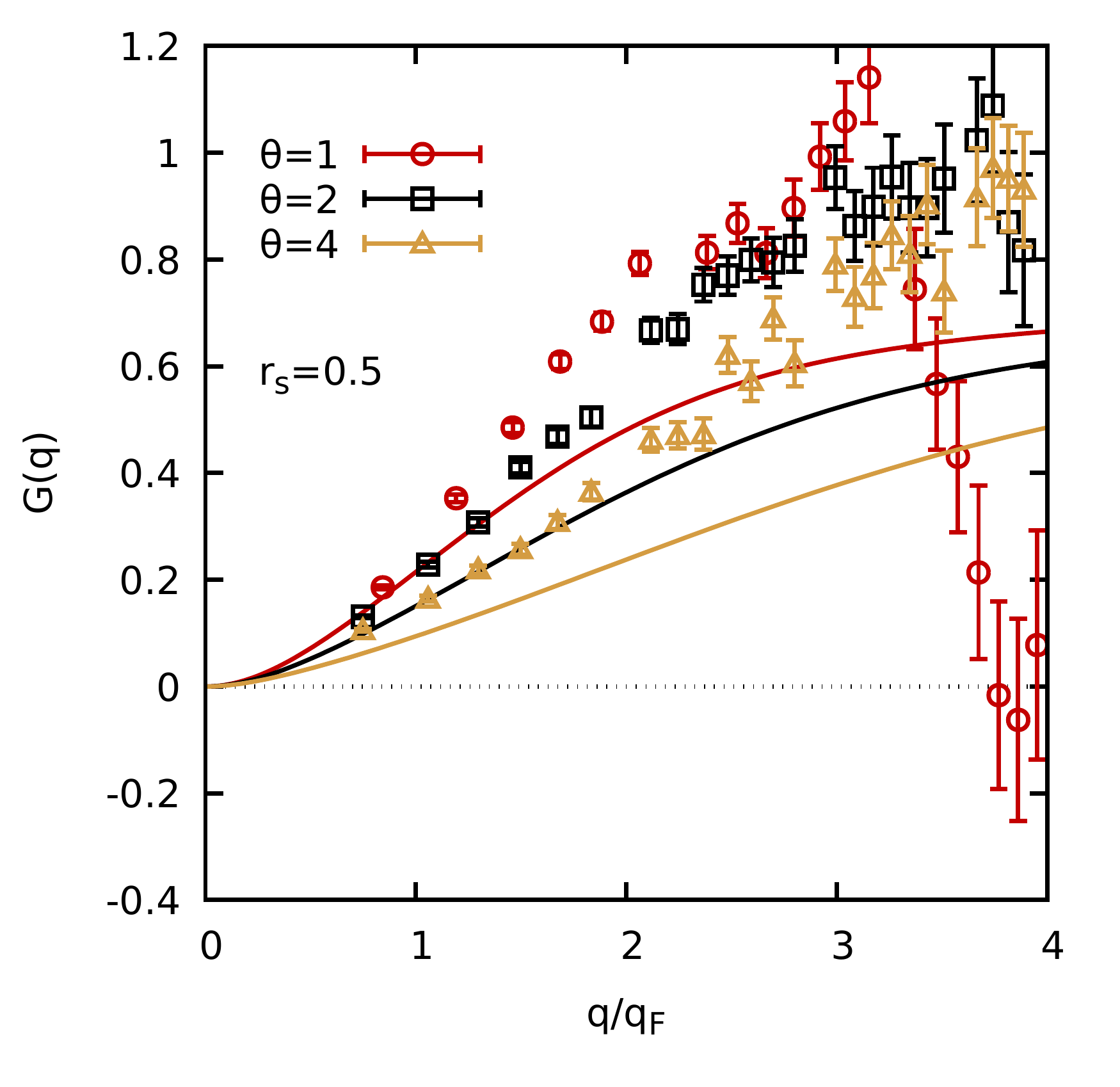}
\includegraphics[width=0.47\textwidth]{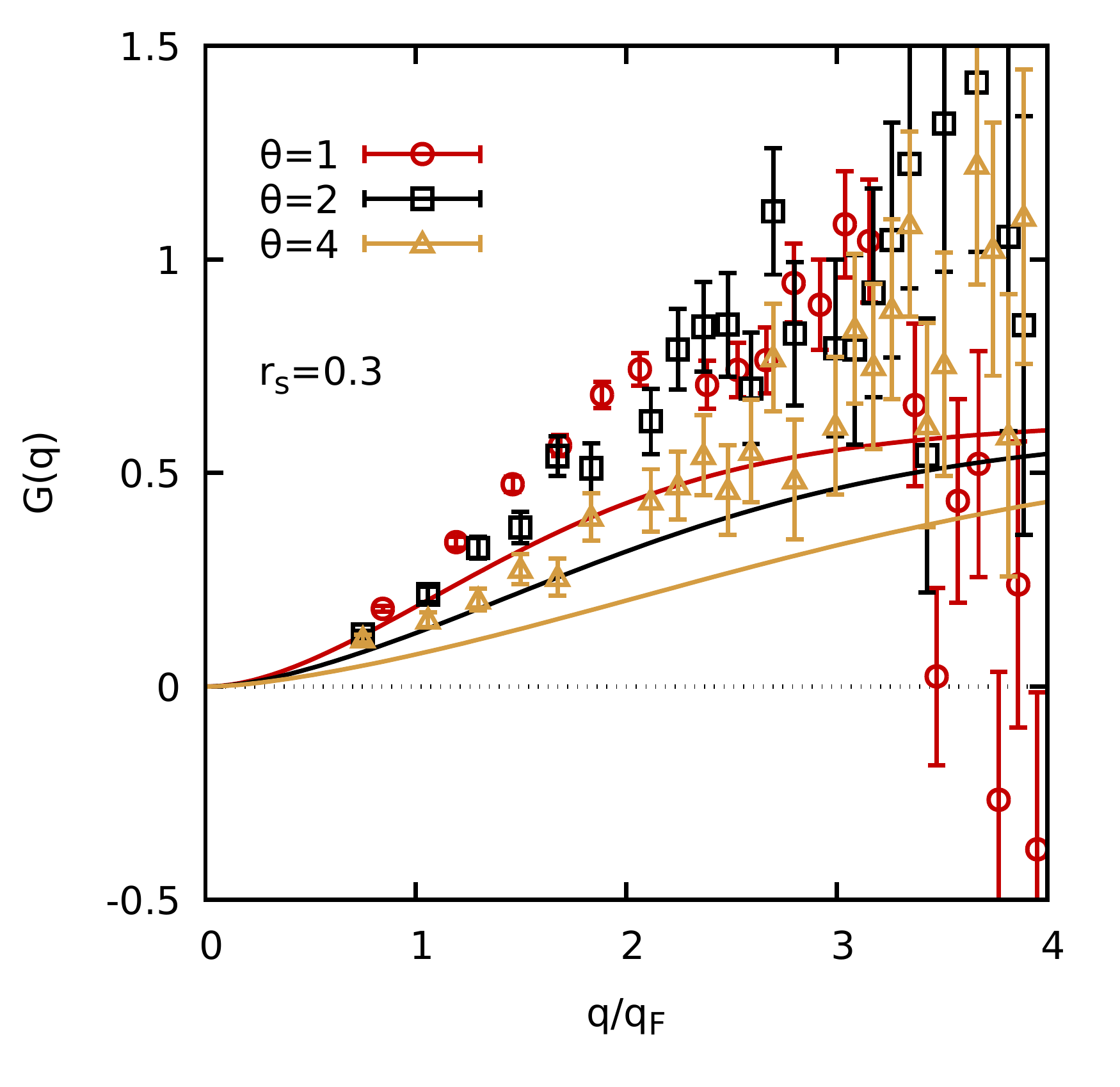}
\caption{\label{fig:LFC_PANEL}
Wave-number dependence of the static local field correction $G(q)$ for $r_s=0.5$ (left) and $0.3$ (right). The red circles, black squares, and yellow triangles depict our finite-size corrected PIMC data for $\theta=1,2,$ and $4$, respectively. The solid lines show the corresponding STLS results at the same conditions.
}
\end{figure}

Moreover, we observe that the dielectric theories always underestimate the magnitude of the density response ($\chi_\textnormal{PIMC}(q)-\chi_\textnormal{RPA/STLS}(q)$ is negative), which is consistent to previous findings reported in Refs.~\cite{dynamic_folgepaper,dornheim_ML}. In addition, the approximate static LFC from STLS leads to a substantial improvement over RPA for all $q$-values (see Figs.~\ref{fig:real_LFC_rs0.5_theta2} and \ref{fig:LFC_PANEL} for a discussion of $G(q)$ itself).

\begin{figure}\centering
\includegraphics[width=0.47\textwidth]{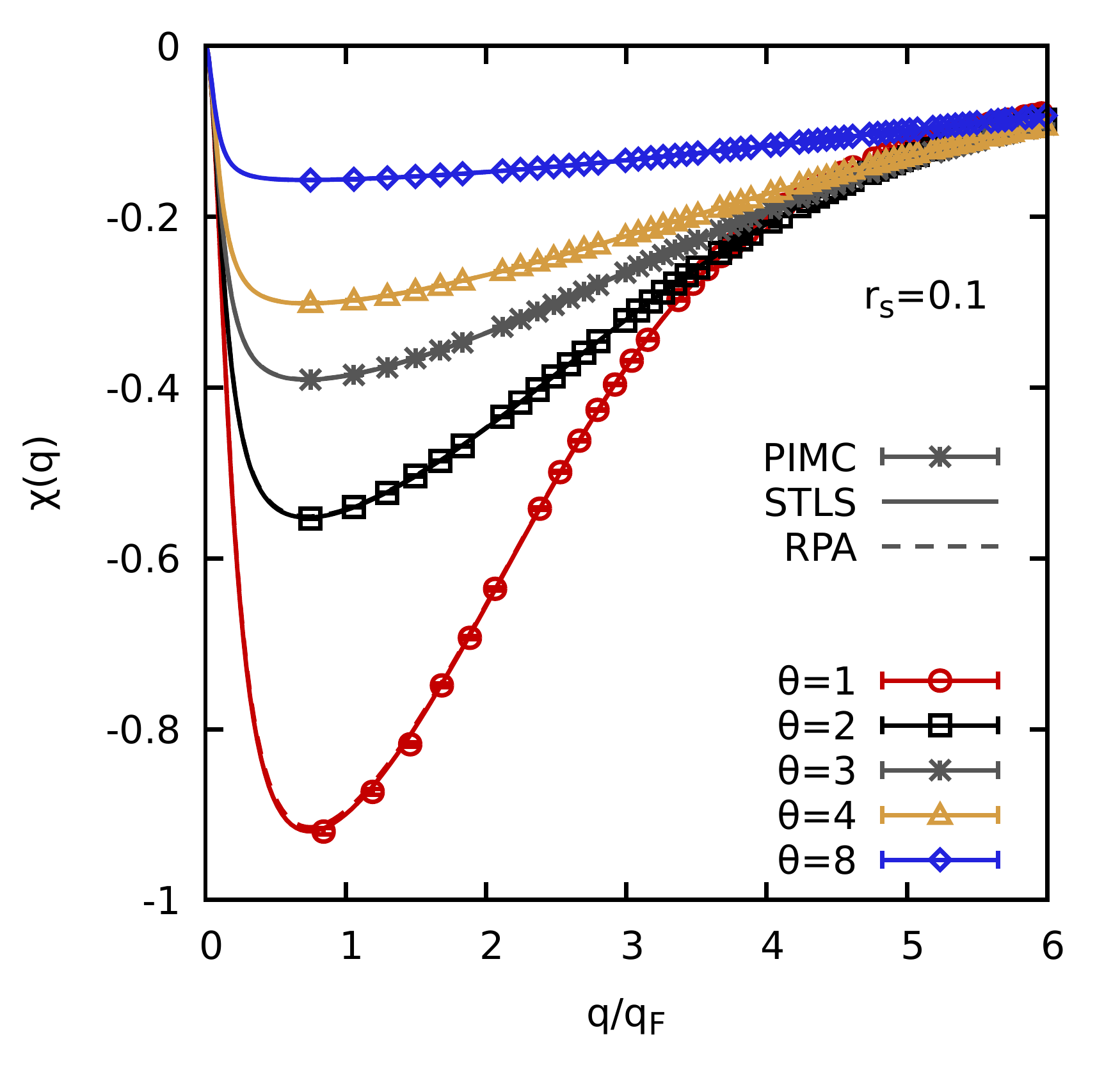}
\includegraphics[width=0.47\textwidth]{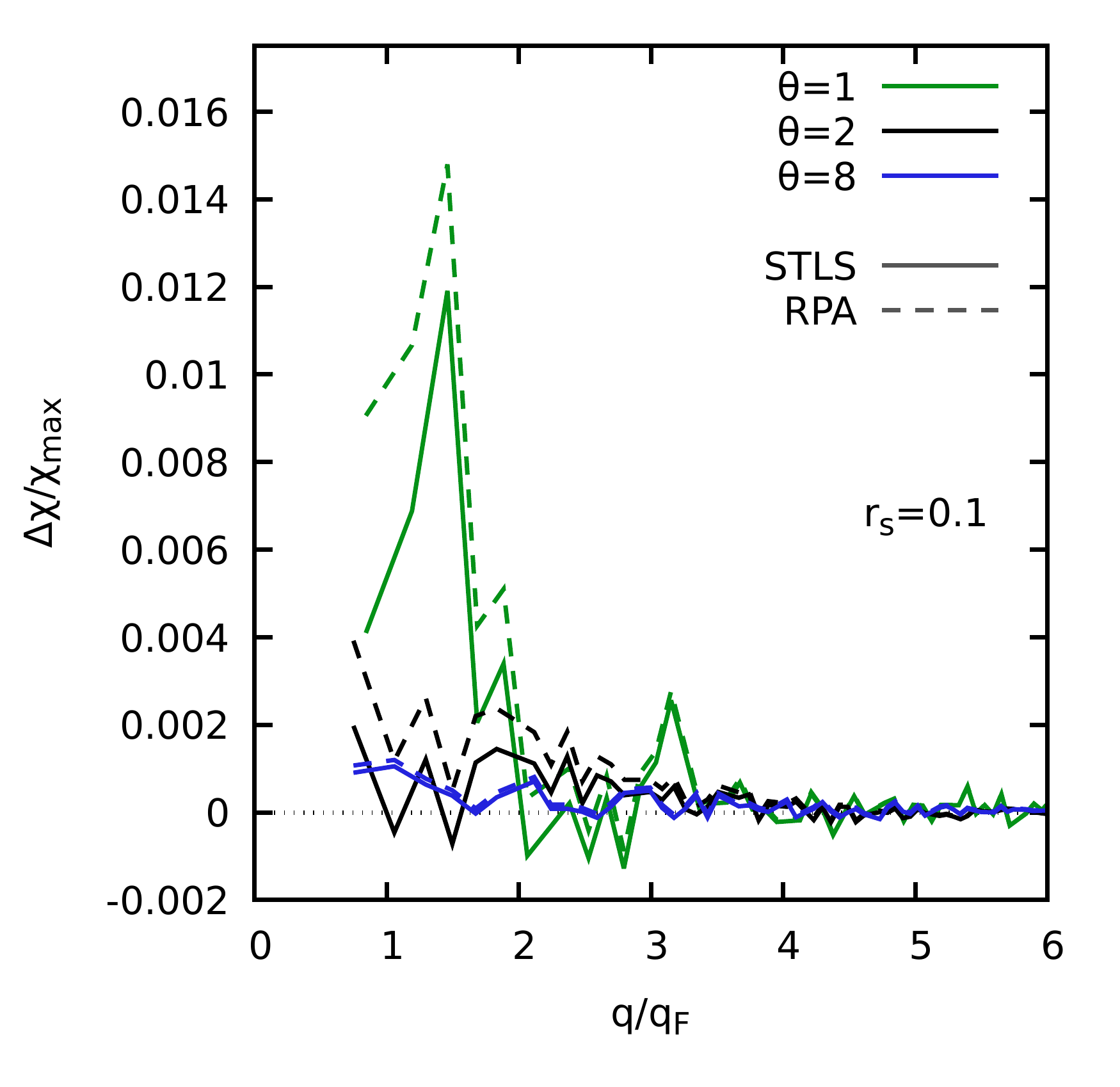}
\includegraphics[width=0.47\textwidth]{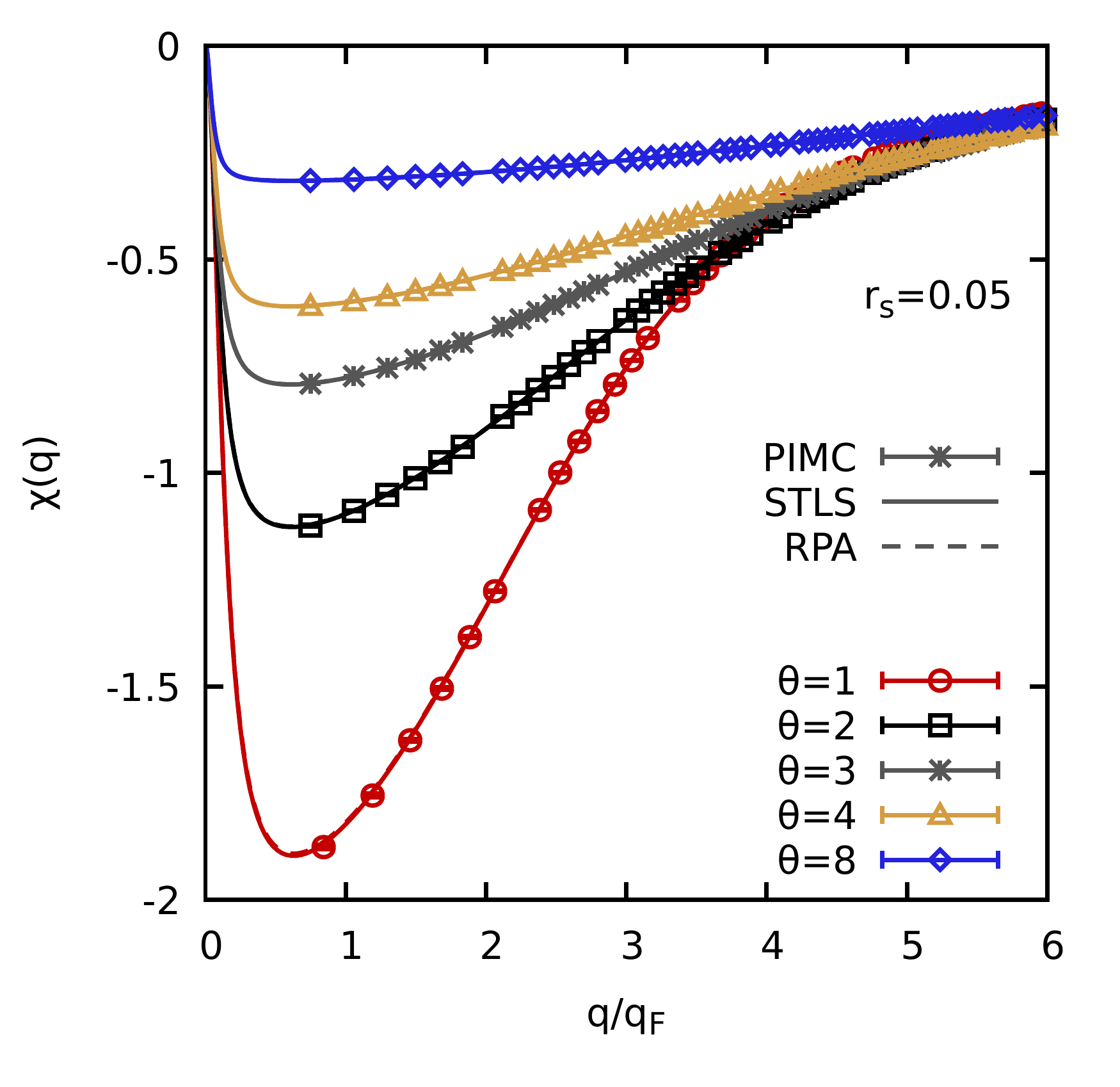}
\includegraphics[width=0.47\textwidth]{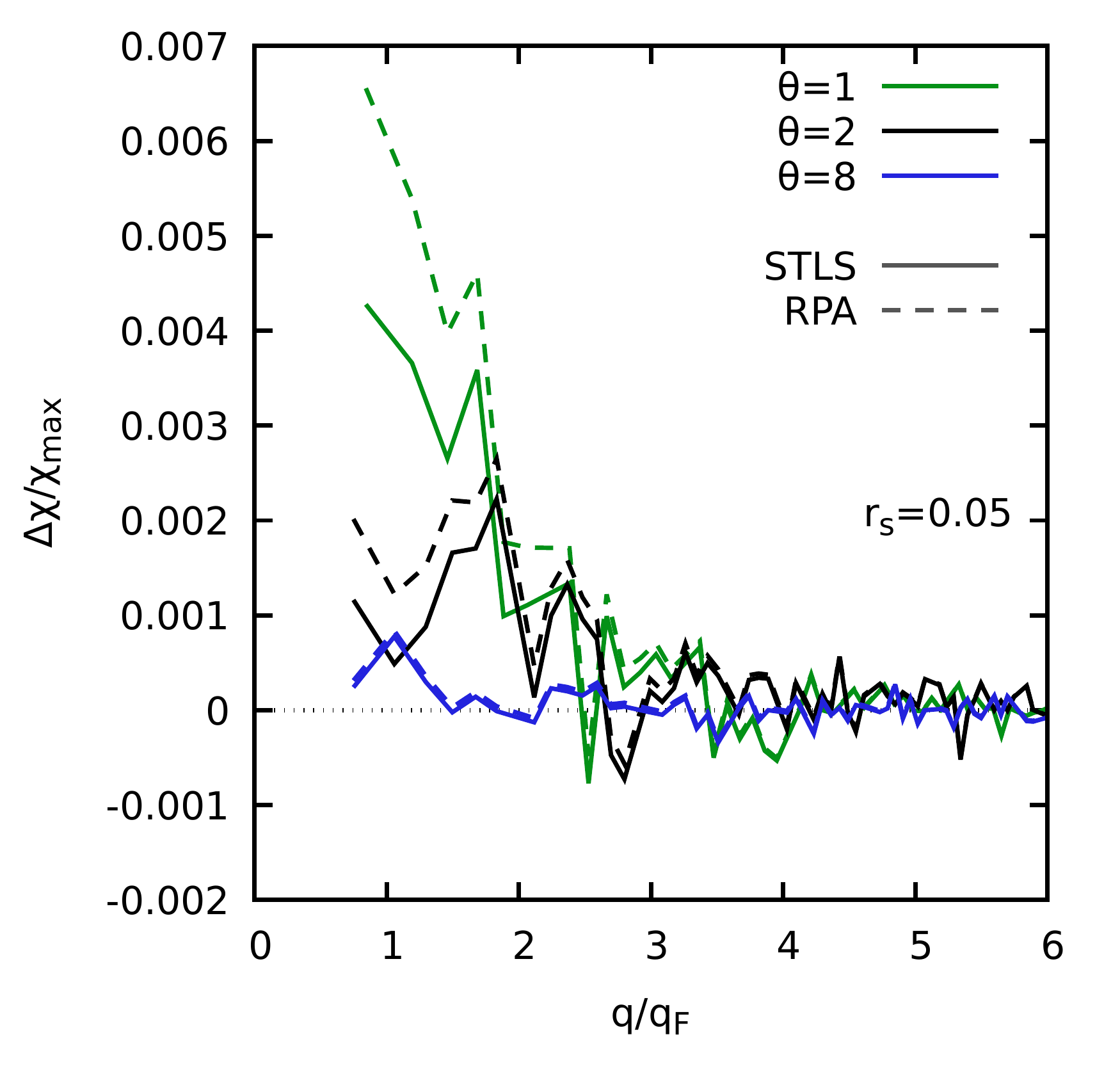}
\caption{\label{fig:CHI_panel_rs_small}
The static density response function for $r_s=0.1$ and $0.05$ for different temperatures. The colored symbols depict our finite-size corrected PIMC data for $\chi(q)$, and the solid (dashed) lines the corresponding results from STLS (RPA) at the same conditions. The right panel shows the relative deviation between our PIMC data and STLS (RPA), we a few temperatures have been omitted for better visibility.
}
\end{figure}

In Fig.~\ref{fig:CHI_panel_rs_small}, we show the same information as in Fig.~\ref{fig:CHI_panel_rs}, but for even larger densities, $r_s=0.1$ (top row) and $r_s=0.05$ (bottom row). Since we find the same qualitative behaviour as for $r_s=0.3$ and $0.5$, we only point out some key differences: i) the PIMC data are limited to even larger values of $q_\textnormal{min}$ due to the smaller simulation cell at higher density; ii) the systematic error in RPA and STLS is significantly smaller due to the reduced coupling strength; iii) the influence of the static LFC from STLS is substantially reduced, and the solid and dashed lines are nearly identical.

\begin{figure}\centering
\includegraphics[width=0.47\textwidth]{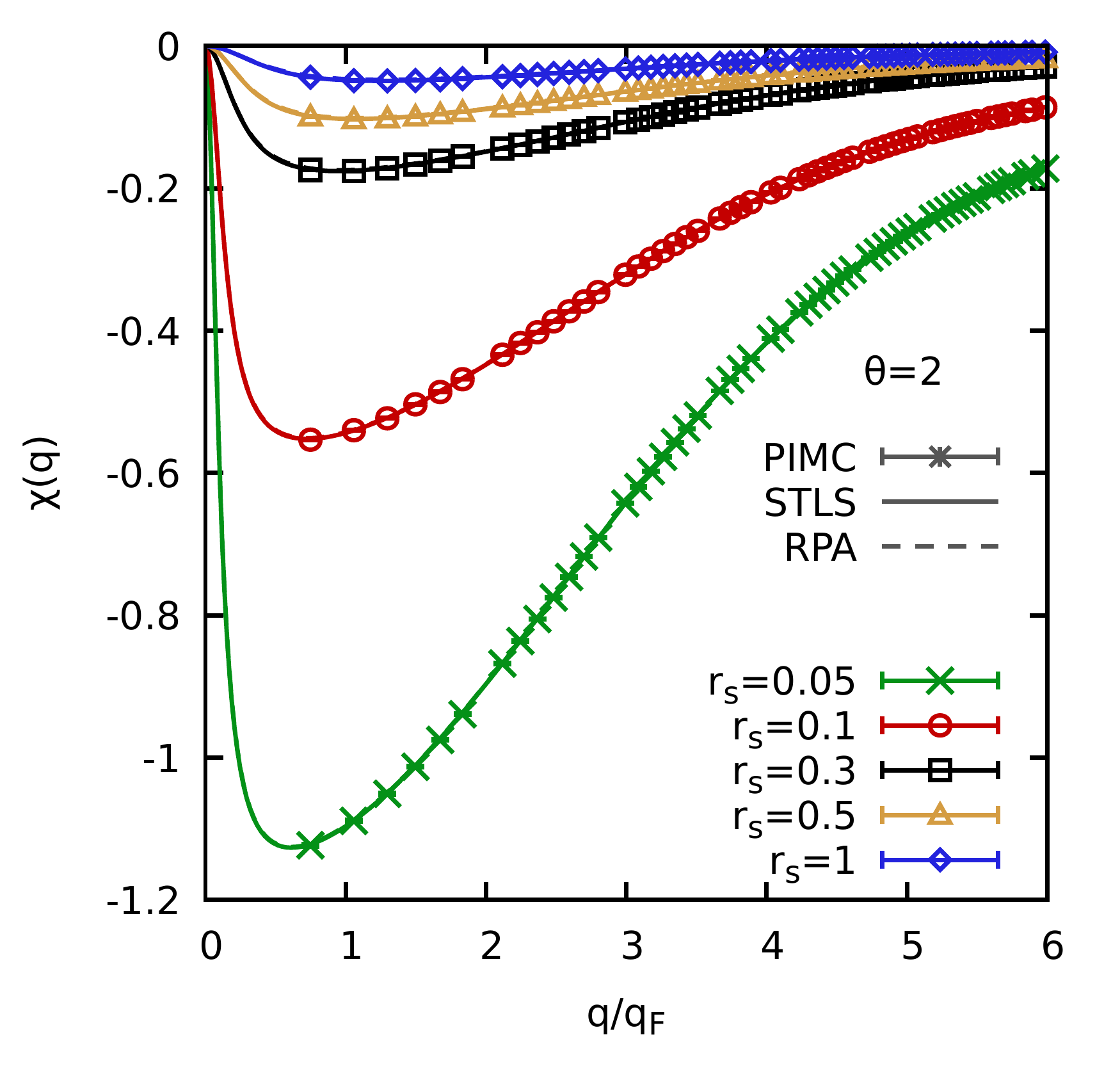}
\includegraphics[width=0.47\textwidth]{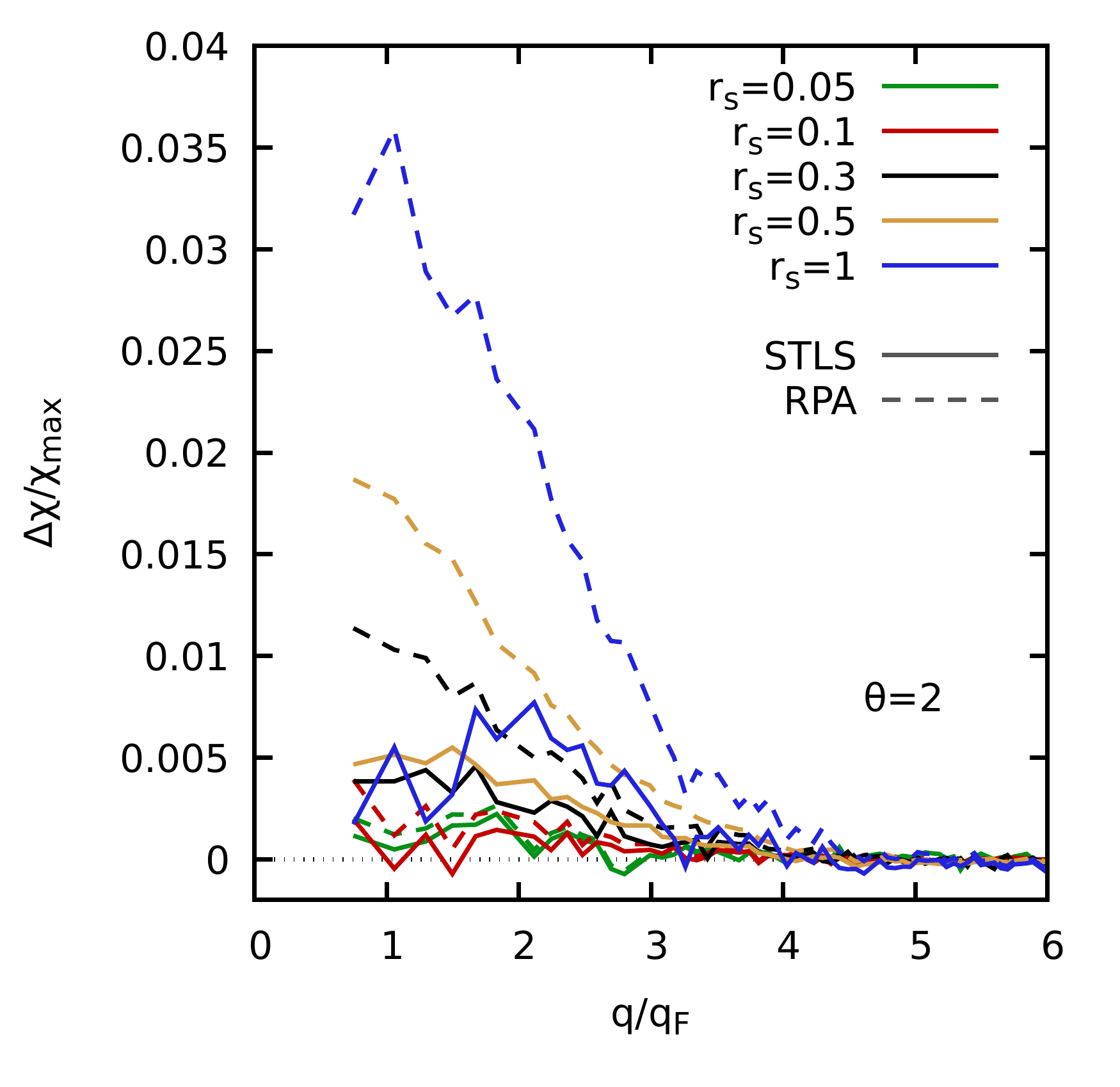}
\caption{\label{fig:CHI_panel_theta}
The static density response function $\chi(q)$ for $\theta=2$ and different $r_s$-values. The results for $r_s=1$ fall into the warm dense matter regime and have been taken from Ref.~\cite{dornheim_ML}.
}
\end{figure}  

Let us conclude the investigation of the static density response function of the UEG in the HED regime by considering its the dependence on the density parameter $r_s$. To this end, we plot $\chi(q)$ for five different $r_s$-values in Fig.~\ref{fig:CHI_panel_theta} at a constant value of the reduced temperature, $\theta=2$. Note that the $r_s=1$ results (blue diamonds) already fall into the WDM regime and have been taken from Ref.~\cite{dornheim_ML}. First and foremost, we note that the density response of the UEG to an external perturbation systematically increases towards small $r_s$. % \textcolor{red}{Insert a convincing explanation here!}.
This can be easily understood by considering the relation~\cite{moroni,dornheim_pre}
\begin{eqnarray}
\Delta n(\mathbf{r}) = 2 A \textnormal{cos}\left( \mathbf{r}\cdot\mathbf{q} \right) \chi(\mathbf{q}) \quad ,
\end{eqnarray}
which states that the induced density $\Delta n(\mathbf{r})$ is proportional to the perturbation strength $A$, with $\chi(\mathbf{q})$ being the pre-factor. Naturally, $\Delta n(\mathbf{r})$ scales with the unperturbed density $n$, and, consequently, so does $\chi(q)$.

Further, we observe that the deviations between dielectric theories and our PIMC data monotonically decrease towards high density, as it is expected, and are hardly significant at $r_s=0.1$ and $r_s=0.05$.

Another question that is interesting in its own right is the investigation of the static LFC in the HED regime. Let us first briefly revisit the issue of finite-size effects. While this was, at least in principle, already considered in Fig.~\ref{fig:LFC_rs0.3_theta4} for $\theta=4$ and $r_s=0.3$, there the relatively large Monte Carlo error bars made the plot less conclusive. A better example for the investigation of $G(q)$ is shown in Fig.~\ref{fig:real_LFC_rs0.5_theta2}, where the static LFC is plotted for $r_s=0.5$ and $\theta=2$ for three different system sizes. We note that these conditions are nearly optimal, as it combines a relatively large coupling strength with a manageable manifestation of the fermion sign problem, and we find $S\approx0.086$ even for $N=34$.

The green crosses, black squares, and purple triangles show finite-size corrected PIMC data for $G(q)$ for $N=34,20,$ and $14$, respectively, and are in excellent agreement with each other over the entire depicted $q$-range. In contrast, the grey triangles correspond to the uncorrected PIMC data for $N=14$ [i.e., by directly using $\chi_0(q)$ in the TDL in Eq.~(\ref{eq:Get_G})] and there appear substantial differences for all wave numbers.

The red line has been obtained using the STLS formalism and even qualitatively disagrees with the PIMC data everywhere. This is quite remarkable, as the density response function $\chi(q)$ from STLS is very accurate and constitutes a substantial improvements over RPA with a maximum deviation of $0.5\%$ around the Fermi wave number, cf.~Fig.~\ref{fig:CHI_panel_rs}. Let us now entangle this seeming contradiction in detail. Firstly, we note that STLS is known to strongly violate the compressibility sum-rule~\cite{stls2},
\begin{eqnarray}\label{eq:CSR}
\lim_{q\to0} G(q,0) = - \frac{q^2}{4\pi} \frac{\partial^2}{\partial n^2} \left( n f_\textnormal{xc} \right) \quad ,
\end{eqnarray}
and, thus, does not give the correct limit in $G(q)$ for small wave numbers. This, however, is of no consequence for $\chi(q)$ itself, as both RPA and STLS become exact for $q\to0$, cf.~Eq.~(\ref{eq:parabola}). At intermediate wave numbers (around $q=q_\textnormal{F}$), the static LFC from STLS is relatively close to the exact PIMC data, which leads to the substantial improvement over RPA regarding $\chi(q)$ observed in Fig.~\ref{fig:CHI_panel_rs}. Upon further increasing $q$, the quality of the LFC from STLS deteriorates (this is also discussed in Ref.~\cite{dornheim_electron_liquid}), but this is not necessarily reflected in $\chi(q)$, as the impact of $G(q)$ on the latter is suppressed by the $4\pi/q^2$ pre-factor in Eq.~(\ref{eq:define_LFC}).

Let us next discuss the qualitative behaviour of the PIMC data, which exhibit a broad maximum around $q\approx 3q_\textnormal{F}$, monotonically decrease thereafter, and even start to attain negative values for $q\gtrsim 5q_\textnormal{F}$. The same behavior has been observed in Ref.~\cite{dornheim_ML} for parts of the WDM regime, and was explained by the negative impact of exchange--correlation effects on the kinetic energy~\cite{militzer_kinetic,kraeft_kinetic} at these conditions.
In contrast, the STLS curve does not capture this feature and instead converges towards a constant value,
\begin{eqnarray}\label{eq:kim}
\lim_{q\to\infty} G(q) = 1 - g(0) \quad ,
\end{eqnarray}
with $g(0)$ being the pair correlation function at zero distance. More specifically, Eq.~(\ref{eq:kim}) follows from the electronic cusp condition~\cite{kimball}, and holds for \textit{static approximations}, i.e., theories where the frequency dependence in Eq.~(\ref{eq:define_LFC}) is neglected. Although our PIMC data for $G(q)$, too, are limited to the static case, they do not fall into this category, as the exact frequency dependence is built into the formalism via the propagation in the imaginary time.

Let us conclude this section with a more systematic comparison of our PIMC data for the static LFC to STLS.
In Fig.~\ref{fig:LFC_PANEL}, we show $G(q)$ for $r_s=0.5$ (left panel) and $r_s=0.3$ (right panel) for three different temperatures. Firstly, we note that the quality of our PIMC results deteriorates with increasing density, as the effect of $G(q)$ on $\chi(q)$ decreases. For this reason, we do not show results for $r_s=0.1$ and $r_s=0.05$, as the data are too noisy. Further, the PIMC data exhibit a maximum in the depicted $q$-range only for $\theta=1$, as this feature is shifted to larger wave numbers for higher temperatures~\cite{dornheim_ML}.
Finally, we find systematic errors in the STLS results for $G(q)$ exceeding $100\%$ even for $r_s=0.3$ and $\theta=4$, where the density response function, the static structure factor, and the interaction energy are extremely accurate. 
In a nutshell, this means that dielectic theories become exact in the high-density/-temperature limit not because of an improved description of the static LFC, but because the impact of the latter on physical observables simply vanishes.

\subsection{Comparison to CPIMC results\label{sec:CPIMC}}

\begin{figure}\centering
\includegraphics[width=0.47\textwidth]{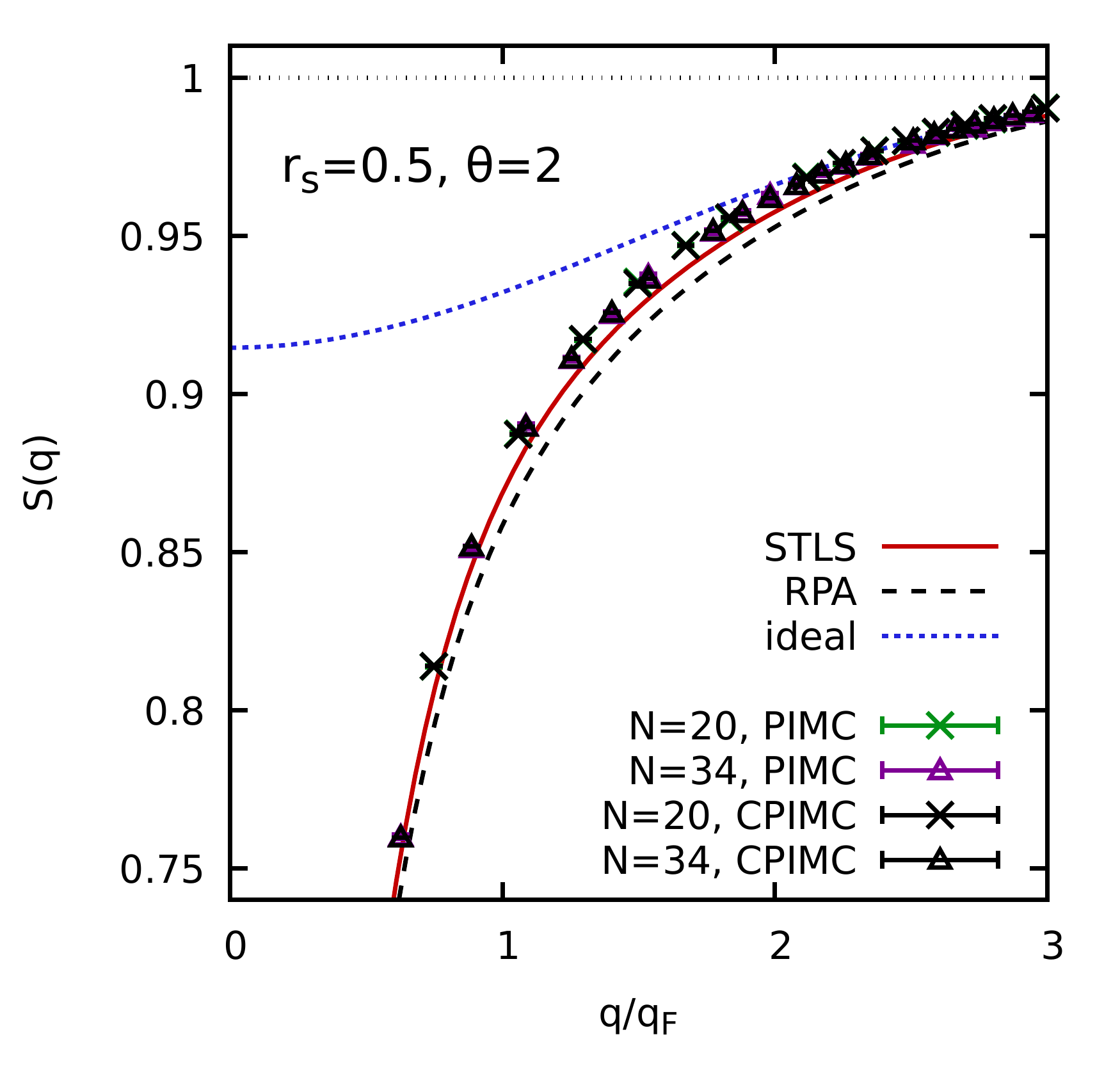}
\includegraphics[width=0.47\textwidth]{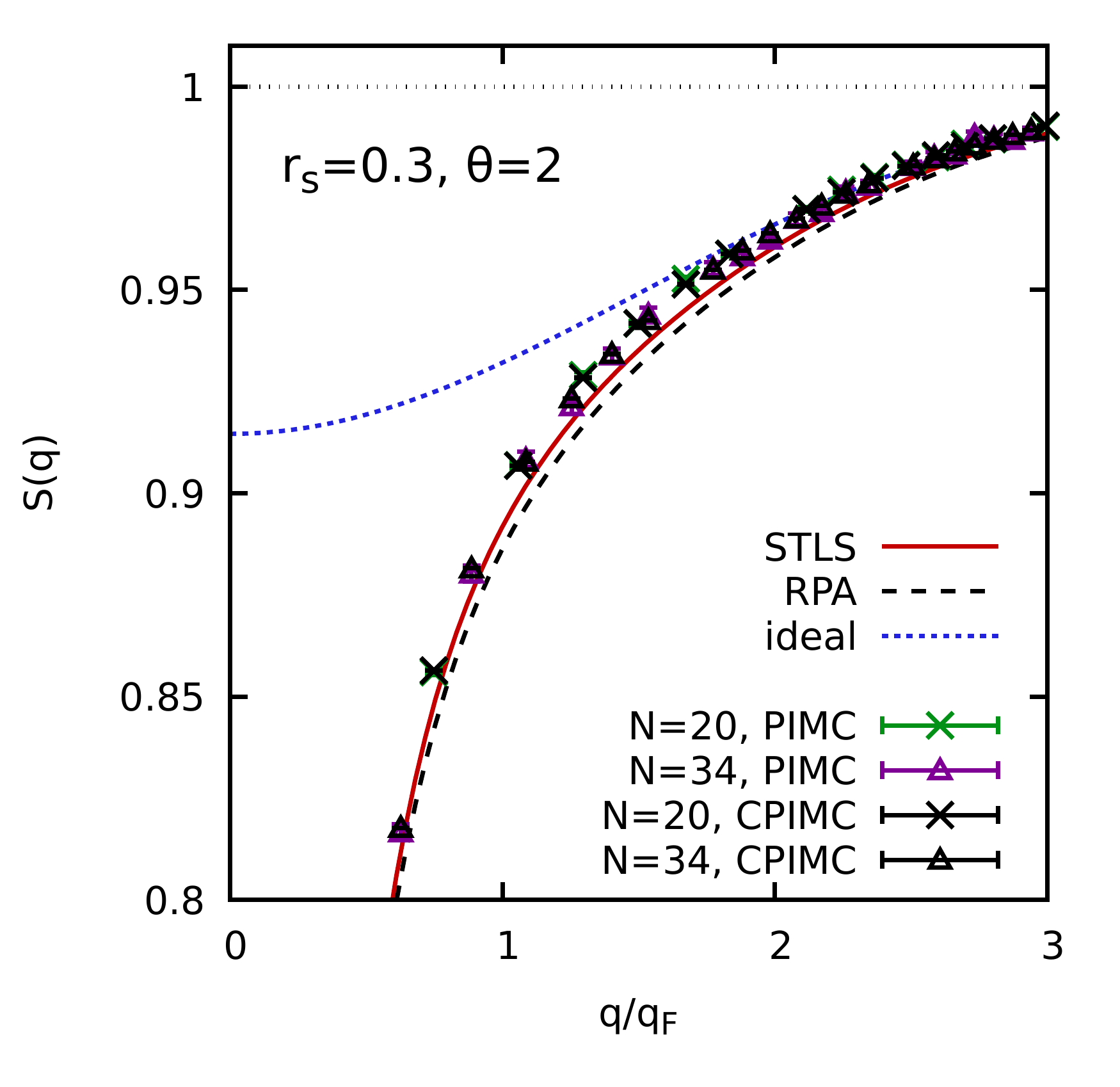}
\includegraphics[width=0.47\textwidth]{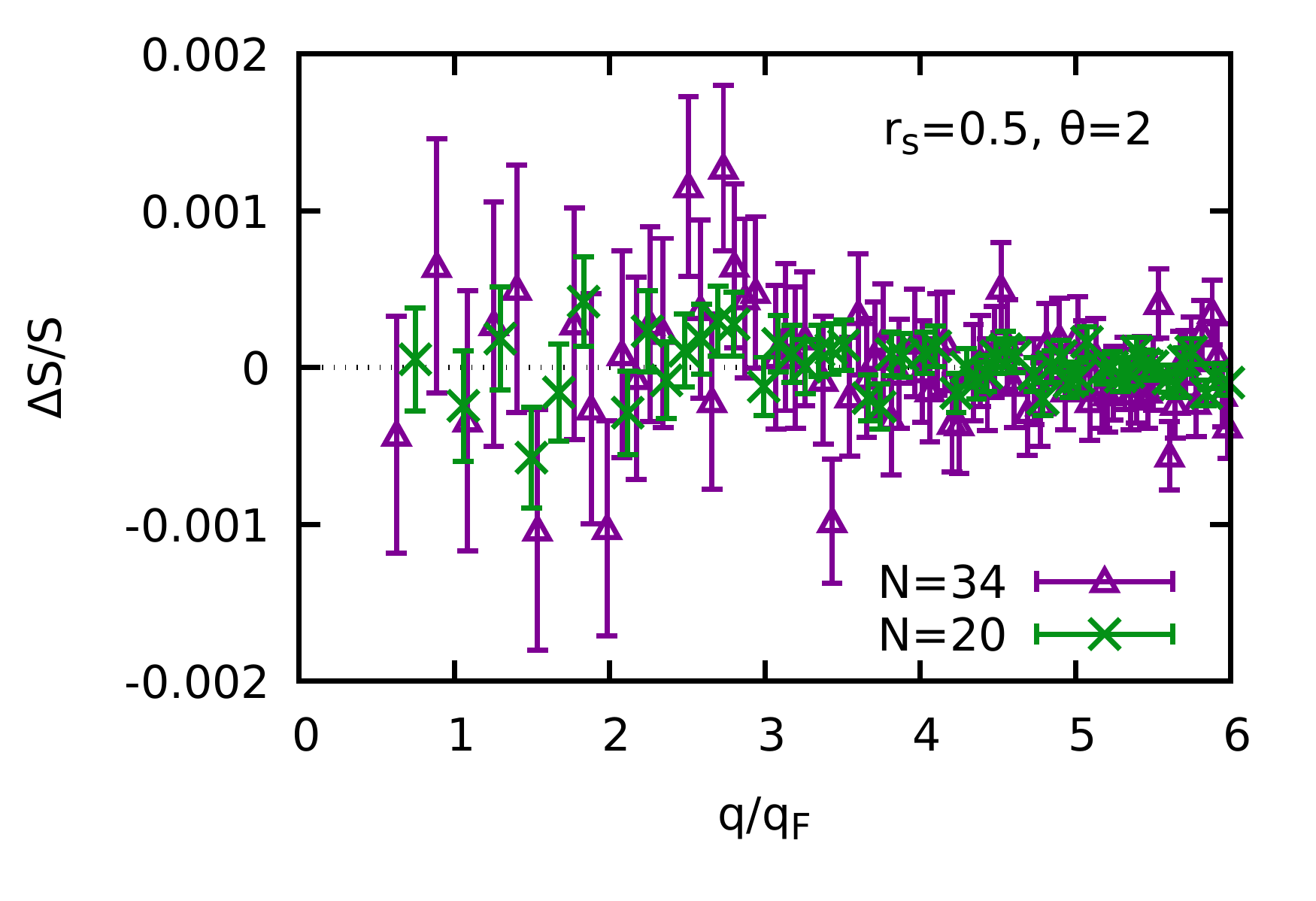}
\includegraphics[width=0.47\textwidth]{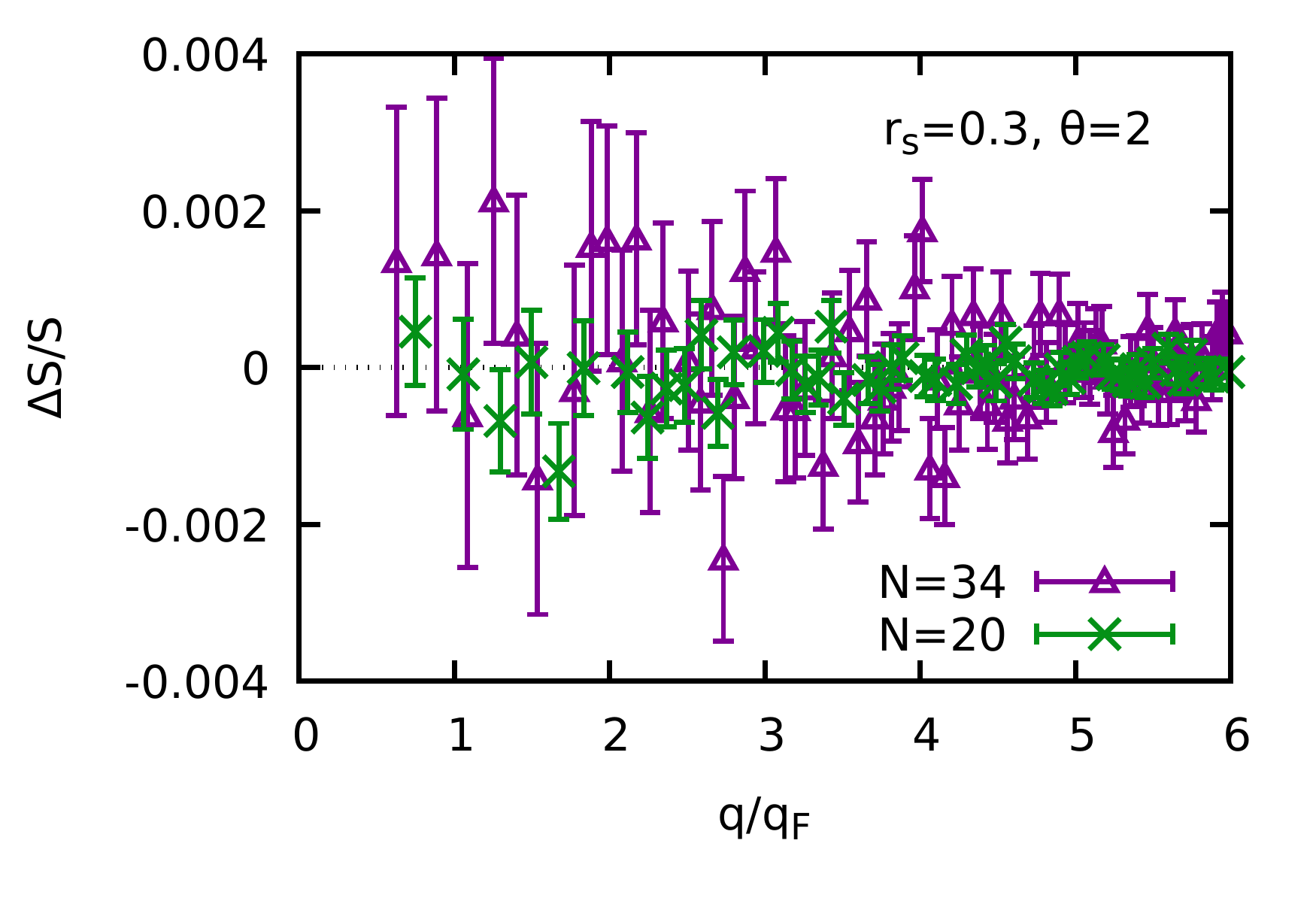}
\caption{\label{fig:Sq_CPIMC}
Comparison of PIMC (colored symbols) and CPIMC data (black symbols) for the static structure factor $S(q)$ at $\theta=2$ and $r_s=0.5$ (left) and $r_s=0.3$ (right). The crosses and triangles correspond to $N=20$ and $N=34$ unpolarized electrons. The bottom row shows the respective relative deviation between the two QMC methods.
}
\end{figure}

Let us conclude the discussion of simulation data with a comparison between the standard PIMC results and previous, highly accurate CPIMC data~\cite{dornheim_cpp,dornheim_prl}. 
This is shown in Fig.~\ref{fig:Sq_CPIMC} for the static structure factor $S(q)$ at $\theta=2$ and $r_s=0.5$ (left column) and $r_s=0.3$ (right column). Here, the colored symbols correspond to PIMC and the corresponding black symbols to CPIMC, and we show results for $N=20$ (crosses) and $N=34$ (triangles) unpolarized electrons. In both cases, the QMC data follow the same trend as RPA and STLS, whereas the ideal curve does not exhibit any screening effects and attains a finite value for $q=0$. More importantly, the respective PIMC and CPIMC data for the same $N$ cannot be distinguished within the given statistical uncertainty for all parameters. This can be seen particularly well in the bottom row, where we show the relative deviation between the data sets from the two different methods.

First and foremost, we stress the high accuracy of both data sets, with a statistical uncertainty of $\Delta S/S\sim10^{-4}$. Moreover, we report perfect agreement between the two independent methods, with no systematic deviations over the entire wave-number range.

\begin{figure}\centering
\includegraphics[width=0.47\textwidth]{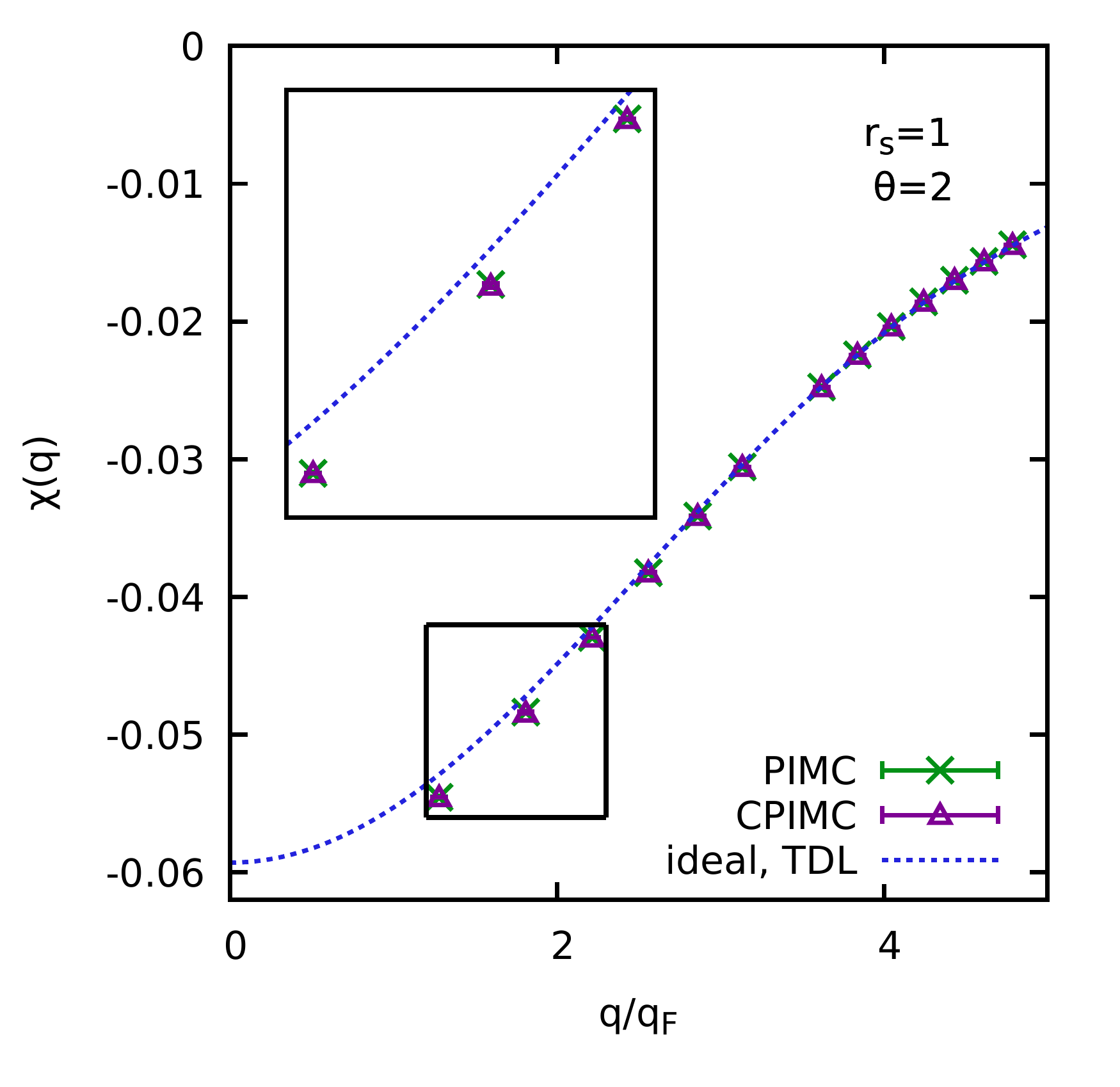}
\includegraphics[width=0.47\textwidth]{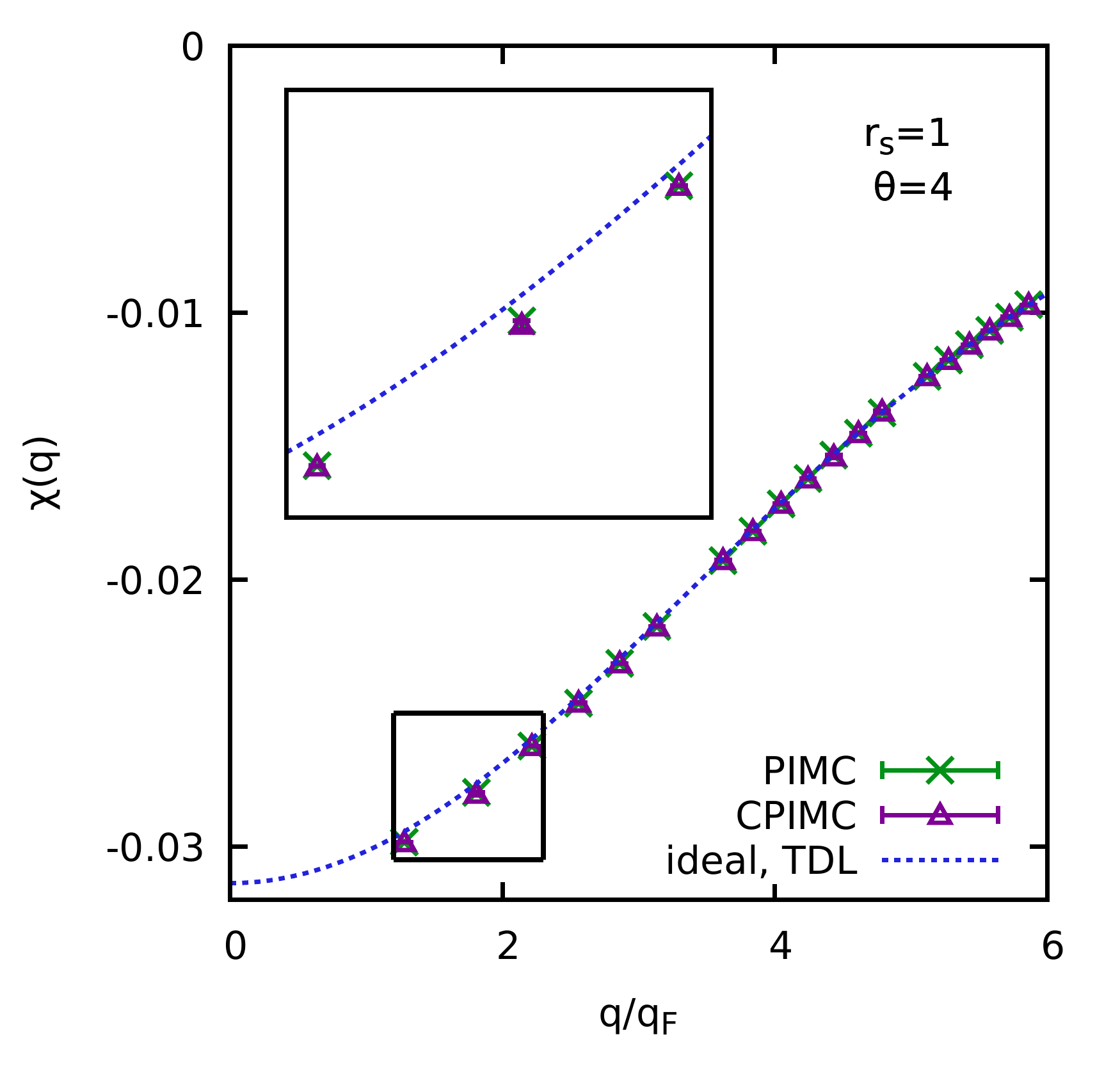}
\caption{\label{fig:CHI_CPIMC}
Comparison of PIMC and CPIMC data for the ideal density response function $\chi_0^N(q)$ for $N=4$ unpolarized electrons at $r_s=1$ and $\theta=2,4$.
}
\end{figure}

\begin{figure}\centering
\includegraphics[width=0.47\textwidth]{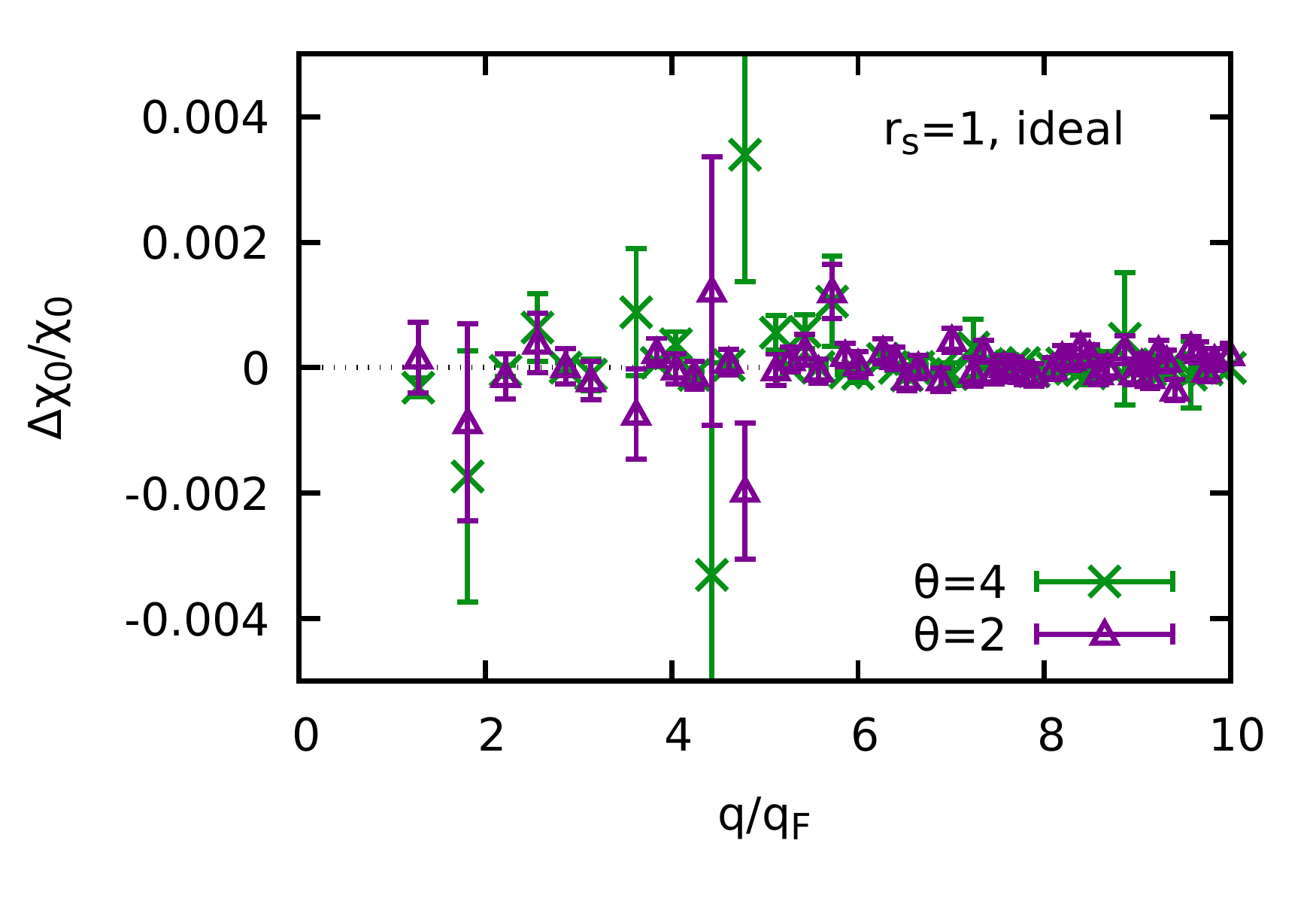}
\caption{\label{fig:DELTA_CHI_CPIMC}
Comparison of PIMC and CPIMC data for the ideal density response function $\chi_0^N(q)$ for $N=4$ unpolarized electrons at $r_s=1$ and $\theta=2,4$.
}
\end{figure}

In Fig.~\ref{fig:CHI_CPIMC}, we show a similar comparison, but for the static density response function $\chi_0^N(q)$ for $N=4$ unpolarized ideal fermions at $r_s=1$ and $\theta=2$ (left) and $\theta=4$ (right). Again, the agreement between PIMC (green crosses) and CPIMC (purple triangles) is perfect for all depicted data points, which constitutes a striking cross validation of our simulation methods: i) the simulations have been carried out by using two completely independent codes, and ii) the results for $\chi_0^N(q)$ have been obtained via different routes. For PIMC, we compute the imaginary-time density--density correlation function $F(q,\tau)$ and subsequently compute the density response function via integration, cf.~Eq.~(\ref{eq:static_chi}). Within CPIMC, on the other hand, we carry out a simulation with a finite but small twist-angle and subsequently insert the occupation numbers into a spectral representation. Yet, the two independent data sets agree with a relative accuracy of $\Delta\chi/\chi\sim10^{-4}-10^{-5}$, as can be seen in Fig.~\ref{fig:DELTA_CHI_CPIMC}.

\section{Summary and Outlook\label{sec:summary}}

In this work, we have presented extensive \textit{ab inito} path integral Monte Carlo data for the static density response of the uniform electron gas in the high-energy-density regime, $r_s\leq0.5$. This was achieved by computing exact results for the imaginary-time density--density correlation function $F(q,\tau)$, which, in turn, provides the full wave-number dependence of the density response from a single simulation of the unperturbed system. Since the fermion sign problem~\cite{dornheim_sign_problem} limits our simulations to relatively small particle numbers $N\leq 34$, we have applied finite-size corrections to our data and, in this way, removed any significant deviations to the respective thermodynamic limit.

Let us briefly summarize our key results as follows: i) we provide highly accurate benchmark data both for $\chi(q)$ and $G(q)$ for 24 density-temperature combinations in the range of $0.85\leq\theta\leq8$ and $0.05\leq r_s \leq 0.5$. These data are available online~\cite{GITHUB} and can be used to benchmark new many-body approximations~\cite{tanaka_hnc,panholzer1,arora}; ii) we have compared the new results against both RPA and STLS. In the HED regime, both dielectric methods provide accurate results with a maximum deviation in $\chi(q)$ of $\sim 4\%$ ($\sim1\%$) for RPA (STLS) at $\theta=0.85$ and $r_s=0.5$. In this way, we have bridged the gap between previous PIMC simulation results in the WDM regime and beyond~\cite{dornheim_ML,dynamic_folgepaper,dornheim_electron_liquid} and perturbative expansions like dielectric theories~\cite{review,stls,stls2}; iii) we have shown that our PIMC approach to the static density response converges towards RPA both for high temperatures and densities, and is in striking agreement to highly accurate CPIMC data, where they are available. This illustrates the consistency of our approach, and further corroborates our current understanding of the UEG at finite temperature.

In addition to being interesting in their own right, our simulation results can be used to extend previous parametrizations of the density response of the warm dense UEG~\cite{dornheim_ML} to also incorporate the previously unexplored high-density limit. This can be particularly important for the construction of advanced functionals for density functional theory simulations based on the adiabatic connection formula and the fluctuation--dissipation theorem~\cite{burke_ac,lu_ac,patrick_ac,goerling_ac}.

Moreover, we mention the possibility to use $F(q,\tau)$ as a starting point for the reconstruction of the dynamic structure factor $S(q,\omega)$, which can be used to benchmark and improve approximations from many-body theory~\cite{kwong,kas1,kas2,kas3}. Furthermore, PIMC simulations can be adapted to estimate the Matsubara Green function, which in turn gives access to the single-particle spectral function $A(q,\omega)$ and the momentum distribution $n(q)$.

Finally, it is possible to modify the PIMC method to include first-order relativistic corrections, e.g., Refs.~\cite{PRC,PLA}. Such simulations could be used to study the onset of relativistic effects of the UEG, which are not considered in current parametrizations~\cite{groth_prl,ksdt,karasiev_status}.

\section*{Acknowledgments}

   This work has been supported by the Norddeutscher Verbund f\"ur Hoch- und H\"ochstleistungsrechnen (HLRN) via grant shp00015 for computing time. 
   
   TD acknowledges support by the Center for Advanced Systems Understanding (CASUS) which is financed by Germany's Federal Ministry of Education and Research (BMBF) and by the Saxon Ministry for Science and Art (SMWK) with tax funds on the basis of the budget approved by the Saxon State Parliament.
   ZM acknowledges support by the Ministry of Education and Science of the Republic of Kazakhstan under Grant No. BR05236730.
   SG acknowledges support by the Deutsche Forschungsgemeinschaft (DFG) via grants BO1366/13 and BO1366/15.


\section*{References}
\begin{thebibliography}{10}




\bibitem{quantum_theory} G.~Giuliani and G.~Vignale, Quantum Theory of the Electron Liquid, Cambridge University Press (2008)

\bibitem{loos} P.-F.~Loos and P.M.W.~Gill, The uniform electron gas, \href{http://onlinelibrary.wiley.com/doi/10.1002/wcms.1257/abstract}{ \textit{Comput.~Mol.~Sci.}~\textbf{6}, 410-429 } (2016)



\bibitem{review} T.~Dornheim, S.~Groth, and M.~Bonitz, The uniform electron gas at warm dense matter conditions,
\href{https://www.sciencedirect.com/science/article/abs/pii/S0370157318300516}{\textit{Phys.~Reports}~\textbf{744}, 1--86} (2018)


\bibitem{mahan} G.D.~Mahan, Many-Particle Physics, Springer Science+Business Media, New York (2000)









\bibitem{wigner} E.~Wigner, On the interaction of electrons in metals, \href{https://journals.aps.org/pr/abstract/10.1103/PhysRev.46.1002}{\textit{Phys.~Rev.}~\textbf{46}, 1002} (1934)



\bibitem{drummond_wigner} N.D.~Drummond, Z.~Radnai, J.R.~Trail, M.D.~Towler, and R.J.~Needs, Diffusion quantum Monte Carlo study of three-dimensional Wigner crystals, \href{https://journals.aps.org/prb/abstract/10.1103/PhysRevB.69.085116}{\textit{Phys.~Rev.~B} \textbf{69}, 085116} (2004)

\bibitem{ichimaru_wigner} K.~Utsumi and S.~Ichimaru, Dielectric formulation of strongly coupled electron liquids at metallic densities. IV. Static properties in the low-density domain and the Wigner crystallization, \href{https://journals.aps.org/prb/abstract/10.1103/PhysRevB.24.3220}{\textit{Phys.~Rev.~B} \textbf{24}, 3220} (1981)














\bibitem{takada1} Y.~Takada and H.~Yasuhara, Dynamical Structure Factor of the Homogeneous Electron Liquid: Its Accurate Shape and the Interpretation of Experiments on Aluminum, \href{https://journals.aps.org/prl/abstract/10.1103/PhysRevLett.89.216402}{\textit{Phys.~Rev.~Lett.}~\textbf{89}, 216402} (2002)

\bibitem{takada2} Y.~Takada, Emergence of an excitonic collective mode in the dilute electron gas, \href{https://journals.aps.org/prb/abstract/10.1103/PhysRevB.94.245106}{\textit{Phys.~Rev.~B}~\textbf{94}, 245106} (2016)





\bibitem{dornheim_dynamic} T.~Dornheim, S.~Groth, J.~Vorberger, and M.~Bonitz, \textit{Ab initio} Path Integral Monte Carlo Results for the Dynamic Structure Factor of Correlated Electrons: From the Electron Liquid to Warm Dense Matter, \href{https://journals.aps.org/prl/abstract/10.1103/PhysRevLett.121.255001}{\textit{Phys.~Rev.~Lett.}~\textbf{121}, 255001} (2018)



\bibitem{dynamic_folgepaper} S.~Groth, T.~Dornheim, and J.~Vorberger, \textit{Ab Initio} Path Integral Monte Carlo Approach to the Static and Dynamic Density Response of the Uniform Electron Gas, \href{https://link.aps.org/doi/10.1103/PhysRevB.99.235122}{\textit{Phys.~Rev.~B} \textbf{99}, 235122} (2019)





















\bibitem{gs2} D.M.~Ceperley and B.J.~Alder, Ground State of the Electron Gas by a Stochastic Method, \href{ http://link.aps.org/doi/10.1103/PhysRevLett.45.566}{ \textit{Phys. Rev. Lett.} \textbf{45}, 566} (1980)


\bibitem{moroni2} S.~Moroni, D.M.~Ceperley, and G.~Senatore, Static Response and Local Field Factor of the Electron Gas, \href{ http://link.aps.org/doi/10.1103/PhysRevLett.75.689}{\textit{Phys.~Rev.~Lett.}~\textbf{75}, 689} (1995)


\bibitem{spink} G.G.~Spink, R.J.~Needs, and N.D.~Drummond, Quantum Monte Carlo study of the three-dimensional spin-polarized homogeneous electron gas, \href{https://journals.aps.org/prb/abstract/10.1103/PhysRevB.88.085121}{\textit{Phys.~Rev.~B}~\textbf{88}, 085121} (2013)

\bibitem{ortiz1} G.~Ortiz and P.~Ballone, Correlation energy, structure factor, radial distribution function, and momentum distribution of the spin-polarized uniform electron gas, \href{https://journals.aps.org/prb/abstract/10.1103/PhysRevB.50.1391}{\textit{Phys.~Rev.~B} \textbf{50}, 1391} (1994)

\bibitem{ortiz2} G.~Ortiz, M.~Harris, and P.~Ballone, Zero Temperature Phases of the Electron Gas, \href{https://journals.aps.org/prl/abstract/10.1103/PhysRevLett.82.5317}{\textit{Phys.~Rev.~Lett.}~\textbf{82}, 5317} (1999)






\bibitem{dft_review} R.O.~Jones, Density functional theory: Its origins, rise to prominence, and future, \href{ https://journals.aps.org/rmp/abstract/10.1103/RevModPhys.87.897 }{ \textit{Rev.~Mod.~Phys.}~\textbf{87}, 897 } (2015)

\bibitem{burke_perspective} K.~Burke, Perspective on density functional theory, \href{https://aip.scitation.org/doi/10.1063/1.4704546}{\textit{J.~Chem.~Phys.}~\textbf{136}, 150901} (2012)





\bibitem{fortov_review} V.E.~Fortov, Extreme states of matter on Earth and in space, \href{https://www.turpion.org/php/paper.phtml?journal_id=pu&paper_id=6821}{\textit{Phys.-Usp.}~\textbf{52}, 615--647} (2009)




















 \bibitem{knudson} M.D.~Knudson, M.P.~Desjarlais, R.W.~Lemke, T.R.~Mattsson, M.~French, N.~Nettelmann, and R.~Redmer, Probing the Interiors of the Ice Giants: Shock Compression of Water to $700$ GPa and $3.8$ g/cm$^3$, \href{https://journals.aps.org/prl/abstract/10.1103/PhysRevLett.108.091102}{ \textit{Phys.~Rev.~Lett.}~\textbf{108}, 091102} (2012) 

\bibitem{militzer3} F.~Soubiran, B.~Militzer, K.P.~Driver, and S.~Zhang, Properties of hydrogen, helium, and silicon dioxide mixtures
in giant planet interiors, \href{https://aip.scitation.org/doi/abs/10.1063/1.4978618}{ \textit{Phys.~Plasmas} \textbf{24}, 041401} (2017) 


\bibitem{manuel} M.~Sch\"ottler and R.~Redmer, Ab Initio Calculation of the Miscibility Diagram for Hydrogen-Helium Mixtures, \href{https://journals.aps.org/prl/abstract/10.1103/PhysRevLett.120.115703}{ \textit{Phys.~Rev.~Lett.}~\textbf{120}, 115703} (2018)













\bibitem{hu_ICF} S.X.~Hu, B.~Militzer, V.N.~Goncharov, and S.~Skupsky, \href{https://journals.aps.org/prb/abstract/10.1103/PhysRevB.84.224109}{\textit{Phys.~Rev.~B} \textbf{84}, 224109} (2011)









\bibitem{lcls1}  L.B.~Fletcher, H.J.~Lee, T.~D\"opppner, E.~Galtier, B.~Nagler, P.~Heimann, C.~Fortmann, S.~LePape, T.~Ma, M.~Millot, A.~Pak, D.~Turnbull, D.A.~Chapman, D.O.~Gericke, J.~Vorberger, T.~White, G.~Gregori, M.~Wei,
B.~Barbrel, R.W.~Falcone, C.-C.~Kao, H.~Nuhn, J.~Welch, U.~Zastrau, P.~Neumayer, J.B.~Hastings, and S.H.~Glenzer, Ultrabright X-ray laser scattering for dynamic warm dense matter physics, \href{https://www.nature.com/articles/nphoton.2015.41}{\textit{Nat.~Photonics} \textbf{9} 274-279} (2015) 



\bibitem{moses} E.I.~Moses, R.N.~Boyd, B.A.~Remington, C.J.~Keane, and R.~Al-Ayat, The National Ignition Facility: Ushering in a new age for high energy density science, \href{https://aip.scitation.org/doi/full/10.1063/1.3116505}{\textit{Phys.~Plasmas}~\textbf{16}, 041006} (2009)


\bibitem{xfel1} T.~Tschentscher, C.~Bressler, J.~Gr\"unert, A.~Madsen, A.P.~Mancuso, M.~Meyer, A.~Scherz, H.~Sinn, and U.~Zastrau,
Photon Beam Transport and Scientific Instruments at the European XFEL, \href{https://www.mdpi.com/2076-3417/7/6/592/htm}{\textit{Appl.~Sci.}~\textbf{7}, 592} (2017)














\bibitem{falk_wdm} K.~Falk, Experimental methods for warm dense matter research, \href{https://www.cambridge.org/core/journals/high-power-laser-science-and-engineering/article/experimental-methods-for-warm-dense-matter-research/7205AE1029BEA0061044F84875F1CEDB}{\textit{High Power Laser Sci. Eng.}~\textbf{6}, e59} (2018)




\bibitem{new_POP} M.~Bonitz, T.~Dornheim, Zh.A.~Moldabekov, S.~Zhang, P.~Hamann, A.~Filinov, K.~Ramakrishna, and J.~Vorberger, Ab initio simulation of warm dense matter, \href{https://arxiv.org/abs/1912.09884}{arxiv:1912.09884} 



\bibitem{wdm_book} F.~Graziani, M.P.~Desjarlais, R.~Redmer, and S.B.~Trickey (eds.), Frontiers and Challenges in Warm Dense Matter, Springer International Publishing (2014)












\bibitem{vanLeeuwen} G.~Stefanucci and R.~van Leeuwen, Nonequilibrium Many-Body Theory of Quantum Systems: A Modern Introduction, Cambridge University Press, Cambridge, United Kingdom, 2013

\bibitem{niclas} N.~Schl\"unzen, J.-P.~Joost, and M.~Bonitz, Achieving the Ultimate Scaling Limit for Nonequilibrium Green Functions Simulations, \textit{Phys.~Rev.~Lett.}~(in press)




\bibitem{kwong} N.-H.~Kwong and M.~Bonitz, Real-Time Kadanoff-Baym Approach to Plasma Oscillations in a Correlated Electron Gas, \href{https://journals.aps.org/prl/abstract/10.1103/PhysRevLett.84.1768}{\textit{Phys.~Rev.~Lett.}~\textbf{84}, 1768} (2000)


\bibitem{kas1} J.J.~Kas and J.J.~Rehr, Finite Temperature Green’s Function Approach for Excited State and Thermodynamic Properties of Cool to Warm Dense Matter, \href{https://journals.aps.org/prl/abstract/10.1103/PhysRevLett.119.176403}{\textit{Phys.~Rev.~Lett.}~\textbf{119}, 176403} (2017)

\bibitem{kas2} J.J.~Rehr and J.J.~Kas, Exchange and correlation in finite-temperature TDDFT, \href{https://link.springer.com/article/10.1140/epjb/e2018-90063-3}{\textit{Eur.~Phys.~J.~B}~\textbf{91}, 153} (2018)

\bibitem{kas3} T.S.~Tan, J.J.~Kas, and J.J.~Rehr, Coulomb-hole and screened exchange in the electron self-energy at finite temperature, \href{https://journals.aps.org/prb/abstract/10.1103/PhysRevB.98.115125}{\textit{Phys.~Rev.~B}~\textbf{98}, 115125} (2018)






\bibitem{wmc_review} W.M.C.~Foulkes, L.~Mitas, R.J.~Needs, and G.~Rajagopal, Quantum Monte Carlo simulations of solids, \href{https://journals.aps.org/rmp/abstract/10.1103/RevModPhys.73.33}{\textit{Rev.~Mod.~Phys.}~\textbf{73}, 33} (2001)




\bibitem{MD} D.~Perez, B.P.~Uberuaga, Y.~Shim, J.G.~Amar, and A.F.~Voter, Accelerated Molecular Dynamics Methods: Introduction and Recent Developments, \href{https://www.sciencedirect.com/science/article/pii/S1574140009005040}{\textit{Ann.~Rep.~Comp.~Chem.}~\textbf{5}, 79-98} (2009)





\bibitem{dornheim_pop} T.~Dornheim, S.~Groth, F.D.~Malone, T. Schoof, T.~Sjostrom, W.M.C.~Foulkes, and M.~Bonitz, \textit{Ab Initio}   Quantum Monte Carlo Simulation of the Warm Dense Electron Gas, \href{ http://aip.scitation.org/doi/full/10.1063/1.4977920}{ \textit{Phys.~Plasmas}  {\bf 24}, 056303} (2017)


\bibitem{brown_chapter} E.~Brown, M.A.~Morales, C.~Pierleoni, and D.~Ceperley, Quantum Monte Carlo Techniques and Applications for Warm Dense
Matter, in F. Graziani, M. P. Desjarlais, R. Redmer, and S. B.
Trickey (Eds.), \textit{Frontiers and Challenges in Warm Dense Matter},
Springer International Publishing (2014)


\bibitem{imada} M.~Takahashi and M.~Imada, Monte Carlo Calculation of Quantum Systems, \href{https://www.jstage.jst.go.jp/article/jpsj1946/53/3/53_3_963/_article/-char/ja/}{\textit{J.~Phys.~Soc.~Jpn.}~\textbf{53}, 963} (1984)

\bibitem{pollock} E.L.~Pollock and D.M.~Ceperley, Simulation of quantum many-body systems by path-integral methods, \href{https://journals.aps.org/prb/abstract/10.1103/PhysRevB.30.2555}{\textit{Phys.~Rev.~B} \textbf{30}, 2555} (1984)

\bibitem{cep} D.M.~Ceperley, Path integrals in the theory of condensed helium, \href{http://link.aps.org/doi/10.1103/RevModPhys.67.279}{\textit{Rev. Mod. Phys.} \textbf{67}, 279-355} (1995)













\bibitem{brown_ethan} E.W.~Brown, B.K.~Clark, J.L.~DuBois, and D.M.~Ceperley, Path-Integral Monte Carlo Simulation of the Warm Dense Homogeneous Electron Gas, \href{https://journals.aps.org/prl/abstract/10.1103/PhysRevLett.110.146405}{\textit{Phys.~Rev.~Lett.}~\textbf{110}, 146405} (2013)



\bibitem{schoof_prl} T.~Schoof, S.~Groth, J.~Vorberger, and M.~Bonitz, \textit{Ab Initio} Thermodynamic Results for the Degenerate Electron Gas at Finite Temperature, \href{https://journals.aps.org/prl/abstract/10.1103/PhysRevLett.115.130402}{\textit{Phys.~Rev.~Lett.}~\textbf{115}, 130402} (2015)


\bibitem{malone1} F.D.~Malone, N.S.~Blunt, J.J.~Shepherd, D.K.K.~Lee, J.S.~Spencer, and W.M.C.~Foulkes, Interaction picture density matrix quantum Monte Carlo, \href{https://aip.scitation.org/doi/abs/10.1063/1.4927434}{\textit{J.~Chem.~Phys.}~\textbf{143}, 044116} (2015)

\bibitem{malone2} F.D.~Malone, N.S.~Blunt, E.W.~Brown, D.K.K.~Lee, J.S.~Spencer, W.M.C.~Foulkes, and J.J.~Shepherd, Accurate Exchange-Correlation Energies for the Warm Dense Electron Gas, \href{https://journals.aps.org/prl/abstract/10.1103/PhysRevLett.117.115701}{\textit{Phys.~Rev.~Lett.}~\textbf{117}, 115701} (2016)





\bibitem{dornheim} T.~Dornheim, S.~Groth, A.~Filinov and M.~Bonitz, Permutation blocking path integral Monte Carlo: a highly efficient approach to the simulation of strongly degenerate non-ideal fermions, \href{ http://iopscience.iop.org/1367-2630/17/7/073017 }{ \textit{New J. Phys.} \textbf{17}, 073017} (2015)

\bibitem{dornheim2} T.~Dornheim, T.~Schoof, S.~Groth, A.~Filinov, and M.~Bonitz, Permutation Blocking Path Integral Monte Carlo Approach to the Uniform Electron Gas at Finite Temperature, \href{ http://scitation.aip.org/content/aip/journal/jcp/143/20/10.1063/1.4936145 }{ \textit{ J. Chem. Phys.} \textbf{143}, 204101} (2015)

\bibitem{groth} S.~Groth, T.~Schoof, T.~Dornheim, and M.~Bonitz, \textit{Ab Initio} Quantum Monte Carlo Simulations of the Uniform Electron Gas without Fixed Nodes, \href{ http://link.aps.org/doi/10.1103/PhysRevB.93.085102 }{ \textit{Phys.~Rev.~B} \textbf{93}, 085102} (2016)


\bibitem{dornheim3} T.~Dornheim, S.~Groth, T.~Schoof, C.~Hann, and M.~Bonitz, \textit{Ab initio} quantum Monte Carlo simulations of the Uniform electron gas without fixed nodes: The unpolarized case, \href{ http://link.aps.org/doi/10.1103/PhysRevB.93.205134 }{ \textit{Phys.~Rev.~B} \textbf{93}, 205134 } (2016)


\bibitem{dubois} J.L.~DuBois,  E.W.~Brown,  and  B.J.~Alder,  Overcoming the
Fermion Sign Problem in Homogeneous Systems, in E.~Schwegler,
B.M.~Rubenstein, and S.B.~Libby (Eds.),
Advances in the Computational Sciences-Symposium in Honor of Dr Berni Alder's
90th Birthday, World Scientific, Singapore (2017)




\bibitem{dornheim_prl} T.~Dornheim, S.~Groth, T.~Sjostrom, F.D.~Malone, W.M.C.~Foulkes, and M.~Bonitz, \textit{Ab Initio}   Quantum Monte Carlo Simulation of the Warm Dense Electron Gas in the Thermodynamic Limit, \href{ http://link.aps.org/doi/10.1103/PhysRevLett.117.156403}{\textit{Phys.~Rev.~Lett.}~\textbf{117}, 156403} (2016)









\bibitem{dornheim_neu} T.~Dornheim, S.~Groth, and M.~Bonitz, Permutation Blocking Path Integral Monte Carlo Simulations of Degenerate Electrons at Finite Temperature, \href{https://onlinelibrary.wiley.com/doi/full/10.1002/ctpp.201800157}{\textit{Contrib.~Plasma Phys.}~\textbf{59}, e201800157} (2019)



\bibitem{blunt} N.S.~Blunt, T.W.~Rogers, J.S.~Spencer, and W.M.C.~Foulkes, Density-matrix quantum Monte Carlo method, \href{https://journals.aps.org/prb/abstract/10.1103/PhysRevB.89.245124}{\textit{Phys.~Rev.~B} \textbf{89}, 245124} (2014)


\bibitem{booth} G.H.~Booth, A.J.W.~Thom, and A.~Alavi, Fermion Monte Carlo without fixed nodes: A game of life, death, and annihilation in Slater determinant space, \href{https://aip.scitation.org/doi/full/10.1063/1.3193710}{\textit{J.~Chem.~Phys.}~\textbf{131}, 054106} (2009)




\bibitem{schoof_cpp} T.~Schoof, M.~Bonitz, A.~Filinov, D.~Hochstuhl, and J.W.~Dufty, Configuration Path Integral Monte Carlo, \href{https://onlinelibrary.wiley.com/doi/abs/10.1002/ctpp.201100012}{\textit{Contrib.~Plasma Phys.}~\textbf{51}, 687-697} (2011)



\bibitem{metropolis} N.~Metropolis, A.W.~Rosenbluth, M.N.~Rosenbluth, A.H.~Teller, and E.~Teller, Equation of State Calculations by Fast Computing Machines, \href{https://aip.scitation.org/doi/abs/10.1063/1.1699114}{\textit{J.~Chem.~Phys.}~\textbf{21}, 1087} (1953)









\bibitem{groth_prl} S.~Groth, T.~Dornheim, T.~Sjostrom, F.D.~Malone, W.M.C.~Foulkes, and M.~Bonitz, \textit{Ab initio} Exchange--Correlation Free Energy of the Uniform Electron Gas at Warm Dense Matter Conditions, \href{https://journals.aps.org/prl/abstract/10.1103/PhysRevLett.119.135001}{\textit{Phys.~Rev.~Lett.}~\textbf{119}, 135001} (2017)










\bibitem{mermin_dft} N.D.~Mermin, Thermal Properties of the Inhomogeneous Electron Gas, \href{https://journals.aps.org/pr/abstract/10.1103/PhysRev.137.A1441}{\textit{Phys.~Rev.}~\textbf{137}, A1441} (1965)

\bibitem{rajagopal_dft} U.~Gupta and A.K.~Rajagopal, Density functional formalism at finite temperatures with some applications, \href{https://www.sciencedirect.com/science/article/pii/0370157382900771}{\textit{Phys.~Reports} \textbf{87}, 259-311} (1982)










\bibitem{kushal} An extensive study of the impact of the exchange-correlation functional from Ref.~\cite{groth_prl} on thermal DFT simulations of hydrogen is currently in preparation for publication.












\bibitem{kugler1} A.A.~Kugler, Theory of the Local Field Correction in an Electron Gas, \href{ http://link.springer.com/article/10.1007/BF01024183}{ \textit{J.~Stat.~Phys.}~\textbf{12}, 35} (1975)











\bibitem{Desjarlais:2017}
M.P.~Desjarlais, P.~Michael C.R.~Scullard, L.X.~Benedict, H.D.~Whitley, and R.~Redmer,
Density-functional calculations of transport properties in
the nondegenerate limit and the role of electron-electron scattering
\href{https://journals.aps.org/pre/abstract/10.1103/PhysRevE.95.033203}{\textit{Phys. Rev. E} \textbf{95}, 033203} (2017)


\bibitem{Veysman:2016} M.~Veysman, G.~R\"opke, M.~Winkel, and H.~Reinholz, Optical conductivity of warm dense matter within a wide
frequency range using quantum statistical and kinetic approaches
\href{https://journals.aps.org/pre/abstract/10.1103/PhysRevE.94.013203}{\textit{Phys. Rev. E},
  \textbf{94}, 013203} (2016)











% for construction of effective potentials

\bibitem{ceperley_potential} G.~Senatore, S.~Moroni, and D.M.~Ceperley, Local field factor and effective potentials in liquid metals, \href{https://www.sciencedirect.com/science/article/pii/S002230939600316X}{\textit{J.~Non-Cryst.~Sol.}~\textbf{205-207}, 851-854} (1996)


\bibitem{zhandos1} Zh.A.~Moldabekov, S.~Groth, T.~Dornheim, H.~K\"ahlert, M.~Bonitz, and T.S.~Ramazanov, Structural characteristics of strongly coupled ions in a dense quantum plasma, \href{https://journals.aps.org/pre/abstract/10.1103/PhysRevE.98.023207}{\textit{Phys.~Rev.~E} \textbf{98}, 023207} (2018)

\bibitem{zhandos2} Zh.A.~Moldabekov, H.~K\"ahlert, T.~Dornheim, S.~Groth, M.~Bonitz, and T.S.~Ramazanov, Dynamical structure factor of strongly coupled ions in a dense quantum plasma, \href{https://journals.aps.org/pre/abstract/10.1103/PhysRevE.99.053203}{\textit{Phys.~Rev.~E} \textbf{99}, 053203} (2019)







% utility for quantum hydrodynamics


\bibitem{diaw1} A.~Diaw and M.S.~Murillo, Generalized hydrodynamics model for strongly coupled plasmas, \href{https://journals.aps.org/pre/abstract/10.1103/PhysRevE.92.013107}{\textit{Phys.~Rev.~E} \textbf{92}, 013107} (2015)

\bibitem{diaw2} A.~Diaw and M.S.~Murillo, A viscous quantum hydrodynamics model based on dynamic density functional theory, \href{https://www.nature.com/articles/s41598-017-14414-9}{\textit{Sci.~Reports}~\textbf{7}, 15352} (2017)

\bibitem{zhandos_hydro} Zh.A.~Moldabekov, M.~Bonitz, and T.S.~Ramazanov, Theoretical foundations of quantum hydrodynamics for plasmas, \href{https://aip.scitation.org/doi/abs/10.1063/1.5003910}{\textit{Phys.~Plasmas} \textbf{25}, 031903} (2018)









% advanced functionals for DFT

\bibitem{burke_ac} A.~Pribram-Jones, P.E.~Grabowski, and K.~Burke, Thermal Density Functional Theory: Time-Dependent Linear Response and Approximate Functionals from the Fluctuation-Dissipation Theorem, \href{https://journals.aps.org/prl/abstract/10.1103/PhysRevLett.116.233001}{\textit{Phys.~Rev.~Lett.}~\textbf{116}, 233001} (2016)

\bibitem{lu_ac} D.~Lu, Evaluation of model exchange-correlation kernels in the adiabatic connection fluctuation-dissipation theorem for inhomogeneous systems, \href{https://aip.scitation.org/doi/10.1063/1.4867538}{\textit{J.~Chem.~Phys.}~\textbf{140}, 18A520} (2014)

\bibitem{patrick_ac} C.E.~Patrick and K.S.~Thygesen, Adiabatic-connection fluctuation-dissipation DFT for the structural properties of solids—The renormalized ALDA and electron gas kernels, \href{https://aip.scitation.org/doi/10.1063/1.4919236}{\textit{J.~Chem.~Phys.}~\textbf{143}, 102802} (2015)

\bibitem{goerling_ac} A.~G\"orling, Hierarchies of methods towards the exact Kohn-Sham correlation energy based on the adiabatic-connection fluctuation-dissipation theorem, \href{https://journals.aps.org/prb/abstract/10.1103/PhysRevB.99.235120}{\textit{Phys.~Rev.~B} \textbf{99}, 235120} (2019)






















\bibitem{dornheim_ML} T.~Dornheim, J.~Vorberger, S.~Groth, N.~Hoffmann, Zh.A.~Moldabekov, and M.~Bonitz, The Static Local Field Correction of the Warm Dense Electron Gas: An ab Initio Path Integral Monte Carlo Study and Machine Learning Representation, \href{https://aip.scitation.org/doi/full/10.1063/1.5123013}{\textit{J.~Chem.~Phys.}~\textbf{151}, 194104} (2019)




\bibitem{dornheim_electron_liquid} T.~Dornheim, T.~Sjostrom, S.~Tanaka, and J.~Vorberger, Strongly Coupled Electron Liquid: ab initio Path Integral Monte Carlo Simulations and Dielectric Theories, \textit{Phys.~Rev.~B} (in press)





















\bibitem{cdop} M.~Corradini, R.~Del Sole, G.~Onida, and M.~Palummo, Analytical Expressions for the Local-Field Factor $G(q)$ and the Exchange-Correlation Kernel ${K}_{\mathrm{xc}}(r)$ of the Homogeneous Electron Gas, \href{ http://link.aps.org/doi/10.1103/PhysRevB.57.14569}{ \textit{Phys.~Rev.~B} \textbf{57}, 14569} (1998)
















\bibitem{stls} S.~Tanaka and S.~Ichimaru, Thermodynamics and Correlational Properties of Finite-Temperature Electron Liquids in the Singwi-Tosi-Land-Sj\"olander Approximation, \href{ http://journals.jps.jp/doi/abs/10.1143/JPSJ.55.2278 }{ \textit{J.~Phys.~Soc.~Jpn.}~\textbf{55}, 2278-2289} (1986)


\bibitem{stls2} T.~Sjostrom, and J.~Dufty, Uniform Electron Gas at Finite Temperatures, \href{ http://link.aps.org/doi/10.1103/PhysRevB.88.115123 }{ \textit{Phys.~Rev.~B} \textbf{88}, 115123} (2013)




\bibitem{tanaka_hnc} S.~Tanaka, Correlational and thermodynamic properties of finite-temperature electron liquids in the hypernetted-chain approximation, \href{https://aip.scitation.org/doi/abs/10.1063/1.4969071}{\textit{J.~Chem.~Phys.}~\textbf{145}, 214104} (2016)










\bibitem{iyetomi_cdw} H.~Iyetomi, K.~Utsumi, and S.~Ichimaru, Dielectric formulation of strongly coupled electron liquids at metallic densities. V. Possibility of a charge-density-wave instability, \href{https://journals.aps.org/prb/abstract/10.1103/PhysRevB.24.3226}{\textit{Phys.~Rev.~B} \textbf{24}, 3226} (1981)


\bibitem{Gudmundsson} E.~H. Gudmundsson, C.~J. Pethick,, and R.~I. Epstein, Structure of neutron star envelopes, Astrophysical Journal {\bf 272}, 286 (1983).
 \bibitem{Potekhin} P. Haensel, A. Y. Potekhin, D. G. Yakovlev, Neutron Stars 1 Equation of State and Structure (Springer, New York 2007).

 \bibitem{Fortov_book} V. E. Fortov, \textit{Extreme States of Matter (High Energy Density Physics, Second Edition)} (Springer, Heidelberg, 2016).
 \bibitem{Atzeni} Stefano Atzeni, J\"urgen Meyer-Ter-Vehn, \textit{The Physics of Inertial Fusion Beam Plasma Interaction Hydrodynamics, Hot Dense Matter} (Oxford University Press, 2004).

\bibitem{Hu} S. X. Hu, B. Militzer, V. N. Goncharov, and S. Skupsky, Strong Coupling and Degeneracy Effects in Inertial Confinement Fusion Implosions, Phys. Rev. Lett. {\bf 104}, 235003 (2010)
 \bibitem{Boehly} T. R. Boehly et al. Initial performance results of the OMEGA laser system, Optics Communications {\bf 133}, 495–506 (1997). 
 \bibitem{Potekhin1999}  A.  Y.  Potekhin,  D.  A.  Baiko,  P.  Haensel,    and  D.  G. Yakovlev, Astronomy and Astrophysics {\bf 346}, 345 (1999).
\bibitem{Reinholz95}
Heidi Reinholz, Ronald Redmer, and Stefan Nagel, Thermodynamic  and transport  properties of dense hydrogen plasmas, Phys. Rev. E  {\bf 52}, 5368 (1995).
\bibitem{Reinholz2000} H. Reinholz, R. Redmer, G. R\"opke, and A. Wierling,
Long-wavelength limit of the dynamical local-field factor and dynamical conductivity of a two-component plasma, \textit{Phys. Rev. E} {\bf 2000}, 62, 5648.
\bibitem{Fortmann2010} C. Fortmann, A. Wierling, and G. R\"opke, Influence of local-field corrections on Thomson scattering in collision-dominated two-component plasmas, \textit{Phys. Rev. E} {\bf 2010}, 81, 026405.






\bibitem{pines} D.~Bohm and D.~Pines, A Collective Description of Electron Interactions: II. Collective vs Individual Particle Aspects of the Interactions, \href{https://journals.aps.org/pr/abstract/10.1103/PhysRev.85.338}{\textit{Phys.~Rev.}~\textbf{85}, 338} (1952)





\bibitem{stls_original} K.S.~Singwi, M.P.~Tosi, R.H.~Land, and A.~Sj\"olander, Electron Correlations at Metallic Densities, \href{ http://link.aps.org/doi/10.1103/PhysRev.176.589}{ \textit{Phys.~Rev.}~\textbf{176}, 589} (1968)










\bibitem{dornheim_cpp} T.~Dornheim, S.~Groth, and M.~Bonitz, Ab initio results for the Static Structure Factor of the Warm Dense Electron Gas, \href{https://onlinelibrary.wiley.com/doi/full/10.1002/ctpp.201700096}{\textit{Contrib.~Plasma Phys.}~\textbf{57}, 468-478} (2017)































\bibitem{chandler} D.~Chandler and P.G.~Wolynes, Exploiting the isomorphism between quantum theory and classical statistical mechanics of polyatomic fluids, \href{https://aip.scitation.org/doi/abs/10.1063/1.441588}{\textit{J.~Chem.~Phys.}~\textbf{74}, 4078} (1981)






\bibitem{dornheim_permutation_cycles} T.~Dornheim, S.~Groth, A.~Filinov, and M.~Bonitz, Path Integral Monte Carlo Simulation of Degenerate Electrons: Permutation-Cycle Properties, \href{https://aip.scitation.org/doi/10.1063/1.5093171}{\textit{J.~Chem.~Phys.}~\textbf{151}, 014108} (2019)














\bibitem{supersolid} S.~Saccani, S.~Moroni, and M.~Boninsegni, 
Excitation Spectrum of a Supersolid, \href{https://journals.aps.org/prl/abstract/10.1103/PhysRevLett.108.175301}{\textit{Phys.~Rev.~Lett.}~\textbf{108}, 175301} (2012)



\bibitem{dornheim_superfluid} T.~Dornheim, A.~Filinov, and M.~Bonitz, Superfluidity of strongly correlated bosons in two- and three-dimensional traps, \href{https://journals.aps.org/prb/abstract/10.1103/PhysRevB.91.054503}{\textit{Phys.~Rev.~B}~\textbf{91}, 054503} (2015)


\bibitem{dynamic_alex2} A.~Filinov, Correlation effects and collective excitations in bosonic bilayers: Role of quantum statistics, superfluidity, and the dimerization transition, \href{https://journals.aps.org/pra/abstract/10.1103/PhysRevA.94.013603}{\textit{Phys.~Rev.~A} \textbf{94}, 013603} (2016)






\bibitem{clark_casula} B.K.~Clark, M.~Casula, and D.M.~Ceperley, Hexatic and Mesoscopic Phases in a 2D Quantum Coulomb System, \href{https://journals.aps.org/prl/abstract/10.1103/PhysRevLett.103.055701}{\textit{Phys.~Rev.~Lett.}~\textbf{103}, 055701} (2009)

\bibitem{dornheim_analyzing} T.~Dornheim, H.~Thomsen, P.~Ludwig, A.~Filinov, and M.~Bonitz, Analyzing Quantum Correlations Made Simple, \href{https://onlinelibrary.wiley.com/doi/abs/10.1002/ctpp.201500120}{\textit{Contrib.~Plasma Phys.}~\textbf{56}, 371-379} (2016)





















\bibitem{boninsegni1} M.~Boninsegni, N.V.~Prokofev, and B.V.~Svistunov, Worm algorithm and diagrammatic Monte Carlo: A new approach to continuous-space path integral Monte Carlo simulations, \href{https://journals.aps.org/pre/abstract/10.1103/PhysRevE.74.036701}{\textit{Phys.~Rev.~E}~\textbf{74}, 036701} (2006)

\bibitem{boninsegni2} M.~Boninsegni, N.V.~Prokofev, and B.V.~Svistunov, Worm Algorithm for Continuous-Space Path Integral Monte Carlo Simulations, \href{https://journals.aps.org/prl/abstract/10.1103/PhysRevLett.96.070601}{\textit{Phys.~Rev.~Lett.}~\textbf{96}, 070601} (2006)











\bibitem{troyer} M.~Troyer and U.J.~Wiese, Computational Complexity and Fundamental Limitations to Fermionic Quantum Monte Carlo Simulations, \href{http://link.aps.org/doi/10.1103/PhysRevLett.94.170201}{\textit{Phys. Rev. Lett.} \textbf{94}, 170201} (2005)

\bibitem{loh} E.Y.~Loh, J.E.~Gubernatis, R.T.~Scalettar, S.R.~White, D.J.~Scalapino and R.L.~Sugar, Sign problem in the numerical simulation of many-electron systems, \href{http://link.aps.org/doi/10.1103/PhysRevB.41.9301}{\textit{Phys. Rev. B} \textbf{41}, 9301-9307} (1990)


\bibitem{dornheim_sign_problem} T.~Dornheim, Fermion sign problem in path integral Monte Carlo simulations: Quantum dots, ultracold atoms, and warm dense matter, \href{https://journals.aps.org/pre/abstract/10.1103/PhysRevE.100.023307}{\textit{Phys.~Rev.~E} \textbf{100}, 023307} (2019)

\bibitem{ceperley_fermions} D.M.~Ceperley, Path Integral Monte Carlo Methods for Fermions, \textit{Monte Carlo and Molecular Dynamics of Condensed Matter Systems}, K.~Binder and G.~Ciccotti (Eds.), Bologna (Italy) (1996)











\bibitem{mezza} F.~Mezzacapo and M.~Boninsegni, Structure, superfluidity, and quantum melting of hydrogen clusters, \href{https://journals.aps.org/pra/abstract/10.1103/PhysRevA.75.033201}{\textit{Phys.~Rev.~A} \textbf{75}, 033201} (2007)








\bibitem{bowen} G.~Sugiyama, C.~Bowen, and B.J.~Alder, Static dielectric response of charged bosons, \href{https://journals.aps.org/prb/abstract/10.1103/PhysRevB.46.13042}{\textit{Phys.~Rev.~B}~\textbf{46}, 13042} (1992)

\bibitem{moroni} S.~Moroni, D.M.~Ceperley, and G.~Senatore, 
Static response from quantum Monte Carlo calculations, \href{https://journals.aps.org/prl/abstract/10.1103/PhysRevLett.69.1837}{\textit{Phys.~Rev.~Lett.}~\textbf{69}, 1837} (1992)





\bibitem{bowen2} C.~Bowen, G.~Sugiyama, and B.J.~Alder, Static Dielectric Response of the Electron Gas, \href{ http://link.aps.org/doi/10.1103/PhysRevB.50.14838}{\textit{Phys.~Rev.~B} \textbf{50}, 14838} (1994)




\bibitem{neutron} M.~Buraczynski and A.~Gezerlis, Static Response of Neutron Matter, \href{https://journals.aps.org/prl/abstract/10.1103/PhysRevLett.116.152501}{\textit{Phys.~Rev.~Lett.}~\textbf{116}, 152501} (2016)








\bibitem{dornheim_pre} T.~Dornheim, S.~Groth, J.~Vorberger, and M.~Bonitz, Permutation Blocking Path Integral Monte Carlo approach to the Static Density Response of the Warm Dense Electron Gas, \href{https://journals.aps.org/pre/abstract/10.1103/PhysRevE.96.023203}{\textit{Phys.~Rev.~E}~\textbf{96}, 023203} (2017)

\bibitem{groth_jcp} S.~Groth, T.~Dornheim, and M.~Bonitz, Configuration Path Integral Monte Carlo approach to the Static Density Response of the Warm Dense Electron Gas, \href{https://aip.scitation.org/doi/abs/10.1063/1.4999907}{\textit{J.~Chem.~Phys.}~\textbf{147}, 164108} (2017)














\bibitem{siegfried_review} S.H.~Glenzer and R.~Redmer, X-ray Thomson scattering in high energy density plasmas, \href{https://journals.aps.org/rmp/abstract/10.1103/RevModPhys.81.1625}{\textit{Rev.~Mod.~Phys.}~\textbf{81}, 1625} (2009)


















\bibitem{berne1} D.~Thirumalai and B.J.~Berne, On the calculation of time correlation functions in quantum systems: Path integral techniques, \href{https://aip.scitation.org/doi/abs/10.1063/1.445597}{\textit{J.~Chem.~Phys.}~\textbf{79}, 5029} (1983)


\bibitem{berne2} E.~Gallicchio and B.J.~Berne, The absorption spectrum of the solvated electron in fluid helium by maximum entropy inversion of imaginary time correlation functions from path integral Monte Carlo simulations, \href{https://aip.scitation.org/doi/abs/10.1063/1.467892}{\textit{J.~Chem.~Phys.}~\textbf{101}, 9909} (1994)

\bibitem{boninsegni_ceperley} M.~Boninsegni and D.M.~Ceperley, Density fluctuations in liquid $^4$He. Path integrals and maximum entropy, \textit{J.~Low.~Temp.~Phys.}~\textbf{104}, 339-357 (1996)

\bibitem{dynamic_alex1} A.~Filinov and M.~Bonitz, Collective and single-particle excitations in two-dimensional dipolar Bose gases, \href{https://journals.aps.org/pra/abstract/10.1103/PhysRevA.86.043628}{\textit{Phys.~Rev.~A} \textbf{86}, 043628} (2012)





\bibitem{groth_thesis} S.~Groth,  Strongly Degenerate Nonideal Fermi Systems: Configuration Path Integral Monte Carlo Simulation, MSc.~thesis, Chrisitan-Albrechts-Universit\"at zu Kiel (2014)




















\bibitem{jarrell} M.~Jarrell and J.E.~Gubernatis, Bayesian inference and the analytic continuation of imaginary-time quantum Monte Carlo data, \href{https://www.sciencedirect.com/science/article/abs/pii/0370157395000747}{\textit{Phys.~Reports}~\textbf{269}, 133-195} (1996)

\bibitem{schoett} J.~Sch\"ott, E.G.C.P.~van Loon, I.L.M.~Locht, M.I.~Katsnelson, and I.~Di Marco, Comparison between methods of analytical continuation for bosonic functions, \href{https://journals.aps.org/prb/abstract/10.1103/PhysRevB.94.245140}{\textit{Phys.~Rev.~B}~\textbf{94}, 245140} (2016)









\bibitem{kugler_bounds} A.A.~Kugler, Bounds for Some Equilibrium Properties of an Electron Gas, \href{https://journals.aps.org/pra/abstract/10.1103/PhysRevA.1.1688}{\textit{Phys.~Rev.~A} \textbf{1}, 1688} (1970)
















\bibitem{militzer_kinetic} B.~Militzer and E.L.~Pollock, Lowering of the kinetic energy in interacting quantum systems, \href{https://journals.aps.org/prl/abstract/10.1103/PhysRevLett.89.280401}{\textit{Phys.~Rev.~Lett.}~\textbf{89}, 280401} (2002)

\bibitem{kraeft_kinetic} W.D.~Kraeft, M.~Schlanges, J.~Vorberger, and H.E.~DeWitt, Kinetic and correlation energies and distribution functions of dense plasmas, \href{https://journals.aps.org/pre/abstract/10.1103/PhysRevE.66.046405}{\textit{Phys.~Rev.~E} \textbf{66}, 046405} (2002)










\bibitem{kimball} J.C.~Kimball, Short-Range Correlations and Electron-Gas Response Functions, \href{https://journals.aps.org/pra/abstract/10.1103/PhysRevA.7.1648}{\textit{Phys.~Rev.~A} \textbf{7}, 1648} (1973)











\bibitem{GITHUB} A link to a Github repository containing all our PIMC data both for $\chi(q)$ and $G(q)$ will be made available upon publication.





\bibitem{panholzer1} M.~Panholzer, R.~Hobbiger, and H.~B\"ohm, Optimized correlations inspired by perturbation theory, \href{https://journals.aps.org/prb/abstract/10.1103/PhysRevB.99.195156}{\textit{Phys.~Rev.~B} \textbf{99}, 195156} (2019)



\bibitem{arora} P.~Arora, K.~Kumar, and R.K.~Moudgil, Spin-resolved correlations in the warm-dense homogeneous electron gas, \href{https://link.springer.com/article/10.1140/epjb/e2017-70532-y}{\textit{Eur.~Phys.~J.~B}~\textbf{90}, 76} (2017)











\bibitem{stls} S.~Tanaka and S.~Ichimaru, Thermodynamics and Correlational Properties of Finite-Temperature Electron Liquids in the Singwi-Tosi-Land-Sj\"olander Approximation, \href{ http://journals.jps.jp/doi/abs/10.1143/JPSJ.55.2278 }{ \textit{J.~Phys.~Soc.~Jpn.}~\textbf{55}, 2278-2289} (1986)


\bibitem{stls2} T.~Sjostrom, and J.~Dufty, Uniform Electron Gas at Finite Temperatures, \href{ http://link.aps.org/doi/10.1103/PhysRevB.88.115123 }{ \textit{Phys.~Rev.~B} \textbf{88}, 115123} (2013)













\bibitem{PRC} V.S.~Filinov, Yu.B.~Ivanov, V.E.~Fortov, M.~Bonitz, and P.R.~Levashov, Color path-integral Monte-Carlo simulations of quark-gluon plasma: Thermodynamic and transport properties, \href{https://journals.aps.org/prc/abstract/10.1103/PhysRevC.87.035207}{\textit{Phys.~Rev.~C} \textbf{87}, 035207} (2013)


\bibitem{PLA} V.S.~Filinov, Yu.B.~Ivanov, M.~Bonitz, V.E.~Fortov, and P.R.~Levashov, Color path-integral Monte Carlo simulations of quark–gluon plasma, \href{https://www.sciencedirect.com/science/article/abs/pii/S0375960112001156}{\textit{Phys.~Lett.~A} \textbf{376}, 1096-1101} (2012)












\bibitem{ksdt} V.V.~Karasiev, T.~Sjostrom, J.W.~Dufty, and S.B.~Trickey, Accurate Homogeneous Electron Gas Exchange-Correlation Free Energy for Local Spin-Density Calculations, \href{https://journals.aps.org/prl/abstract/10.1103/PhysRevLett.112.076403}{\textit{Phys.~Rev.~Lett.}~\textbf{112}, 076403} (2014)


\bibitem{karasiev_status} V.V.~Karasiev, S.B.~Trickey, and J.W.~Dufty, Status of free-energy representations for the homogeneous electron gas, \href{https://journals.aps.org/prb/abstract/10.1103/PhysRevB.99.195134}{\textit{Phys.~Rev.~B} \textbf{99}, 195134} (2019)




















\end{thebibliography}
\end{document}